\documentclass[11pt]{article}
\usepackage{graphicx} % Required for inserting images
\usepackage{amsmath}
\usepackage{lipsum}
\usepackage{pdfcomment}

\usepackage[table,xcdraw]{xcolor} % Enable table colors and HTML color codes
\usepackage{subcaption}
\usepackage[utf8]{inputenc}
\usepackage{longtable}
\usepackage{graphicx}
\usepackage{float}
\usepackage{booktabs}
\usepackage{float} 
\usepackage{geometry}
\usepackage{lineno}
\usepackage{authblk}
\usepackage{tcolorbox}
\usepackage{rotating}
\usepackage{booktabs}
\usepackage{threeparttable}
\usepackage{geometry} % Optional: for full page layout control
\usepackage{lscape}
\usepackage{siunitx}  % For aligning numbers on the decimal point
\usepackage{adjustbox} % To adjust table width if needed
\usepackage{pgfplots}
\usepackage{caption}
\usepgfplotslibrary{groupplots}

\usepackage[numbers,sort&compress]{natbib}

\usepackage{hyperref}
\hypersetup{
  colorlinks=true,
  linkcolor=blue,       % 章节跳转链接（如目录、交叉引用）
  citecolor=blue,       % 文献引用
  urlcolor=blue         % 网页链接
}

\usepackage{cite}

\usepackage{array}
\usepackage{multirow}

\usepackage{amssymb}
\usepackage{pifont}

\usepackage[colorinlistoftodos]{todonotes}

\usepackage{wrapfig}

\usepackage{cite}

\usepackage{amsthm}
\usepackage{titlesec}
\usepackage{titletoc}
\titleclass{\subsubsubsection}{straight}[\subsection]
\newcounter{subsubsubsection}[subsubsection]
\renewcommand\thesubsubsubsection{\thesubsubsection.\arabic{subsubsubsection}}
\titleformat{\subsubsubsection}{\normalfont\normalsize\bfseries}{\thesubsubsubsection}{1em}{}
\titlespacing*{\subsubsubsection}{0pt}{3.25ex plus 1ex minus .2ex}{1.5ex plus .2ex}
\titleformat*{\section}{\fontsize{11}{12}\selectfont\bfseries}
\titleformat*{\subsection}{\fontsize{11}{12}\selectfont\bfseries}
\titleformat*{\subsubsection}{\fontsize{11}{12}\selectfont\bfseries}
\titleformat*{\paragraph}{\fontsize{11}{12}\selectfont\bfseries}
\titleformat*{\subparagraph}{\fontsize{11}{12}\selectfont\bfseries}

\titlecontents{subsubsubsection}
[8em] % adjust this value to fit your needs
{\small}
{\hyperlink{toc.\thecontentslabel}{\thecontentslabel.} }
{}
{\ \titlerule*[.5pc]{.}\contentspage}

\titlespacing*{\subsubsubsection}{0pt}{3.25ex plus 1ex minus .2ex}{1.5ex plus .2ex}

\usepackage{caption}
\usepackage{microtype}

\usepackage{enumitem}
\setlist{itemsep=0em}

\usepackage{fancyhdr} % For custom headers/footers
\usepackage{lastpage} % For total number of pages

\usepackage{listings}
\usepackage{xcolor}

\definecolor{codegreen}{rgb}{0,0.6,0}
\definecolor{codegray}{rgb}{0.5,0.5,0.5}
\definecolor{codepurple}{rgb}{0.58,0,0.82}
\definecolor{backcolour}{rgb}{0.95,0.95,0.92}

\lstdefinestyle{mystyle}{
    backgroundcolor=\color{backcolour},   
    commentstyle=\color{codegreen},
    keywordstyle=\color{magenta},
    numberstyle=\tiny\color{codegray},
    stringstyle=\color{codepurple},
    basicstyle=\ttfamily\footnotesize,
    breakatwhitespace=false,         
    breaklines=true,                 
    captionpos=b,                    
    keepspaces=true,                 
    numbers=left,                    
    numbersep=5pt,                  
    showspaces=false,                
    showstringspaces=false,
    showtabs=false,                  
    tabsize=2
}

\title{SYNTHESIS OF SERVICE LIFE PREDICTION FOR BRIDGES IN TEXAS}

\author[1]{Lu Gao, Ph.D.}
\author[1]{Yi-Lung Mo, Ph.D.}
\author[1]{Shalaka Dhonde}
\author[1]{Daisy Saldarriaga}
\author[1]{Lingguang Song, Ph.D.}
\author[1]{Ahmed Senouci, Ph.D.}

\affil[1]{Department of Civil and Environmental Engineering, University of Houston}
\date{}

\begin{document}
\maketitle    

\section*{Abstract}
In procurement requirements for design-build project contracts for bridge structures, the Texas Department of Transportation (TxDOT) may implement a 100-year service life requirement. However, there are no indicated measures or any technical recommendations that provide directions to satisfy the given requirement of service life. In addition, TxDOT and consultants use TxDOT recommendations for durability to improve performance during service life of design-bid-build and design-build projects but no quantitative methods or codified guidance is available to validate how the enhanced service life requirements are met. Further, the state of Texas has large number of existing old bridge thus the evaluation of the remaining service life of these bridges is a very important economic issue for TxDOT. The replacement of all these bridges is not possible since the available financial resources are limited. Therefore, it is very essential to prioritize the repair works based on the estimated remaining service life. As a result, this research study has been conducted to obtain information about state-of-the-art and state-of practice of bridge service life prediction. The research team gathered and analyzed the relevant information on various topics related to service life prediction which can be utilized as guidelines while dealing with the determination of service life of old as well as new bridges.

The extensive literature survey conducted by the research team provides valuable information for TxDOT which can be helpful for service life prediction of bridges in the state of Texas. By utilizing the collected information under the scope of the project, the following benefits can be achieved: 1) This project would provide guidance on managing available funds efficiently for the required repair activities using the data on condition of the bridges. 2). The review of the available information obtained from different sources would provide better understanding of various deterioration models used for predicting service life, inspection checks and methods, maintenance practices and rehabilitation or replacement requirements for bridges, and would enhance the knowledge on achieving and extending the service life of bridges in Texas.  3). The output from this research project would be beneficial for maintaining the existing bridges in a good condition, improving their service life to make them economically efficient and determining strategies to achieve design service life for newly constructed bridges.

\section{Introduction}
The state of Texas has over 50,000 bridges, of which a large number are existing old bridges. The evaluation of the remaining service life of these bridges is a very important economic issue for the Texas Department of Transportation (TxDOT). Therefore, it is necessary to prioritize the repair works based on the estimated remaining service life. The research team has conducted a thorough review of the state-of-the-art and state-of-practice of bridge service life prediction models and the results are summarized in this report. The purpose of this report is to provide TxDOT with a comprehensive summary of prediction model to estimate service life of bridges. Chapter 2 and 3 contain information regarding the durability tests for concrete and steel bridges respectively. Chapter 4 presents empirical models for predicting bridge service life, including a total of 20 models used by other states and countries. A preliminary machine learning modeling of NBI data collected in Texas is also presented in Chapter 4. Chapter 5 presents mechanistic models for predicting bridge service life, including corrosion, carbonation, sulfate attack, freeze-thaw, alkali silica reaction, and fatigue. Corrosion control methods are presented in Chapter 6 and include mechanical barrier and electrochemical methods, as well as a review of alternative materials resistant to corrosion. Chapter 7 contains a review of 12 reports regarding service life prediction models for old bridges. Chapter 8 presents service life prediction for newly constructed bridge under design-build contracts. Chapter 9 presents a review of remaining service life verification at handoff.

\section{Durability Related Tests for Concrete Bridge}

In the United States, the most common Non-Destructive Tests (NDT) includes Acoustic Emission, Electrical Resistivity Method, Delamination Detection Machinery, Ground Penetrating Radar, Electromagnetic Methods, Impact-Echo Testing, Infrared Thermography, Ultrasonic Pulse Echo Testing, Magnetic Method, Neutron Probe, Nuclear Method, Pachometer, Smart Concrete and Rebound and Penetration. \citet{gucunski2013nondestructive} graded the performance of each NDT technology for four deterioration types: delamination, corrosion, cracking and concrete deterioration. The final grade values obtained for each NDT method for all deterioration types are showed in the following table. The authors concluded that even though some technologies showed potential for detecting and assessing each deterioration type, no single technology can potentially evaluate all deterioration types. Three technologies were identified as having a good potential for corrosion detection (half-cell potential, electrical resistivity, and galvanostatic pulse measurement) and six technologies were identified as having a good potential for delamination detection (impact echo, ultrasonic pulse echo, impulse response, chain dragging and hammer sounding, ground penetrating radar and infrared thermography).

\begin{table}[H]
	\caption{Overall Value and Ranking of NDT Technologies \citep{gucunski2013nondestructive}}
	\label{fig:table1}	
	\centering
	\includegraphics[width=0.7\linewidth]{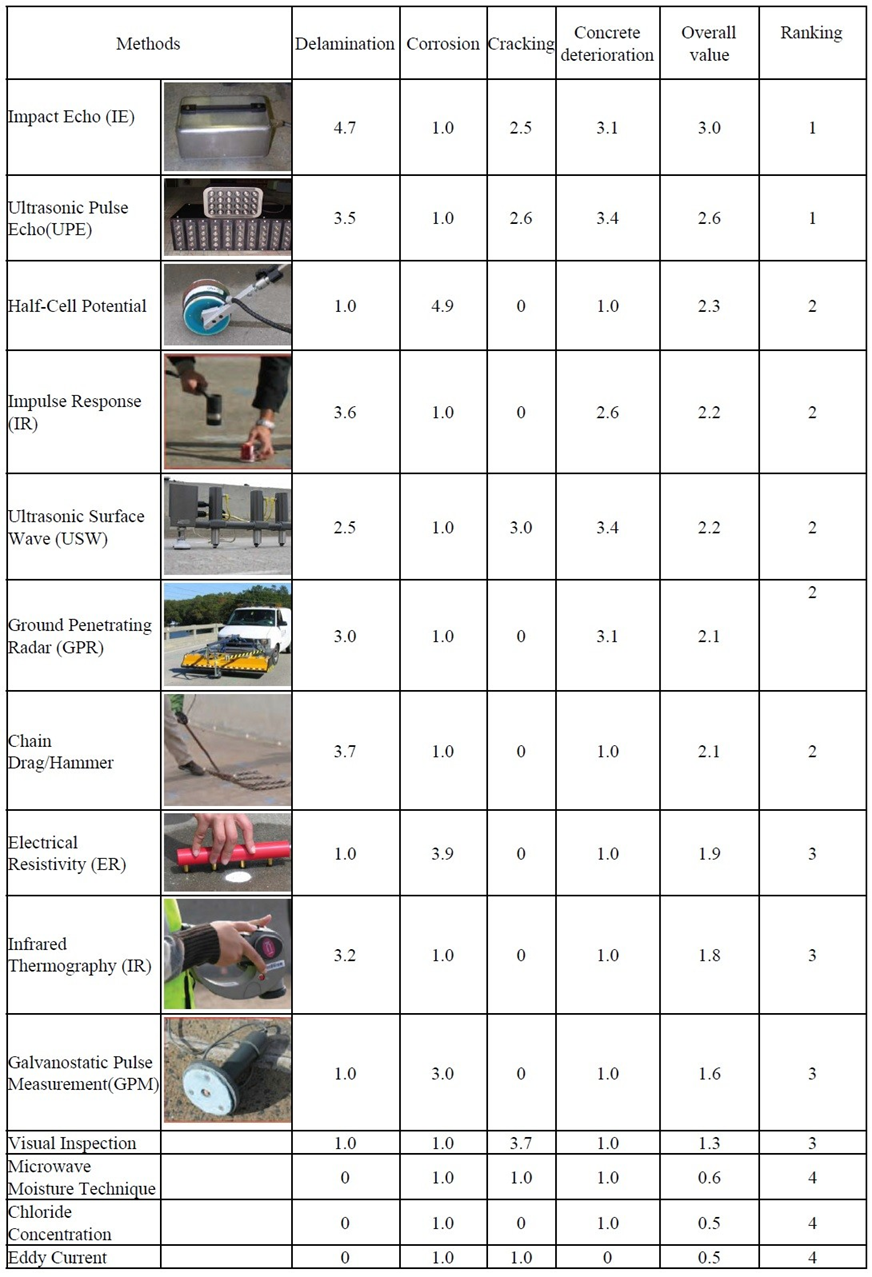}
\end{table}

The following table shows the accuracies of the measured parameters in the service-life estimation of reinforced concrete bridges. These accuracies were collected from different sources in the existing literature \citep{barnes2008research,hasan2016experimental,gudimettla2015field,helal2015non,lo2002curing,her2014non}.

\begin{table}[H]
	\caption{Service life estimation parameters to measure and NDT tests and accuracy}
	\label{fig:table2}	
	\centering
	\includegraphics[width=0.7\linewidth]{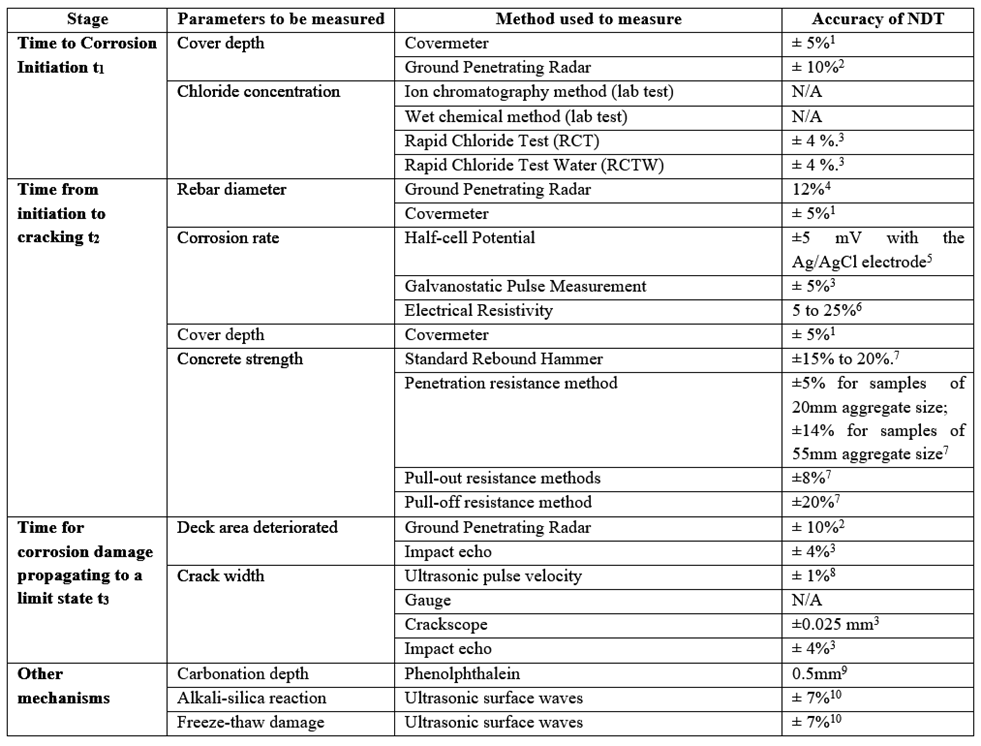}
\end{table}

\subsection{Concrete Condition}
\subsubsection{Surface Defects  (e.g., Crack, Delamination, Spalling)}
\paragraph{Impact-Echo Testing}
The impact echo (IE) method is a seismic or stress wave–based method used in the detection of defects in concrete, especially delamination \citep{sansalone1989detecting}. The main purpose of impact echo testing is to detect and characterize wave reflectors or “resonators” in a concrete bridge deck. This is achieved by striking the surface of the tested object and measuring the response at a nearby location. The following figure shows the application of the impact echo testing technology to a bridge deck.

\begin{figure}[H]
	\centering
	\includegraphics[width=0.7\linewidth]{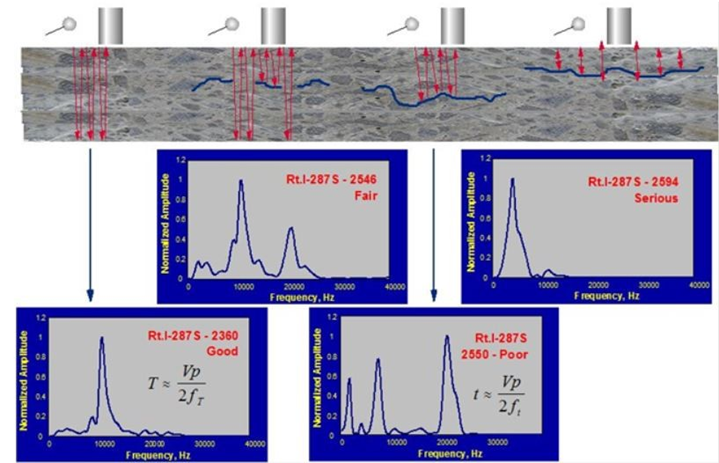}
	\caption{Physical Principle of IE \citep{gucunski2013nondestructive}}
	\label{fig:figure1}
\end{figure}

Applications:
\begin{itemize}
	\item Characterization of surface-opening cracks (vertical cracks in bridge decks).
	\item Detection of ducts, voids in ducts, and rebars.
	\item Material characterization. 
\end{itemize}

Limitations:
\begin{itemize}
	\item In case of a deck with asphalt concrete overlay, detection of delamination is possible only when the asphalt concrete temperature is sufficiently low so that the material is not highly viscous, or when the overlay is intimately bonded to the deck.
	\item It is necessary to conduct data collection on a very dense test grid to define the boundary conditions for delaminated areas accurately. 
\end{itemize}

\paragraph{Ground Penetrating Radar (GPR)}

Ground Penetrating Radar is one of the rapid NDT methods which uses electromagnetic (EM) waves to locate objects buried inside the structure and produce contour maps of subsurface features such as steel reinforcement, wire mesh, etc. In this type of NDT method, GPR antenna transmits high EM waves into bridge deck and then a portion of energy is reflected back to the surface from reflector present inside. This energy is further received by antenna. GPR uses electromagnetic waves to locate objects buried inside the structure and to produce contour maps of subsurface features (steel reinforcements, wire meshes, or other interfaces inside the structures).

Applications:
\begin{itemize}
	\item Condition assessment of bridge decks and tunnel linings
	\item Pavement profiling
	\item Mine detection
	\item Archaeological and geophysical investigations
	\item Borehole inspection and building inspection
	\item Evaluation of the deck thickness
	\item Measurement of the concrete cover and rebar configuration
	\item Characterization of delamination potential
	\item Characterization of concrete deterioration
	\item Description of concrete as a corrosive environment
	\item Estimation of concrete properties 
\end{itemize}

Limitations:
\begin{itemize}
	\item Inability to directly image and detect delamination of bridge deck unless they are epoxy impregnated or filled with water
	\item Negatively affected by cold conditions
	\item Unable to provide any information regarding mechanical properties of concrete
	\item Cannot provide information about presence of corrosion, corrosion rates or rebar section loss
\end{itemize}

\paragraph{Ultrasonic Pulse Echo (UPE)}

UPE uses ultrasonic (acoustic) stress waves to detect objects, interfaces, and anomalies. These waves are generated by exciting a piezoelectric material with a short-burst, high amplitude pulse that has very high voltage and current. This test concentrates on measuring transit time of ultrasonic waves that are passing through material and reflected to the surface of tested medium. Based on transit time, this technology can be used to indirectly detect internal flaws such as cracking voids, delamination, etc. A UPE test concentrates on measuring the transit time of ultrasonic waves traveling through a material and being reflected to the surface of the tested medium. Based on the transit time or velocity, this technique can also be used to indirectly detect the presence of internal flaws. An ultrasonic wave is generated by a piezoelectric element. As the wave interfaces with a defect, a small part of the emitted energy is reflected back to the surface.

\begin{figure}[H]
	\centering
	\includegraphics[width=0.7\linewidth]{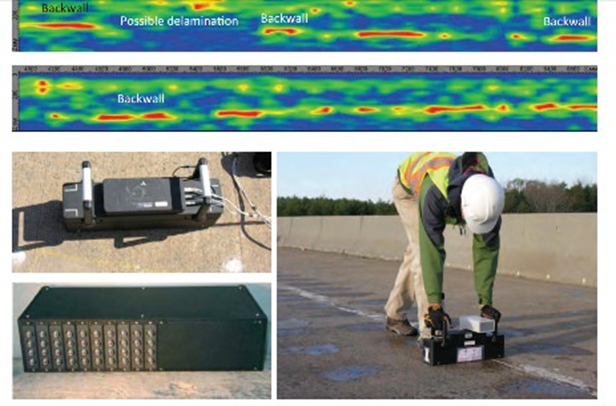}
	\caption{Bridge Deck Survey Using MIRA Ultrasonic System. B-scans (top) and Equipment and Data Collection (bottom) (Gucunski et al. , 2013)}
	\label{fig:figure2}
\end{figure}

Applications:
\begin{itemize}
	\item Condition assessment for evaluating probable material damage from Aggregate Silica Reaction (ASR), freeze–thaw, and other deterioration processes.
	\item Used in material quality control and quality assurance of concrete and hot-mix asphalt, to evaluate material modulus and strength.
	\item Measurement of the depth of vertical (surface) cracks in bridge decks or other elements. 
\end{itemize}

Limitations:
\begin{itemize}
	\item Unable to provide reliable modulus values on deteriorated sections of a concrete deck, such as de-bonded or delaminated sections.
	\item Plays only a supplemental role in deterioration detection, and experience is required for understanding and interpreting test results.
	\item The USW modulus evaluation becomes more complicated for layered systems, such as decks with asphalt concrete overlays, where the moduli of two or more layers differ significantly.
\end{itemize}

\paragraph{Infrared Thermography Technology}
The infrared thermography technology is used before applying hammer sounding tests to detect possible subsurface deterioration including delamination or spall of concrete through the monitoring of temperature variations on a concrete surface using a high-end infrared camera. In the process of crack detection using High Resolution Digital Imaging, the sections of concrete bridge elements are photographed using motion-controlled digital camera. These digital images are then analyzed by image processing to determine structure’s current condition including crack size, location, and distribution. 

Applications include:

\begin{itemize}
	\item To detect voids and delamination in concrete.
	\item To detect delamination and de-bonding in pavements, voids in shallow tendon ducts (small concrete cover), cracks in concrete, and asphalt concrete segregation for quality control.
\end{itemize}

Limitations:

\begin{itemize}
	\item Does not provide information about the depth of the flaw.
	\item Deep flaws are also difficult to detect.
	\item Affected by surface anomalies and boundary conditions.
\end{itemize}

\paragraph{Impulse Response}

The impulse response method is a dynamic response method that evaluates the dynamic characteristics of a structural element to a given impulse. The typical frequency range of interest in impulse response testing is 0 to 1 kHz. The basic operation of an impulse response test is to apply an impact with an instrumented hammer on the surface of the tested element and to measure the dynamic response at a nearby location using a geophone or accelerometer \citep{clausen2012onsite}.

\begin{figure}[H]
	\centering
	\includegraphics[width=0.5\linewidth]{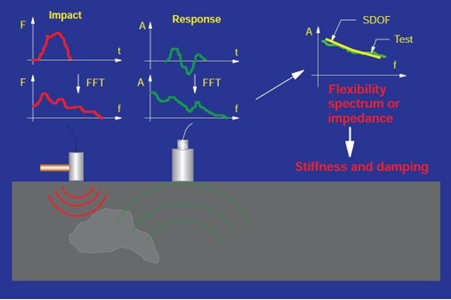}
	\caption{Principle of Impulse Response Testing (Gucunski et al. , 2013)}
	\label{fig:figure3}
\end{figure}

\paragraph{Ultrasonic Surface Waves (USW)}

The ultrasonic surface waves (USW) method is a branch of the spectral analysis of surface waves (SASW) method used to evaluate material properties (elastic moduli) in the near surface zone. The SASW uses the phenomenon of surface wave dispersion (i.e., velocity of propagation as a function of frequency and wavelength, in layered systems to obtain the information about layer thickness and elastic moduli). The USW and SASW tests are the same, but the frequency range of interest is limited to a narrow high-frequency range in which the surface wave penetration depth does not exceed the thickness of the tested object. The surface wave velocity can be precisely related to concrete modulus in bridge decks, using either the measured or assumed mass density, or Poisson ratio of the material. A USW test consists of recording the response of the deck, at two receiver locations, to an impact on the surface of the deck, as illustrated in the following figure. 

\begin{figure}[H]
	\centering
	\includegraphics[width=0.5\linewidth]{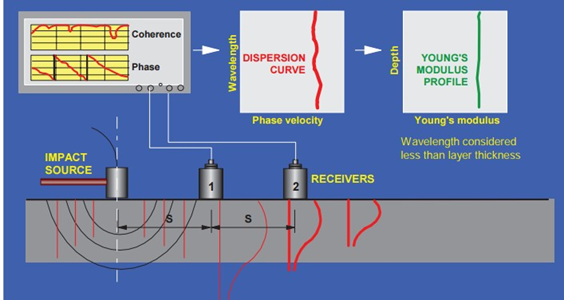}
	\caption{Evaluation of a Layer Modulus by SASW (USW) Method \citep{gucunski2013nondestructive}}
	\label{fig:figure4}
\end{figure}

\paragraph{Infrared Thermography}

To detect subsurface defects, IR thermography keeps track of electromagnetic wave surface radiations related to temperature variations in the infrared wave- length.

\begin{figure}[H]
	\centering
	\includegraphics[width=0.7\linewidth]{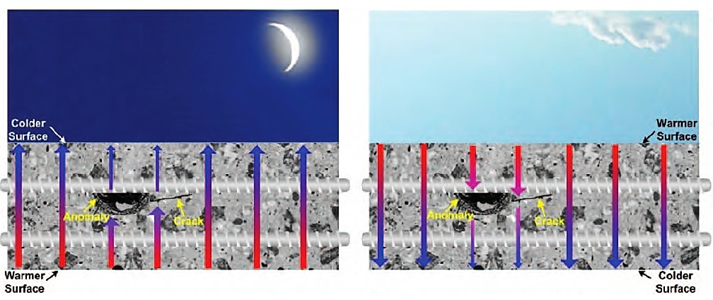}
	\caption{Principle of Passive Infrared Thermography \citep{gucunski2013nondestructive}}
	\label{fig:figure5}
\end{figure}

\paragraph{Chain Dragging and Hammer Sounding }

Chain dragging and hammer sounding are the most common inspection methods used by state DOTs and other bridge owners for the detection of delaminations in concrete bridge decks. The objective of dragging a chain along the deck or hitting it with a hammer is to detect regions where the sound changes from a clear ringing sound (sound deck) to a somewhat muted and hollow sound (delaminated deck). Chain dragging is a relatively fast method for determining the approximate location of a delamination. The speed of chain dragging varies with the level of deterioration in the deck. Hammer sounding is much slower and is used to accurately define the boundaries of a delamination. It is also a more appropriate method for the evaluation of smaller areas \citep{gucunski2013nondestructive}.

\paragraph{Acoustic Emission}
The phenomenon of acoustic sound generation in structures under stress is called Acoustic Emission (AE). Acoustic emission works by detecting how acoustic waves in materials propagate due to the presence of structural flaws. Under an applied load, a stress acts on the material and produces local plastic deformation. This stress produces an elastic wave that travels outward from the source, moving through the body until it arrives at sensors attached to the surface of the structure. An AE test covers a large area with one test and it can also be used for continuous monitoring \citep{rehman2016nondestructive}.

\subsubsection{Concrete Resistivity and Permeability}

\paragraph{Electrical Resistivity}

The electrical resistivity (ER) technique is used to find moisture in the concrete which can be linked to the presence of cracks. The presence and amount of water and chlorides in concrete are important parameters in assessing its corrosion state or describing its corrosive environment. Damaged and cracked areas, resulting from increased porosity, are preferential paths for fluid and ion flow. The higher the ER of the concrete is, the lower the current passing between anodic and cathodic areas of the reinforcement will be \citep{gucunski2013nondestructive}.

\begin{figure}[H]
	\centering
	\includegraphics[width=0.5\linewidth]{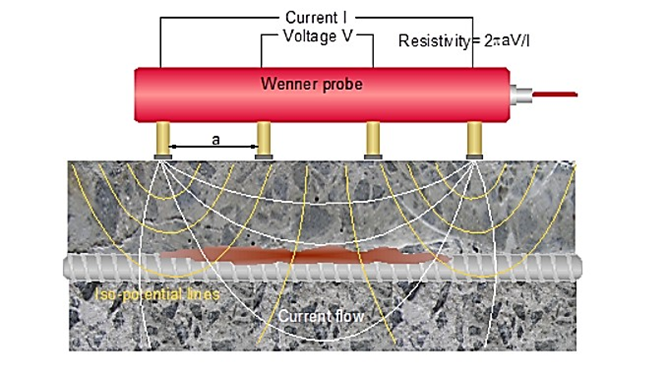}
	\caption{Electrical Resistivity Principle \citep{gucunski2013nondestructive}}
	\label{fig:figure6}
\end{figure}

The electrical resistivity ($\rho$) or conductivity ($\sigma$) of concrete indicates the resistance of concrete against the flow of electrical current. The determination of electrical resistivity of concrete has become an established non-destructive measurement technique in the assessment of the durability of concrete structures. Electrical resistivity of concrete is affected by a number of factors such as pore structure (continuity and tortuosity), pore solution composition, moisture content, and temperature. Pore structure of concrete varies with water to cementitious material (w/cm) ratio, degree of hydration, and use of mineral admixtures such as blast furnace slag, fly ash, and silica fume. 
Concrete pore solution contains K$^+$, Na$^+$, Ca$^{2+}$, SO$_4^{2-}$, and OH$^-$. Chloride ion may also appear due to the deicing salt or seawater. The use of mineral admixture could change the composition and concentration of ions in pore solution. However, it has been found that changes in pore structure exerted a greater influence on the measured resistivity than changes in pore solution composition and concentration. Degree of hydration affects resistivity as further hydration reduces the concrete porosity. When concrete resistivity is measured, the electrical current is mainly due to the ion mobility, ion-ion, and ion-solid interactions. Moisture content plays an important role in concrete resistivity as electrical current in the concrete is carried by the pore water. Electrical resistivity increases with decreasing moisture content. Temperature change was found to have a significant effect on electrical resistivity of concrete, and usually, an increase in temperature leads to a decrease in resistivity. Temperature affects resistivity by changing the ion mobility, ion-ion, and ion-solid interactions, as well as the ion concentration in pore solution. Various techniques have been developed to measure the resistivity of concrete. Two-electrode method and four-electrode method are the most used methods. Resistivity can be measured by the following formula:

\begin{equation}
	\rho = R \cdot \frac{A}{L}
\end{equation}

\noindent where $R$ is the measured resistance, $A$ is the cross-sectional area, and $L$ is the distance between electrodes, as shown in the following figure. 

\begin{figure}[H]
	\centering
	\includegraphics[width=0.3\linewidth]{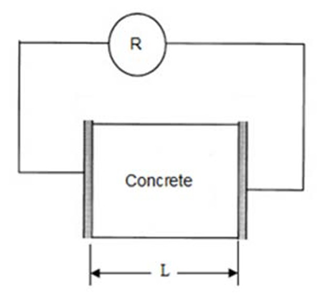}
	\caption{Schematic illustration of concrete resistivity measurement by two-plate method (Presuel-Moreno et al. , 2013)}
	\label{fig:figure7}
\end{figure}

\subparagraph{Correlation between Concrete Resistivity and Diffusivity }
During the chloride diffusion process, diffusivity is the controlling parameter which determines the time it takes for chloride ions to diffuse into concrete and reach the critical chloride threshold for corrosion limitation. However, most test methods, such as the Rapid Chloride Migration (RCM) test, Rapid Chloride Permeability Test (RCPT) or Bulk Diffusion (BD) method, are either expensive or time-consuming for determining the concrete permeability properties, which limits their use as a routine quality control tool. Recently, electrical resistivity of concrete has been applied as an indirect method to evaluate concrete chloride permeability.

The Florida Department of Transportation (FDOT) performed experiments to study the correlation between resistivity and Rapid Chloride Permeability results \citep{kessler2005resistivity}. In this investigation, resistivity was measured using the Wenner method. This research reported a good correlation between RCP test and resistivity results for specimens that were wet cured in a controlled environment or cured in lime water. Based on this correlation, FDOT developed a surface resistivity method (FDOT FM5-578 and then an AASHTO test method TP-95) to characterize concrete permeability and proposed a relationship between resistivity and chloride permeability.

\begin{table}[H]
	\centering
	\includegraphics[width=0.7\linewidth]{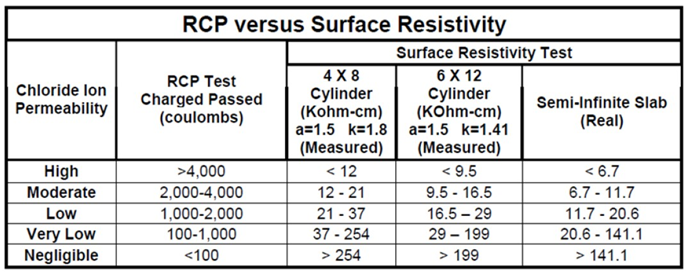}
	\caption{Correlation between surface resistivity and chloride ion permeability \citep{presuel2013analysis}}
	\label{fig:table3}
\end{table}

Besides investigations carried out on laboratory specimens, research has also been performed on field results to correlate electrical resistivity and apparent diffusivity coefficients (Dca). As Dca is usually obtained after a long period of exposure ranging from months to years and even longer, the aging effect needs to be considered as concrete diffusivity changes with time.

\subparagraph{Correlation between Concrete Resistivity and Corrosion Rates }

During the crack initiation stage, rebar is depassivated and corrosion has initiated. In this stage, the most important parameter is corrosion rate which determines how fast the reinforced concrete structure is deteriorating. The propagation stage of concrete structures could be significantly increased by reducing the corrosion rate.

Once corrosion is initiated by chloride ions, corrosion rate is dependent on numerous parameters such as relative humidity (RH), oxygen availability, ratio of anodic/cathodic area, concrete resistivity and so on. When concrete is under water or concrete cover is thick, corrosion rate of steel in concrete is usually considered to be under cathodic control, that is, corrosion rate is dependent on the availability of O2. When concrete is under aerated condition, such as the splash zone, the O2 flux into concrete is usually enough to support the anodic current. In this condition, cathodic control no longer exists and the factor limiting the corrosion rate is the flow of ionic current through concrete, that is, the electrical resistivity of concrete. Resistive control describes the relationship between corrosion rate and electrical resistivity of concrete (or mortar), which has been studied by various investigations.

The correlation between the corrosion rate of depassivated steel and concrete resistivity has been reported in various research works \citep{alonso1988relation,bertolini1997concrete,andrade1996corrosion}. Most of these investigations found a linear relationship between corrosion rate and concrete conductivity. 

An empirical equation describing relation between corrosion rate and resistivity was proposed by \citet{andrade2004test}:

\begin{equation}
	I_{\text{corr}} = \frac{3 \times 10^3}{\rho}
\end{equation}

with $I_{\text{corr}}$ in $\mu$A/cm$^2$ and $\rho$ (electrical resistivity) in $\Omega\cdot$cm.

\subsubsection{Permeation Test Method}

The permeability of aggressive substances into concrete is the main cause for concrete deterioration. Permeability represents the governing property for estimating the durability of concrete structures. Permeation tests are non-destructive testing methods that measure the near-surface transport properties of concrete. The three categories of measuring concrete permeability are:
\begin{itemize}
	\item hydraulic permeability which is the movement of water through concrete;
	\item gas permeability which is the movement of air through concrete;
	\item chloride-ion permeability which involves the movement of electric charge.
\end{itemize}

The measuring of chloride penetrability is the most commonly used non-destructive method that provides an indication of concrete permeability through established correlations. The standard guideline on the application and interpretation of chloride penetrability is ASTM C 1202: Standard Test Method for Electrical Indication of Concrete’s Ability to Resist. The test involves coring a standard sized cylinder from the in-situ concrete. The sample is then trimmed, sealed with an epoxy coating from two sides, saturated in water and then placed in a split testing device filled with a sodium chloride solution with an applied voltage potential (Concrete Institute of Australia, 2008). The charge passing through the concrete is then measured where:
\begin{itemize}
	\item A value of between 100 and 1000 Coulombs represents low permeability
	\item A value greater than 4000 Coulombs represents high permeability
\end{itemize}

\subsubsection{Concrete Cover and Rebar Distribution}

\paragraph{Pachometer}
Also, known as a cover meter, a pachometer is used to detect the presence of ferromagnetic materials (e.g. steel or iron) embedded in concrete. Primarily, a pachometer measures the depth of concrete cover to the reinforcing steel. It operates by generating a magnetic field and measuring the interaction between the field and the metal. The intensity of the response is a function of the location and size of the embedded material. A pachometer is very useful for determining if a section of a road deck has inadequate concrete cover due to erosion on the surface \citep{ryan2006bridge}.

\subsubsection{Concrete Strength}
\paragraph{Rebound and Penetration}

In use since the 1950s, this method is very simple and easy to use. A standardized hammer strikes the surface of the concrete, and the amount of rebound is measured. The amount of rebound is related to the strength of the concrete that was struck. However, because of the high variability of concrete mixes, there is no absolute scale for concrete strength based on the measured rebound. Thus, this method can only be used to determine relative concrete strength throughout a concrete bridge \citep{ryan2006bridge}.

\paragraph{Penetration Resistance Method }

Penetration resistance methods are invasive NDT procedures that explore the strength properties of concrete using previously established correlations. These methods involve driving probes into concrete samples using a uniform force. Measuring the probe’s depth of penetration provides an indication of concrete compressive strength by referring to correlations. Due to the insignificant effect of the penetration resistance methods on the structural integrity of the probed sample, the tests are considered to be non-destructive despite the disturbance of the concrete during penetration. The most commonly used penetration resistance method is the Windsor probe system. The system consists of a powder-actuated gun, which drives hardened allow-steel probes into concrete samples while measuring penetration distance via a depth gauge. The penetration of the Windsor probe creates dynamic stresses that lead to the crushing and fracturing of the near-surface concrete \citep{hellier2003handbook}.

\begin{figure}[H]
	\centering
	\includegraphics[width=0.4\linewidth]{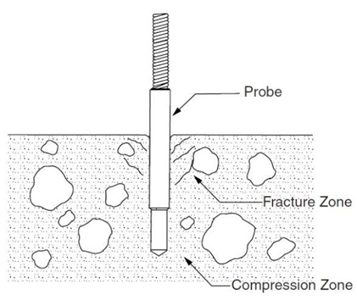}
	\caption{Schematic diagram of typical concrete failure mechanism during probe penetration \citep{hellier2003handbook}}
	\label{fig:figure8}
\end{figure}

\paragraph{Pull-out Resistance Methods }
Pull-out resistance methods measure the force required to extract standard embedded inserts from the concrete surface. Using established correlations, force required to remove the inserts provides an estimate of concrete strength properties. The two types of inserts, cast-in, and fixed-in-place, define the two types of pull-out methods. Cast-in tests require an insert to be positioned within the fresh concrete prior to its placement. Fixed-in-place tests require less foresight and involve positioning an insert into a drilled hole within hardened concrete. Pull-out resistance methods are non-destructive yet invasive methods which are commonly used to estimate compressive strength properties of concrete. The most commonly used pull-out test method is the LOK test developed in 1962 by Kierkegarrd-Hansen. The test requires an insert embedment of 25mm to insure sufficient testing of concrete with coarse aggregates. The force required to remove the insert is referred to as the “lok-strength”, which in other pull-out resistance methods is referred to as the pull-out force \citep{helal2015non}.

\begin{figure}[H]
	\centering
	\includegraphics[width=0.4\linewidth]{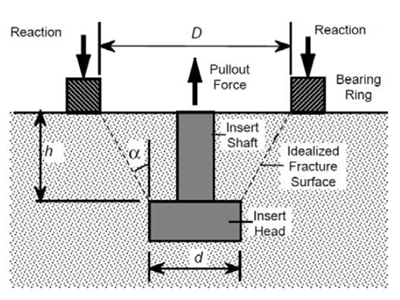}
	\caption{Schematic diagram of typical pull-our resistance methods \citep{hellier2003handbook}}
	\label{fig:figure9}
\end{figure}

\paragraph{Pull-off Resistance Method }

The pull-off test is an in-situ strength assessment of concrete which measure the tensile force required to pull a disc bonded to the concrete surface with an epoxy or polyester resin. The pull-off force provides an indication of the tensile and compressive strength of concrete by means of established empirical correlation charts. The most commonly used pull-off test is the 007 Bond Test. The test consists of a hand operated lever, bond discs, an adjustable alignment plate, and force gauges. The disc is bonded to the concrete surface by a high strength adhesive and is attached to the hand operated lever by a screw. After leveling the adjustable alignment plate, tension force is applied by the lever and measured by the force gauge. The pull-off tensile strength is calculated by dividing the tensile force at failure by the disc area and is used to determine the compressive strength of concrete by using previously established empirical correlations. The main advantage of pull-off test methods is that they are simples, quick and could be used to test a wide range of construction settings. A significant limitation is the curing time required for the adhesive, which is generally around 24 hours. Another limitation relates to the human error in surface preparation which may cause the adhesive to fail \citep{hallberg2005development}.

\begin{figure}[H]
	\centering
	\includegraphics[width=0.2\linewidth]{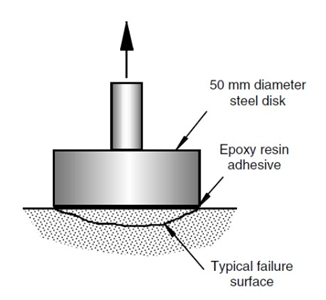}
	\caption{Schematic diagram of pull-off resistance NDT method \citep{helal2015non}}
	\label{fig:figure10}
\end{figure}

\paragraph{Maturity Test Method}

The maturity method is a NDT technique for determining strength gain of concrete based on the measured temperature history during curing. The maturity function is presented to quantify the effects of time and temperature. The resulting maturity factor is then used to determine the strength of concrete based on established correlations. The maturity method has various applications in concrete construction such as formwork removal and post-tensioning. Temperature versus time is recorded using thermocouples inserted into fresh concrete. The measured time history could be used to compute a maturity index which provides a reliable estimate of early age concrete strength as a function of time. The standard guideline on the testing and interpretation of the maturity method is ASTM C 1074-11: Standard Practice for Estimating Concrete Strength by Maturity Method. The factors that lead to variability in testing are aggregate properties, cement properties, water-cement ratio and curing temperature (Concrete Institute of Australia, 2008). Before attempting to estimate in-situ strength of concrete, laboratory testing on concrete samples of similar characteristics must be performed to develop the correct maturity function while minimizing the effect of the aforementioned factors. Temperature probe locations must be carefully selected to measure a representative temperature of the entire concrete section \citep{helal2015non}.

\begin{figure}[H]
	\centering
	\includegraphics[width=0.4\linewidth]{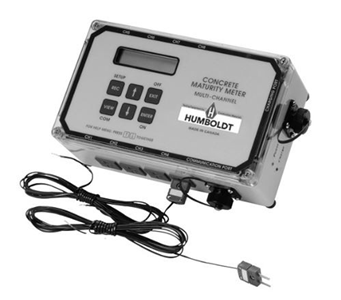}
	\caption{Maturity test apparatus with thermocouple \citep{helal2015non}}
	\label{fig:figure11}
\end{figure}

\subsubsection{Sulfate Resistance Test}

According to ASTM C 150, Type II cement contains less than 8\% C3A, and Type V cement contains less than 5\%. Additionally, the use of Type MS (moderate sulfate resistant) cement and Type HS (high sulfate resistant) cement can provide sulfate resistance. Tests to examine sulfate resistance include petrographic examination of sulfate-exposed specimens over time and ASTM C 1157. ASTM C 1157 utilizes a physical test (ASTM C 1012) for sulfate resistance by evaluating expansion of mortar prisms made with the cement and requiring them to have expansion below a certain limit without specifying the cement composition limits \citep{ferraris2006sulfate}.

\subsubsection{Alkali-Silica Reaction Tests}

There are various methods to identify ASR distress. To determine that ASR is the cause of damage, the presence of ASR gel must be verified. Petrographic examination (ASTM C 856 or AASHTO T 299) is the most positive method for identifying ASR distress in concrete. Silica gel appears as a darkened area in the aggregate particle or around its edges. Another method to detect ASR gel in concrete is the uranyl-acetate treatment procedure. The concrete surface is sprayed with a solution of uranyl acetate, rinsed with water, examined under ultraviolet light and ASR gel appear as bright yellow or green areas \citep{natesaiyer1992protected}. Another method for detecting gel in concrete structures is the Los Alamos staining method, which is used in the field as well as the laboratory. In this method, the substance is applied to a fresh concrete surface and viewed for yellow staining, which indicates gel containing potassium. A second substance called rhodamine B, is applied to the rinsed surface and allowed to react, and then the surface is rinsed with water. The rhodamine B stain produces a dark pink stain in the area of the yellow stain. The stain corresponds to calcium-rich ASR gel. It is important to note that presence of gel identified by both the uranyl-acetate treatment and the Los Alamos staining method does not necessarily mean that destructive ASR has occurred. To confirm the diagnostic, additional tests are necessary. Additionally, the ultrasonic surface waves test could be used for evaluating probable material damage from alkali-silica reaction.

\subsection{Rebar Corrosion Condition}

\subsubsection{Corrosion Potential}

\paragraph{Half-Cell Potential}

The half-cell potential (HCP) measurement is an electrochemical technique to evaluate active corrosion in reinforced steel and prestressed concrete structures. The method can be used at any time during the life of a concrete structure and in any climate, as long as the temperature is higher than 2 oC \citep{gucunski2013nondestructive}. This method can measure the potential difference between a standard portable half-cell, normally a copper/copper sulphate standard reference electrode placed on the surface of the concrete with the steel reinforcement underneath. The reference electrode is connected to the positive end of the voltmeter and the steel reinforcement to the negative. As a result, the test shows the probability of corrosion activity taking place at the point where the measurement of potentials is taken from a half-cell, typically a copper-copper sulphate half-cell. An electrical contact is established with the exposed steel and the half-cell is moved across the surface of concrete for measuring the potentials.

\begin{figure}[H]
	\centering
	\includegraphics[width=0.7\linewidth]{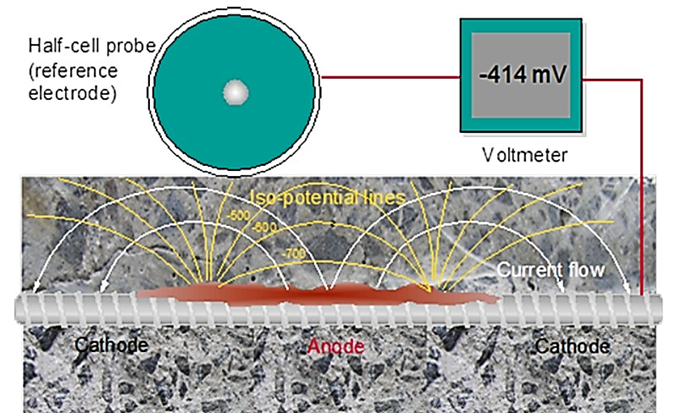}
	\caption{HCP Principle \citep{gucunski2013nondestructive}}
	\label{fig:figure12}
\end{figure}

\subsubsection{Corrosion Rate}

\paragraph{Linear Polarization Resistance (LPR)}

Linear polarization resistance (LPR) is a non- destructive testing technique used in steel corrosion rate measurement. There is a direct relationship between the measured corrosion current and the mass of steel consumed by Faraday’s law. Corrosion current can be derived indirectly throughout the following expression:

\begin{equation}
	i_{\text{corr}} = \frac{B}{R_p}
\end{equation}

\noindent where:
\begin{itemize}
	\item $i_{\text{corr}}$ = the corrosion current density, representing the change in current (mA/ft$^2$);
	\item $B$ = a constant relating to the electrochemical characteristics of steel in concrete;
	\item $R_p$ = the polarization resistance, expressed as:
	\[
	R_p = \frac{\Delta E}{I_{\text{app}}}
	\]
	where $\Delta E$ is the change in potential and $I_{\text{app}}$ is the applied current.
\end{itemize}

\paragraph{Galvanostatic Pulse Measurement (GPM)}

Galvanostatic pulse measurement (GPM) is an electrochemical NDT method used for rapid assessment of rebar corrosion, based on the polarization of rebars using a small current pulse.

\begin{figure}[H]
	\centering
	\includegraphics[width=0.4\linewidth]{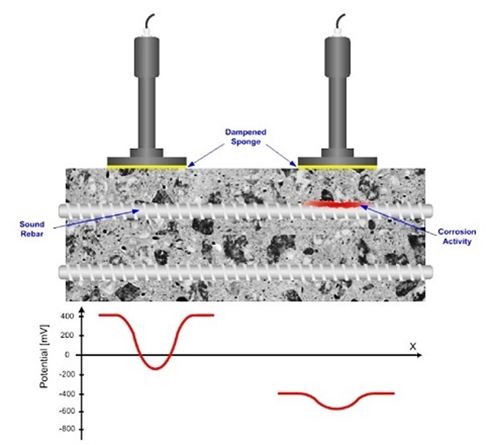}
	\caption{GPM Principle \citep{gucunski2013nondestructive}}
	\label{fig:figure13}
\end{figure}

\subsubsection{Chloride Related Test}

For bridge deck evaluation, measurements and samples are to be taken from the critical failure zone of the bridge deck and equally distributed throughout the length of the deck in non-damaged areas. Critical failure zone is typically the right traffic lane wheel path areas. Damage is defined as spalled, delaminated, and patch areas (asphalt or concrete). Thus, a damage condition survey is to be performed before cover depth measurements and chloride sampling \citep{williamson2007bridge}. Due to the high alkalinity (pH >12.5) of the concrete pore solution, a passive oxide film is formed on the rebar surface. This passive layer initially protects the rebar from corrosion \citep{presuel2013analysis}. However, the presence of chloride ions could destroy the passive layer even at high alkalinity once it exceeds a certain concentration threshold (CT). Once CT is exceeded, corrosion initiates and then propagates.

Chloride diffusivity into concrete is usually considered the most important parameter that determines the service life of reinforced concrete structures. Time to corrosion initiation is strongly related to the chloride ion permeability of concrete. The corrosion propagation period is the time from corrosion initiation to the end service of structures, which is controlled by the corrosion rate. Transport of chloride ions into concrete involves complex physical and chemical processes. Diffusion is the main mechanism to transport chlorides into water-saturated concrete from the concrete surface to the rebar surface. The corrosion rate is usually mainly controlled by the electrical resistivity of concrete once corrosion has initiated.

Various test procedures have been developed to evaluate the chloride penetration resistance of concrete. These tests are classified into three categories: 1) diffusion tests including AASHTO T259 (salt ponding test), NT BUILD 433 (bulk diffusion test) and other natural long-term full- immersion tests; 2) migration tests, including ASTM C1202 (rapid chloride permeability test) and NT Build 492 (chloride migration test); 3) indirect tests, such as electrical resistivity measurement. Duration of the test methods ranges from minutes (resistivity method) to several years (diffusion test) \citep{presuel2013analysis}.

\paragraph{Bulk Diffusion Test (ASTM C1556)}

Bulk diffusion test, designated as NT Build 433 or ASTM C1556, is a test method used to determine the apparent chloride diffusion coefficient of concrete (Build, 1995; ASTM, 2003). In this method, chloride ions penetrate into concrete only through diffusion, as shown in the following figure. The exposure time for this test is at least 35 days for low quality concrete and 90 days for high quality concrete. Longer exposure times up to 1 to 3 years are also used.

\begin{figure}[H]
	\centering
	\includegraphics[width=0.5\linewidth]{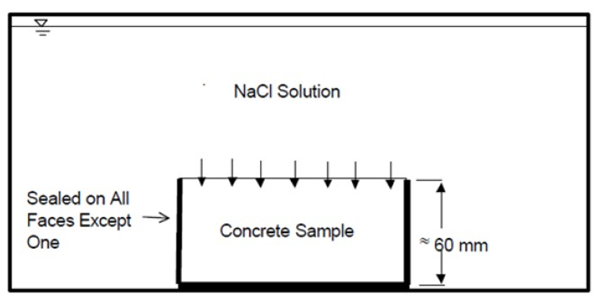}
	\caption{Schematic illustrations of bulk diffusion test \citep{presuel2013analysis}}
	\label{fig:figure14}
\end{figure}

\paragraph{Rapid Chloride Migration Test}

Rapid Chloride Migration (RCM) test is designed according to NT Build 492 (Build, 1999). A potential ranging from 10\,V--60\,V is used to accelerate the penetration of chlorides and the test period ranges from 6 to 96 hours. The duration and applied voltage depends on the quality of concrete. The averaged chloride penetration depth is obtained by splitting the specimen and spraying 0.1\,N AgNO$_3$ as a color indicator at the cross section. Non-steady-state migration coefficient ($D_{\text{nssm}}$) can be obtained with the following equation:

\begin{equation}
	D_{\text{nssm}} = \frac{0.0239(273 + T)L}{(U - 2)t} \left( x_d - 0.0238 \sqrt{\frac{(273 + T)Lx_d}{U - 2}} \right)
\end{equation}

where,\\
$D_{\text{nssm}}$ = non-steady-state migration coefficient, $10^{-12}$\,m$^2$/s;\\
$U$ = absolute value of the applied voltage, V;\\
$T$ = average value of the initial and final temperatures in the anolyte solution, $^\circ$C;\\
$L$ = thickness of the specimen, mm;\\
$x_d$ = average value of the penetration depths, mm;\\
$t$ = test duration, hour.

\begin{figure}[H]
	\centering
	\includegraphics[width=0.7\linewidth]{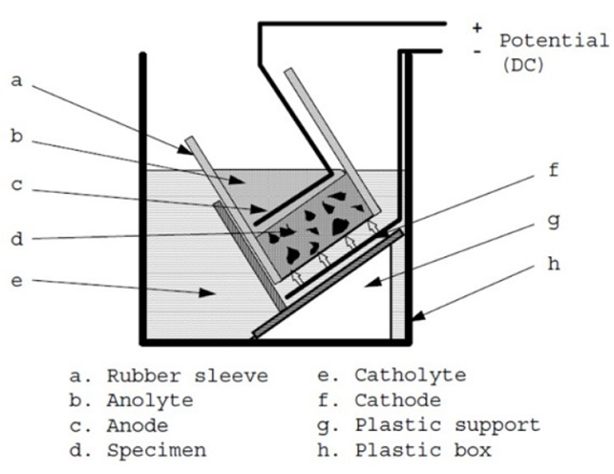}
	\caption{Schematic illustrations of RCM test setup \citep{presuel2013analysis}}
	\label{fig:figure15}
\end{figure}

Based on results from RCM test, the resistance to chloride penetration can be assessed by the relationship shown in the following table. 

\begin{table}[H]
	\centering
	\caption{Relationship between non-steady-state migration coefficients and resistance to chloride penetration \citep{presuel2013analysis}}
	\begin{tabular}{|c|c|}
		\hline
		\textbf{$D_{\text{nssm}}$ ($10^{-12}$ m$^2$/s)} & \textbf{Resistance to Chloride Penetration} \\
		\hline
		$> 15$     & Low \\
		10--15     & Moderate \\
		5--10      & High \\
		2.5--5     & Very high \\
		$< 2.5$    & Extremely high \\
		\hline
	\end{tabular}
\end{table}

\paragraph{Neutron Probe for Detection of Chlorides}
Also known as Prompt Gamma Neutron Activation (PGNA), this NDT method determines the composition of light elements (Ca, Si, Fe, Cl, S, Al) in concrete. The amounts of these elements present in concrete provide an assessment of the concrete’s general structural condition. A given section of concrete is irradiated with neutrons by a portable californium neutron source. When irradiated, each element produces a characteristic gamma ray which is detected and counted by a highly pure germanium detector \citep{lee2014non}.

\subsubsection{Carbonation Depth Measurement Test}
To physically measure the extent of carbonation on concrete, a freshly exposed surface of the concrete is sprayed with a 1\% phenolphthalein solution. The indicator solution turs pink when pH is above 8.6, and where the solution remains colorless the pH of the concrete is below 8.6, suggesting carbonation. The 1\% phenolphthalein solution is made by dissolving 1g of phenolphthalein in 90cc of ethanol. The solution is then made up to 100cc by adding distilled water. On freshly extracted cores the core is sprayed with phenolphthalein solution, the depth of the uncolored layer from the external surface is measured to the nearest mm at 4 or 8 positions, and the average taken. In drilled holes, the dust is first removed from the hole and again the depth of the uncolored layer measured at 4 or 8 positions and the average taken. If the concrete still retains its alkaline characteristics the color of the concrete will change to purple. If carbonation has taken place the pH will have changed to 7 and there will be no color change (IAEA, 2002).

\subsection{Case Studies}

\subsubsection{NCHRP Project 558 \citep{sohanghpurwala2006manual}}

NCHRP Report 558 developed field evaluation procedures for bridge superstructure members. The evaluation consists of the following steps.
\begin{enumerate}
	\item Grid stationing. In this step, the structure is usually marked by a grid with space of 2 feet or 5 feet. The distance can be measured by tools such as a land wheel.
	\item Visual survey. The visual survey should be conducted according to the ACI 201.1 R-92 “Guide for Making a Condition Survey of Concrete in Service”.
	\item Delamination survey. In this step, the delamination survey is recommended to be carried out based on ASTM D-4580-86 “Standard Practice for Measuring Delaminations in Concrete Bridge Decks by Sounding”. The survey can be done by using a hammer, chain, or chain drag.
	\item Cover depth measurements. A minimum of 30 measurements per span is required to measure the cover depths. 5 out of the 30 measurements need to be done by excavating cores. The rest measurements can be obtained from covermeter.
	\item Continuity testing. Prior to conducting corrosion potential and corrosion rate testing, electrical continuity of the reinforcing bar must be confirmed. Continuity between locations can be measured by a multimeter and low resistance copper wire.
	\item 	Chloride ion distribution core sampling. The cores need to be extracted based on ASTM C42/C42M-99 “Standard Test Method for Obtaining and Testing Drilled Cores and Sawed Beams of Concrete”.
\item	Corrosion potential survey. The corrosion potential survey should be carried out based on ASTM C-876 “Standard Test Method for Half-Cell Potentials of Uncoated Reinforcing Steel in Concrete”.
\item	Corrosion rate measurement. Corrosion rates can be measured using linear polarization device, which measures the corrosion current density (CCD) of the reinforcing steel at a test location.
\item Concrete resistivity test. Resistivity can be measured by two type of tests four-point and single point tests \citep{balakumaran2016chloride}.
\end{enumerate}

\begin{figure}[H]
	\centering
	\includegraphics[width=0.5\linewidth]{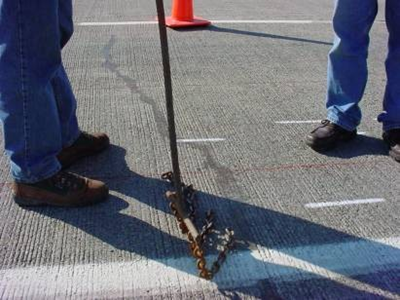}
	\caption{Chain Drag}
	\label{fig:figure16}
\end{figure}

\begin{figure}[H]
	\centering
	\includegraphics[width=0.5\linewidth]{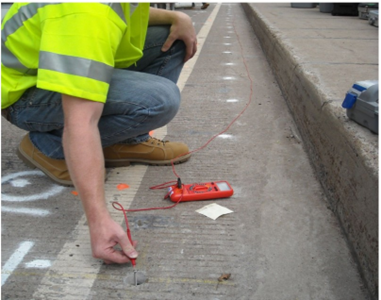}
	\caption{Continuity Test}
	\label{fig:figure17}
\end{figure}

\begin{figure}[H]
	\centering
	\includegraphics[width=0.7\linewidth]{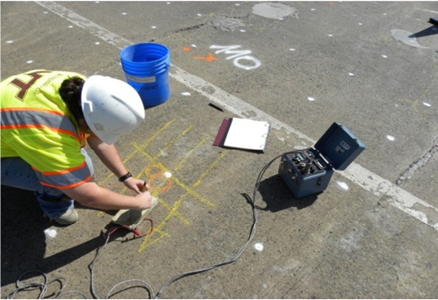}
	\caption{Corrosion Rate Test}
	\label{fig:figure18}
\end{figure}

\subsubsection{VDOT \citep{williamson2007bridge}}

Virginia DOT developed service life estimates of concrete bridge decks and costs for maintaining concrete bridge decks for 100 years. With respect to service life estimates, a probability based chloride corrosion service life model was used to estimate the service life of bridge decks built under different concrete and cover depth specifications between 1969 and 1971 and 1987 and 1991. In addition, the influence of using alternative reinforcing steel as a secondary corrosion protection method was also evaluated. Life cycle costs were estimated for maintaining bridge decks for 100 years considering the present age of the deck.

The research surveyed 37 bridge decks. The distribution of bridge deck types was as follows: 10 bare steel with w/c = 0.47, 16 with w/c = 0.45 and 11 with w/cm = 0.45. The authors stated that bridge deck rehabilitation decisions are based on the deterioration of the worst-span lane of the deck. The right-hand lane normally receives more traffic and therefore deteriorates at a faster rate. For that reason, and due to safety and traffic control issues, only the right-hand lanes were surveyed. The deck survey included a visual survey, non-destructive testing, and the collection of 15 - 4 in concrete cores per deck. The following data were gathered for each bridge deck during the visual survey:

\begin{itemize}
	\item The length and width of the right traffic lane were measured.
	\item Patched areas within the right-hand lane were measured and recorded. 
\end{itemize}

The following non-destructive tests were conducted during the field survey:
\begin{itemize}
	\item Cover depth determinations for the top mat of reinforcing steel. 40-80 measurements were taken per span at 4-foot intervals in the wheel paths using a Profometer 3 cover depth meter. If the span length did not allow for 40 measurements to be taken at 4-foot intervals the interval was reduced to 2 feet.
	\item The right-hand lane was sounded using the chain drag method to determine delaminated areas.
	\item 
\end{itemize}

The concrete cores were taken within the wheel paths on the deck as that is the critical deterioration area. Three of the cores were taken for petrographic analysis purposes and did not contain reinforcing steel. Of the remaining 12, all contained reinforcing steel and 3 were taken directly over cracks in the deck. After the cores were removed the water on the surface resulting from the coring process was allowed to evaporate. The cores were then wrapped in two layers of 4-mil polyethylene and one layer of aluminum foil. The cores were then wrapped in a protective layer of duct tape to preserve the in-place moisture content of the concrete samples until they could be analyzed in the lab. The following tests were conducted in the laboratory on the cores taken from the bridge decks:

\begin{itemize}
	\item Chloride concentrations were determined in accordance to ASTM C 1152-97 at the following seven depths for each concrete core sample: 0.5 in, 0.75 in, 1.0 in, 1.25 in, 1.5 in, at the depth of the reinforcement, and below the reinforcement.
	\item Chloride titration data for diffusion constant (Dca) and surface concentration (C0), cover depth measurements, and the deck damage survey was used to estimate service lives.
\end{itemize}

\section{Durability Related Tests for Steel Bridge}

\subsection{Fatigue}

\subsubsection{Acoustic Emission (AE)}

AE is also used to inspect steel structures and operates on the same principles as described in the concrete structures section. As steel deforms under stress, it produces energy in the form of elastic waves. Damage in the steel will correspond to detectable fluctuations in the elastic wave, which are picked up by sensors attached to the surface of steel members \citep{nair2010acoustic}. AE is particularly effective at detecting fatigue in steel members and is commonly used to inspect fracture critical sections of a bridge.

\subsubsection{Smart Paint}
Smart paint uses microencapsulated dyes that outline fatigue cracks as the crack forms and propagates. The paint contains a small resin layer which conducts electricity, and electrodes are attached to measure the depth and size of a crack. This promising method could enable engineers to monitor vibrations throughout the lifetime of a structure, allowing for very accurate predictions of when fatigue will become a problem \citep{lee2014non}.

\subsubsection{Penetrant Test (Dye Penetrant)}
The penetrant test is used to detect surface flaws in steel members. Based on the principle of capillary action, the surface tension of the dye allows it to penetrate into small openings in steel. These capillary forces are very strong and can act against the force of gravity, which is particularly useful when inspecting steel members which are suspended \citep{hellier2003handbook}. A penetrant test requires no specialized equipment and can be performed rapidly. Various substances can be used and may be applied in many ways, from simple application with aerosol spray cans to more sophisticated means, such as dipping in large tanks on an automatic basis. More sophisticated methods require tanks, spraying, and drying equipment. A quantitative analysis, dye penetrant testing is simple to do and is a good way to detect surface-breaking cracks in nonferrous metals. It’s suitable for automatic testing, but with the same limitations that apply to automatic defect recognition in magnetic particle inspection. Disadvantages of dye penetrant testing method are: it restricted to surface-breaking defects only and it is less sensitive than some other methods and uses a considerable number of consumables.

\subsubsection{Magnetic Particle}

The magnetic particle method is used to detect surface or near surface defects. The steel member to be inspected is placed under a magnetic field and a fine powdered ferrous material is sprayed or blown onto the member. The concentration of the ferrous particles indicates the presence of a crack of flaw. For this method to work, the magnetic field must be aligned perpendicular to an expected discontinuity, requiring that the inspector have some idea of where the crack is located beforehand \citep{lee2014non}. Basically, magnetic crack detection equipment takes two forms. First, for test pieces that are part of a large structure, or for pipes and heavy castings, for example, that cannot be moved easily, the equipment takes the form of just a power pack to generate a high current. For factory applications on smaller, more manageable test pieces, bench-type equipment with a power pack, an indicating ink system that recirculates the fluid, and facilities to grip the workpiece and apply the current flow or magnetic flux flow in a methodical, controlled manner is preferred. The advantage of magnetic particle inspection is that it is generally is simple to operate and apply. This testing is quantitative, and it can be automated, apart from viewing. However, modern developments in automatic defect recognition can be used in parts with simple geometries, such as billets and bars. In this case, a special camera captures the defect indication image and processes it for further display and action. This type of nondestructive testing is restricted to ferromagnetic materials, as well as to surface or near-surface flaws. Magnetic particle inspection is not fail-safe; lack of indication can mean that no defects exist, or that the process wasn’t carried out properly.

\subsection{Corrosion}

\subsubsection{Corrosion Sensors}

This method evaluates direct measurements of the electrical resistance of steel members to detect the extent and rate of corrosion in the member. A continuous monitoring tool, corrosion sensors provide a direct measurement of metal loss \citep{reading1996critical}.

\subsubsection{Robotic Inspection}

Designed to eliminate the need for human inspectors, robotic inspection is essentially an automated inspector who does not require rest or pay. A specially designed car holds the robot in a specially designed multi-linkage system, and allows it to move freely underneath a bridge. The robot is equipped with a variety of cameras and sensors that allow it to locate cracks or other flaws, without putting a human inspector in danger \citep{oh2009bridge}.

\subsubsection{Radiology Neutron Method (Thermal Neutron Radiography)}

Neutron radiography utilizes transmission of radiation to obtain information on the structure and/or inner processes of a given object. The basic principle of NR is very simple. The object under examination is placed in the path of the incident radiation, and the transmitted radiation is detected by a two-dimensional imaging system. The NR arrangement consists of a neutron source, a pin-hole type collimator which forms the beam, and a detecting system which registers the transmitted image of the investigated object. The most important characteristic technical parameter of an NR facility is the collimation ratio L/D, where L is the distance between the incident aperture of the collimator and the imaging plane, D is the diameter of the aperture. This important parameter describes the beam collimation and will limit the obtainable spatial resolution by the inherent blurring independently from the properties of the imaging system. This unsharpness Ubeam can be related to the distance between the object and the detector plane l2 and the L/D ratio \citep{balasko1996dynamic}:

\begin{equation}
	U_{\text{beam}} = \frac{l_s}{L/D}
\end{equation}

\begin{figure}[H]
	\centering
	\includegraphics[width=0.5\linewidth]{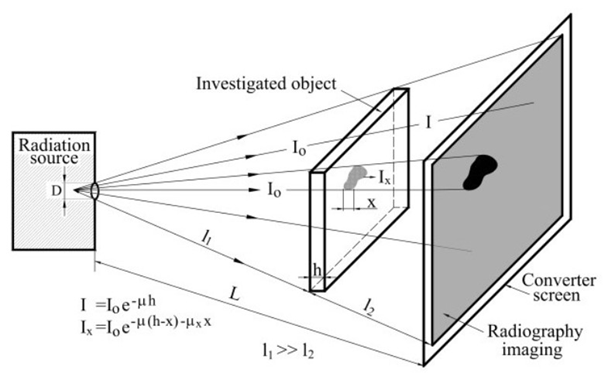}
	\caption{General principle of radiography \citep{balasko1996dynamic}}
	\label{fig:figure19}
\end{figure}

\subsubsection{Optical Holography}

Optical Holographic techniques can be used for nondestructive testing of materials (HNDT). Non-optical Holography techniques include Acoustical, Microwave, X-Ray and Electron beam Holography. HNDT essentially measures deformations on the surface of the object. However, there is sufficient sensitivity to detect sub-surface and internal defects in metallic and composite specimens. In HNDT techniques, the test sample is interferometrically compared with the sample after it has been stressed (loaded). A flaw can be detected if by stressing the object it creates an anomalous deformation of the surface around the flaw. Optical holography is an imaging method, which records the amplitude and phase of light reflected from an object as an interferometric pattern on film. It thus allows reconstruction of the full 3-D image of the object. In HNDT, the test sample is interferometrically compared in two different stressed states. Stressing can be mechanical, thermal, vibration, etc. The resulting interference pattern contours the deformation undergone by the specimen in between the two recordings. Surface, as well as sub-surface defects, show distortions in the otherwise uniform pattern. Also, the characteristics of the component, such as vibration modes, mechanical properties, residual stress, etc. can be identified through holographic inspection.

\subsection{Other Defects}

\subsubsection{Radiographic Testing}

Using either gamma or x-rays, this testing method is used to detect cracks, voids, separations, and inclusions in a steel member. When high energy radio waves pass through a metal object, they are absorbed differently by flaws. To capture the location of flaws, photographic film is placed on the opposite side of the member from which the radio waves are being emitted. The photographic film creates a permanent record of defects with a 1:1 ratio to the actual geometry of the member \citep{hellier2003handbook}. Various radiographic and photographic accessories are necessary, including radiation monitors, film markers, image quality indicators, and darkroom equipment. Radiographic film and processing chemicals also are required. In radiographic testing, information is presented pictorially. A permanent record is provided, which can be viewed at a time and place distant from the test. Radiography is not suitable for several types of testing situations. For example, radiography is inappropriate for surface defects and automation, unless the system incorporates fluoroscopy with an image intensifier or other electronic aids. Radiography generally can’t cope with thick sections, and the testing itself can pose a possible health hazard. Film processing and viewing facilities are necessary, as is an exposure compound. With this method, the beam needs to be directed accurately for 2-D defects. Also, radiographic testing does not indicate the depth of a defect below the surface.

\subsubsection{Ultrasonic Testing (UT)}

In addition to concrete structures, UT can be used to inspect steel members and operates on the same principle as described previously. It is primarily used to detect cracks, loss of cross section, and measure the thickness of steel members. Additionally, UT can be used to measure real time strain of a steel member, particularly useful when trying to evaluate structural response due to dynamic loads such as road traffic \citep{fuchs1998ultrasonic}.

\subsubsection{Eddy Current}

Eddy currents are created through a process called electromagnetic induction. An alternating electric current is applied to a small coil which creates a fluctuating magnetic field. When this field comes in contact with a conducting material, such as a steel member, eddy currents are induced in the material. When structural flaws are present the eddy currents are disrupted \citep{hellier2003handbook}. Most eddy current electronics have a phase display that allows the operator to identify defect conditions. Some units can inspect a product simultaneously at two or more different test frequencies. These units allow specific, unwanted effects to be electronically canceled to give improved defect detection. Most automated systems are for components with simple geometries. Eddy current testing is suitable for automation and can determine a range of conditions of the conducting material, such as defects, composition, hardness, conductivity, and permeability. Information can be provided in simple terms, often go or no-go. Phase display electronic units can be used to obtain greater product information. Compact, portable testing units are available, and this type of testing does not require consumables, except for probes, which sometimes can be repaired. This technique is flexible because of the many probes and test frequencies that can be used for different applications. A disadvantage of eddy current is that many parameters can affect the responses. This means that the signal from a desired material characteristic (for example, a crack) may be masked by an unwanted parameter, such as hardness change. Careful probe and electronics selection is necessary in some applications. Also, tests generally are restricted to surface-breaking conditions and slightly subsurface flaws.

\section{Empirical Models for Predicting Bridge Service Life}

\subsection{Factors Affecting the Deterioration and Service Life of Concrete Bridges}

For concrete bridges, the factors affecting life expectancy include climatic conditions (freeze index and cumulative precipitation), geometric (including span length and number of spans), age, construction technique, wearing surface type, bond strength of overlay with bridge deck, highway functional class, repair history, deck distressed area, traffic volume, wheel locations, and accumulated truck loads \citep{ford2012estimating}. The deterioration of concrete bridges are also linked to corrosion, fatigue, temperature, and/or collision causing changes in strength and stiffness \citep{lin1995reliability}. Primarily, concrete deterioration is caused by corrosion of reinforcement steel, which is determined by the chloride concentration, diffusion coefficient, average depth of bar cover, size and spacing of reinforcement, concrete type, type of curing, amount of air entrainment, carbonation, and water-to-cement ratio \citep{sohanghpurwala2006manual}.

\subsection{Factors Affecting the Deterioration and Service Life of Steel Bridges}

In the case of steel bridges, deterioration and life expectancy have been analyzed on factors including bridge age, volume of truck traffic, truck size distributions, truck axle configuration and weight, cumulative precipitation, freeze index, road classification, type of wearing surface, degradation of individual component, fatigue durability, span length, and high temperatures \citep{rodriguez2005factors}.

\subsection{Service Life Prediction Models}

Both empirical and mechanistic models have been applied in the literature of infrastructure asset life expectancy estimation. Some literature focused on the life of the asset while other literature focused on asset's component longevity. While empirical modeling techniques relies on historical asset condition data (mostly via visual inspection) \citep{gao2019evaluation,gao2019impacts,zhang2018nested,gao2017bayesian,saldarriaga2017evaluation,qiao2016transportation,gao2015milled,prozzi2012maintenance,gao2012peer,gao2012bayesian,gao2011performance,gao2007using}, mechanistic deterioration models focus on damage mechanisms (such as corrosion, fatigue, overstress) of the asset through field or laboratory tests.

For these reasons, empirical models are usually used at network-level, and mechanistic models are applied for project-level analysis. In this report, empirical service life estimation models will be briefly discussed. However, the focus of this report is on the mechanistic deterioration models. Various researchers have considered deterioration of highway bridges and tried to track change over time for various types of bridge and service conditions (i.e., type of roadway) by using National Bridge Inventory (NBI) condition numbers or a similar state-specific index \citep{azizinamini2014bridges}.

In the literature, the following empirical models have been used for bridge service life prediction. Each type of model is discussed in the following sections.

\begin{itemize}
	\item Linear/nonlinear regression models
	\item Times series models
	\item Panel model/dynamic panel model
	\item Discrete choice models
	\item Duration/Survival/Reliability models
	\item Markov chain based models
	\item Machine learning models
\end{itemize}

\subsection{Bridge Condition Data}

The 1968 Federal-Aid Highway Act required that bridge data are regularly collected. Examples of bridge database include FHWA’s NBI database and the Long-Term Bridge Performance (LTBP) program. The NBI database contains inspection data from all 50 states and Puerto Rico. This data has been available since 1992. Inspections are typically conducted biannually, pending special exemptions. Various performance measures exist in the NBI database that can be used in life determination, including Sufficiency Rating, Inventory Rating, Structural Evaluation, Deck Geometry, Bridge Posting, Scour Critical Bridges, Deck, Substructure, or Superstructure \citep{reading1996critical}. The following are the condition definitions of the three NBI data items of deck, superstructure, and substructure \citep{ford2012estimating}.

\begin{enumerate}[label=\textbf{\arabic*.}]
	\item \textbf{FAILED CONDITION} – Out of service – beyond corrective action.
	
	\item \textbf{``IMMINENT'' FAILURE CONDITION} – Major deterioration or section loss present in critical structural components, or obvious vertical or horizontal movement affecting structure stability. Bridge is closed to traffic, but corrective action may allow light service.
	
	\item \textbf{CRITICAL CONDITION} – Advanced deterioration of primary structural elements. Fatigue cracks in steel or shear cracks in concrete may be present, or scour may have removed substructure support. Unless closely monitored, it may be necessary to close the bridge until corrective action is taken.
	
	\item \textbf{SERIOUS CONDITION} – Loss of section, deterioration, spalling, or scour have seriously affected primary structural components. Local failures are possible. Fatigue cracks in steel or shear cracks in concrete may be present.
	
	\item \textbf{POOR CONDITION} – Advanced section loss, deterioration, spalling, or scour.
	
	\item \textbf{FAIR CONDITION} – All primary structural elements are sound but may have minor section loss, cracking, spalling, or scour.
	
	\item \textbf{SATISFACTORY CONDITION} – Structural elements show some minor deterioration.
	
	\item \textbf{GOOD CONDITION} – Some minor problems.
	
	\item \textbf{VERY GOOD CONDITION} – No problems noted.
	
	\item \textbf{EXCELLENT CONDITION} – Like new; no problems.
\end{enumerate}

Another standard of bridge condition inspection is the ”AASHTO Guide for Commonly recognized (CoRe) Structural Elements” \citep{thompson2000aashto}, which was created in 1992 as a basis for states to describe bridge element condition at an appropriate level of detail for maintenance management (AASHTO, 2010).

\subsection{Models From Previous Studies}

\subsubsection{Linear/Non-linear Regression Models}

Linear and non-linear regression models are the most commonly applied technique by agencies for infrastructure asset performance modeling, due to their ease of application and interpretation. For example, a linear regression model can be expressed as

\begin{equation}
	y_i = x_i' \beta + \varepsilon_i,\quad i = 1, \dots, N
\end{equation}

\noindent
where $y_i$ represents the condition of the $i$th bridge and $x_i$ represents the variables of traffic, material, environment, and other factors. Such models can be applied to:
\begin{enumerate}
	\item predict a continuous performance measure as a function of age and other variables, or
	\item directly predict service life as a function of the explanatory variables.
\end{enumerate}

\noindent
Linear regression models can be extended to various model subtypes by changing the error term assumptions (e.g., ordinary, indirect, generalized, two-stage and three-stage least squares, instrumental variables, limited and full information maximum likelihood, and seemingly unrelated regression).

\paragraph{\citep{bolukbasi2004estimating}} 
Historic NBI rating data for 2,601 bridges from Illinois was used to define regression equations relating the condition rating of the deck, superstructure, and substructure to the bridge age. Bridges with sudden rating increase were not included in the study. The regression equations propose the rating and a corresponding service life if no maintenance is performed. Equations are provided for nine bridge categories, and each category has equations to estimate the rating for the deck, superstructure, and substructure. Third degree polynomial equations were developed using regression analysis in the following form:

\begin{equation}
	Y_i(T) = C_0 + C_1 T + C_2 T^2 + C_3 T^3
\end{equation}

\noindent
where $Y_i(T)$ is the condition rating of the $i$th element at a given age $T$, and $C_0$, $C_1$, $C_2$, and $C_3$ are constants.

\medskip

\noindent
The following table summarizes the deterioration equations and sums of squares of residuals for each group of bridges across different components.

\begin{table}[H]
	\centering
	\includegraphics[width=0.7\linewidth]{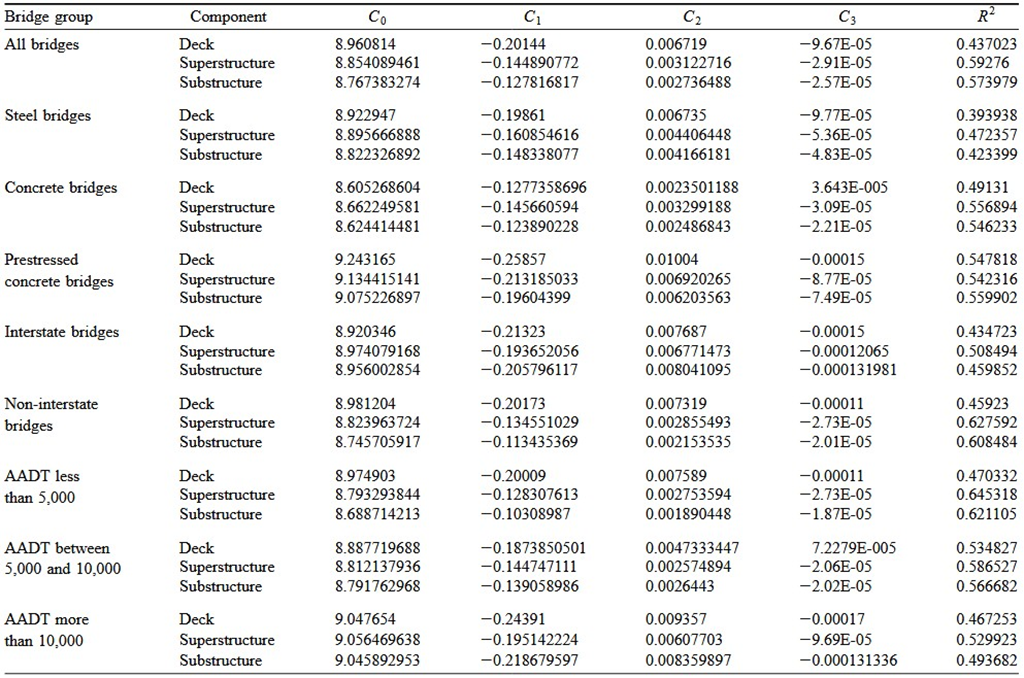}
	\caption{Deterioration of Bridge Components \citep{bolukbasi2004estimating}}
	\label{fig:table5}
\end{table}

According to \citet{bolukbasi2004estimating}, the end of service life of a bridge is typically defined as when a rating of 3 is obtained, and thus maintenance would be required to continue the use of the structure. 

\paragraph{\citep{agrawal2008bridge}}

\citet{agrawal2008bridge} developed regression equations relating condition rating to age for common bridge components in the state of New York. The result of Agrawal and Kawaguchi’s work was a computer program based on summarized Pontis data that calculates the deterioration rates of bridge components. The program uses a cascade algorithm to classify bridges based on different factors. These factors, which are shown below, were used to create a class of bridges with similar characteristics.

\begin{itemize}
	\item Element design type
	\item New York State Department of Transportation (NYSDOT) Region
	\item Bridge ownership
	\item Superstructure design type
	\item Superstructure material type
	\item AADT
	\item Salt usage
	\item Snow accumulation
	\item Climate groups
	\item Functional class
	\item Feature under
\end{itemize}

Within a specific component, multiple equations may be provided for different materials or types of components. As an example, for deck curb, four equations were provided, where T is the time in years:

\begin{equation}
	\text{Granite/Granite/Stone:} \quad CR = 7 - 0.0605424\,T + 0.0001089\,T^2 - 1.0 \times 10^{-7}\,T^3
\end{equation}

\begin{equation}
	\text{Steel Plate:} \quad CR = 7 - 0.0577393\,T - 0.0001956\,T^2 - 1.7 \times 10^{-6}\,T^3
\end{equation}

\begin{equation}
	\text{Timber:} \quad CR = 7 - 0.0584921\,T - 0.0003144\,T^2 - 2.4 \times 10^{-6}\,T^3
\end{equation}

\begin{equation}
	\text{Concrete:} \quad CR = 7 - 0.0507576\,T - 0.0002625\,T^2 - 1.9 \times 10^{-6}\,T^3
\end{equation}

The ratings in the state of New York goes from 1 to 7, where 7 indicates perfect condition, 5 indicates minor deterioration but still functioning as designed, 3 indicates serious deterioration or not functioning as designed, and 1 indicates a failed condition. Ratings 2, 4, and 6 are used to assign a middle ground between the odd numbered ratings. If the failure state is defined as a condition rating 3 and the component no longer functioning as designed, then the service life of each component can be estimated (Board, 2015).

\paragraph{\citet{stukhart1991study}}

\citet{stukhart1991study} presented numerous equations predicting the condition rating for bridge decks, superstructures, and substructures to the Texas Department of Transportation (TxDOT). The majority of the equations either use NBI data for Texas bridges or the opinion of expert engineers. Part of the equations were determined using regression analysis by the Transportation Systems Center and is a function of age and ADT. Additional equations developed using the NBI data for Texas bridges, related age and ADT to condition ratings. Linear, piecewise linear, and nonlinear equations were proposed. Further, more equations were developed based on expert opinion considering the worst case scenario, the most likely scenario, and the best-case scenario. The nonlinear equations are terminated at the minimum value, and beyond this point, the condition rating would appear to increase, which is not possible without maintenance. As only bridges without maintenance, repair, or rehabilitation were used in the analysis, the rating should not increase with increasing age. Piecewise linear equations were determined for different functional classifications for the deck, superstructure, and substructure condition ratings. The coefficients B0, B1, B2, and B3 used in the piecewise linear equations are presented in the following table. The condition rating CR is described by three linear equations that are applicable during certain times (t) of the bridge life:

\begin{equation}
	CR =
	\begin{cases}
		B_0 + B_1 t, & \text{if } t \leq t_1 \\
		B_0 + B_1 t_1 + B_2 (t - t_1), & \text{if } t_1 < t < t_2 \\
		B_0 + B_1 t_1 + B_2 (t_2 - t_1) + B_3 (t - t_2), & \text{if } t \geq t_2
	\end{cases}
\end{equation}

\noindent
The authors defined $t_1$ and $t_2$ as 25 and 45 years, respectively. A nonlinear regression analysis was performed to determine the parameters for the best-fit exponential decay curve. Parameters were determined for bridge decks and superstructures based on functional classification using the multiyear data set. The best-fit parameters and equations were used to estimate the service life of bridge decks and superstructures.

\medskip

\noindent
The basic equation used to estimate the service life is shown in the following equation:

\begin{equation}
	CR = \beta_1 e^{t / \beta_2}
\end{equation}

\noindent
The estimated service life is approximately equal to the absolute value of $\beta_2$ for this set of data. Looking at the values of $\beta_2$, the only reasonable values are for decks of prestressed concrete bridges on the state and farm-to-market highway systems.

\medskip

\noindent
Finally, more equations were developed based on a survey of Texas bridge engineers’ opinions. They were asked to provide estimates of the worst-case, the most likely, and the best-case expected remaining service life based on expert opinion. The predicted remaining service life was based on a given condition rating: 9 for new, 7 for good, 5 for fair, and 3 for poor condition. From these responses, an estimated condition rating deterioration rate was determined. As is usual with opinion-based surveys, there was significant variation in the responses.

\subsubsection{Time Series Models}

Time series models are used to predict future condition values based on previously observed conditions. The general representation of an autoregressive model, well known as AR($p$), is:

\begin{equation}
	y_t = \alpha_0 + \alpha_1 y_{t-1} + \alpha_2 y_{t-2} + \cdots + \alpha_p y_{t-p} + \varepsilon_t
\end{equation}

\noindent
in which $y_t$ represents the condition of a bridge at time period $t$, $y_{t-p}$ represents the condition of the bridge at time period $t - p$, $\alpha$ are the coefficients to be estimated, and $\varepsilon_t$ is the error term.

\subsubsection{Panel Data Models/Dynamic Panel Models}

Panel data contain observations of multiple individuals, $N$, obtained over multiple time periods, $T$, where $N \gg T$. Panel data models have been applied to bridge deterioration modeling \citep{madanat1997probabilistic} (e.g., Madanat et al., 1997; Bulusu and Sinha, 1997). A general panel model has the form:

\begin{equation}
	y_{it} = \alpha + x_{it}' \beta + u_{it}, \quad i = 1, \dots, N; \quad t = 1, \dots, T
\end{equation}

\noindent
where $i$ is the individual dimension and $t$ is the time dimension. Different assumptions can be made on the precise structure of this general model, including the fixed effects model and the random effects model.

\medskip

\noindent
The fixed effects model assumes the error term has the following form:

\begin{equation}
	u_{it} = u_i + v_{it}
\end{equation}

\noindent
in which $v_{it}$ is a normally distributed random error term. $u_i$ is the unobserved individual-specific, time-invariant effect, which is assumed to be correlated with the explanatory variables.

\medskip

\noindent
The random effects model assumes $u_i$ is a random variable uncorrelated with the explanatory variables, and $u_i \sim \mathcal{N}(0, \sigma^2)$. Typical estimation methods for panel data models include the first difference (FD) estimator and the within-group (WG) estimator.

\medskip

\noindent
Unlike traditional panel data models, dynamic panel data models include lagged levels of the dependent variable as regressors:

\begin{equation}
	y_{it} = \alpha + x_{i,t-1} + x_{it}' \beta + u_{it}, \quad i = 1, \dots, N; \quad t = 1, \dots, T
\end{equation}

\noindent
Since lags of the dependent variable are correlated with the lagged error terms, traditional estimators such as FD and WG estimators are inconsistent. This is because the lagged dependent variables are correlated with the error terms. Therefore, instrumental variables from deeper lags need to be used to obtain consistent estimators.

\subsubsection{Discrete Choice Models}

Discrete choice models are usually used to estimate discrete condition indicators (e.g., NBI condition ratings) \citep{zhang2018nested}. Discrete choice models can be expressed mathematically using utility functions. Let $U_{ni}$ be the utility that the $n$th bridge (or bridge component) receives from being in condition state $i$. The behavior of the bridge is utility-maximizing: the $n$th bridge chooses the condition state that provides the highest utility. 

The condition state of the $n$th bridge is designated by a set of binary variables $y_{ni}$ for each condition state:

\begin{equation}
	y_{ni} =
	\begin{cases}
		1, & \text{if } U_{ni} > U_{nj}, \quad \forall j \neq i \\
		0, & \text{otherwise}
	\end{cases}
\end{equation}

The utilities $U$ are further defined as functions of explanatory variables in a linear form:

\begin{equation}
	U_{ni} = \beta x_{ni} + \varepsilon_{ni}
\end{equation}

\noindent
where $x_{ni}$ is a vector of observed explanatory variables relating to condition state $i$, $\beta$ is the corresponding vector of coefficients, and $\varepsilon_{ni}$ captures the impact of all unobserved factors that affect the choice.

\medskip

\noindent
The probability of being in different condition states is then given by:

\begin{equation}
	P(y_{ni} = 1) = P(U_{ni} > U_{nj}, \quad \forall j \neq i)
\end{equation}

\noindent
Different distributional assumptions of $\varepsilon_{ni}$ lead to different forms of discrete choice models, including Binary Logit, Binary Probit, Multinomial Logit, Conditional Logit, Multinomial Probit, Nested Logit, Generalized Extreme Value Models, Mixed Logit, and Exploded Logit.

\subsubsection{Duration Models or Survival Models or Reliability Models}

Duration models, sometimes referred to as reliability models or survival models, are a probabilistic approach for predicting the likelihood of a continuous dependent variable passing beyond or ``surviving'' at any given unit of time \citep{ng2009optimal,klatter2003life,van2004use,hearn2007service,agrawal2008bridge}. These models often involve statistical analysis under data censoring, where the failure time is unknown for some of the bridge facilities.

\medskip

\noindent
A key distinguishing feature of survival analysis from classical regression is its ability to handle censored data. The only information available for some bridge structures is that they have survived up to a certain time, but are no longer followed afterwards. This type of censoring is called \textit{right censoring}.

\medskip

\noindent
For right-censored data, the actual information for the $i$th bridge structure, $i = 1, \dots, n$, is contained in the pair $(t_i, d_i)$, where $t_i$ is the failure time and $d_i$ is the censoring indicator. The censoring indicator takes the value one if the failure event has been observed, and zero otherwise (i.e., censored). It can be defined as:

\begin{equation}
	d_i =
	\begin{cases}
		1, & \text{if } t_i \leq c_i \\
		0, & \text{if } t_i > c_i
	\end{cases}
\end{equation}

\noindent
where $c_i$ is the censoring time.

\medskip

\noindent
For a random time to failure $T$, the probability density function (PDF) is defined as $f(t)$, and the cumulative distribution function (CDF) is defined as:

\[
F(t) = P(T \leq t)
\]

Two other commonly used functions in survival analysis are:

\begin{itemize}
	\item \textbf{Survival function:} $S(t) = P(T > t) = 1 - F(t)$
	\item \textbf{Hazard function:} $h(t) = \dfrac{f(t)}{S(t)}$
\end{itemize}

\noindent
The hazard function can be interpreted as the instantaneous rate of failure given survival up to time $t$. The assumptions about the form of the hazard function lead to different reliability models, such as the Weibull and lognormal models.

\medskip

\noindent
The hazard function is usually modeled as a function of explanatory variables representing traffic and environmental factors.

\paragraph{\citet{ng1996prediction}}
A bridge deterioration model is an essential component of a computerized bridge management system (BMS). Existing BMSs use Markov chain theory to model the deterioration process as a decay of condition ratings over time. An alternative approach based on time-dependent reliability theory is proposed by the authors. The new approach is in principle a generalization of the Markov chain models. Rather than addressing the stochastic nature of condition rating, the proposed approach seeks to model the random time using survival analysis. The ratings are random variables. Therefore, the probability of these ratings may reach or exceed a certain threshold value within the time interval [0,t]. The authors define the reliability function S(t) as the probability of survival of a system within the time [0,t]. In other words, it is the probability that the time to failure exceeds the time, t:

\begin{equation}
	S(t) = P[T > t], \quad t \geq 0
\end{equation}

\noindent
where $T$ is a non-negative random variable representing the time to failure and is commonly known as the failure time or lifetime. The threshold value for failure was specified at the condition rating of 3. The reliability function $S(t)$ expresses the reliability of a new bridge at any point in time.

\medskip

\noindent
For an in-service bridge, the authors used an equivalent function known as the \textit{hazard function}, $h(t)$. The hazard function specifies the instantaneous rate of failure at time $t$, given that the individual has survived up until time $t$. Therefore:

\begin{equation}
	h(t) = \frac{f(t)}{S(t)}
\end{equation}

\noindent
where $f(t)$ is the probability density function of $T$. Given the distribution of $T$ in any of these forms, information about the remaining life and future bridge performance can be determined.

\paragraph{\citet{klatter2003life}}

A methodology for a probabilistic life-cycle cost approach to bridge management was applied to the concrete highway bridges in the Netherlands. The Dutch national road network contains over 3,000 highway bridges, most of which are 30 years old or more. The annual maintenance cost of these bridges is a substantial part of the total maintenance cost. The question arises of when to carry out bridge replacements. A fundamental solution is to take a life-cycle cost approach with costs of maintenance and replacement and service lifetime as key elements. Maintenance strategieswere drawn up for groups of similar elements, such as concrete elements, preserved steel, extension joints, and bearings. The structures were categorized into generic types, each with its own maintenance characteristics. For each structure, the maintenance cost was estimated on the basis of the life-cycle cost analyses of the underlying elements. After aggregation over the entire stock, this process eventually led to the maintenance cost on a network level. To calculate the life-cycle cost, lifetime distributions for concrete bridges were determined, and the expected cost of replacing the bridge stock was computed. The uncertainty in the lifetime of a bridge can best be represented with a Weibull distribution, which can be fit on the basis of aggregating the lifetimes of demolished bridges (complete observations) and the ages of current bridges (rightcensored observations). Using renewal theory, the future expected cost of replacing the bridge stock can then be determined while taking into account current bridge ages and the corresponding uncertainties in future replacement times. The proposed methodology has been used to estimate the cost of replacing the Dutch stock of concrete bridges as a function of time.

\paragraph{\citet{van2004use}}
This authors proposed a new method to determine lifetime distributions for concrete bridges and to compute the expected cost of maintaining and replacing a bridge stock. The uncertainty in the lifetime of a bridge can be represented with a Weibull distribution. It is recommended to fit this Weibull distribution on the basis of aggregating the lifetimes of demolished bridges (complete observations) and the ages of current bridges (right-censored observations). Using renewal theory, the future expected cost of replacing the bridge stock can then be determined while taking account of the current bridge ages and the corresponding uncertainties in the future replacement times. The same approach has been applied to the maintenance times of bridge elements. The proposed method is used to estimate the expected cost of maintaining and replacing the Dutch stock of concrete viaducts and bridges as a function of time (Van Noortwijk and Klatter, 2004).

\paragraph{\citet{hearn2007service}}

This study examines costs and performance of four types of reinforced concrete bridge decks currently in service on Colorado DOT (CDOT) highway bridges. These four types allow a comparison between bare decks and decks with waterproofing membranes, and between decks with uncoated steel reinforcement and decks with epoxy-coated steel reinforcement. Histories of deck condition ratings are used to estimate deck service life and to generate population models of service life. Decks with waterproofing membrane have longer service life than bare decks. Condition data indicate longer service life for decks with uncoated reinforcing steel, but this outcome may be due to the limited extent of condition data for decks having epoxy-coated reinforcement. Costs for bridge decks are evaluated as initial costs, present values, and annualized costs. By all present value and annualized cost measures, decks with waterproofing membrane are least expensive. This outcome is not sensitive to the value of the discount factor. Service life is taken as the time required, in years, for a new bridge deck to reach NBI condition rating 5. Individual estimates of deck service life are used to compute discrete, cumulative probability of reaching condition rating five.

\begin{equation}
	D(x_i) = \frac{n_i}{N}
\end{equation}

\noindent
where $N$ is the total number of decks in a population, $x_i$ is a sorted list of service life values running from least time to greatest time, $n_i$ are index values ($1, 2, 3, \dots, N$) corresponding to the service life values $x_i$, and $D(x_i)$ are discrete fractional values of probability.

\medskip

\noindent
The error between discrete probability values and the population models is computed as:

\begin{equation}
	\text{Err} = \sum (D(x_i) - F(x_i))^2
\end{equation}

\noindent
Error is evaluated at every data point and summed for overall error between the discrete data and each population model.

\subsubsection{Weibull Survival Models}

Weibull survival models are used to model the probability of failure. These models are useful for modeling any change in state and can be applied based on age. They utilize the actual scatter in duration data for a particular state by considering the duration as a random variable. Weibull models are considered more reliable for calculating deterioration rates.

\begin{equation}
	S_i(t) = e^{-\left( \frac{t}{\eta_i} \right)^{\beta_i}}, \quad t > 0,\ \beta_i > 0,\ \eta_i > 0
\end{equation}

\noindent
where $\beta_i$ and $\eta_i$ are a pair of shape and scale parameters.

\subsubsection{Markov Models}

Markov Models describe probability of transition between discrete states for fixed time. They can estimate how likely a structure is to degrade a particular amount over a given period of time. In these models, the next condition depends only on the current condition of the structure and does not relate to the historical condition of a bridge or structure. The deterioration progress is divided into discrete states which are derived from periodical inspection information.

\subsubsection{Markov Chain Based Models}

Markovian chain theory is one of the most widely used methods in bridge deterioration modeling \citep{jiang1989bridge,estes2001bridge,ertekin2008lcca}. In this approach, the condition of a bridge is first discretized into $n$ states in terms of its condition index. Hence, bridge condition can be represented by a condition state probability vector:

\begin{equation}
	X_t = [X_{1t},\ X_{2t},\ \dots,\ X_{nt}]'
\end{equation}

\noindent
where: \\
$X_t$ = condition state probability vector of a bridge at year $t$;\\
$X_{it}$ = probability that a bridge is in condition state $i$ at year $t$, for $i = 1, \dots, n$, where larger $i$ corresponds to a worse condition state, and
\[
\sum_{i=1}^n X_{it} = 1
\]

\medskip

\noindent
The deterioration process of a bridge can be expressed by the change in the elements of the condition state probability vectors. A Transition Probability Matrix (TPM) is used to represent this change. A typical TPM for bridge deterioration can be expressed as:

\begin{equation}
	P_t =
	\begin{bmatrix}
		P_{11t} & P_{12t} & \cdots & P_{1nt} \\
		\vdots & \ddots & \ddots & \vdots \\
		0 & 0 & \cdots & 1
	\end{bmatrix}
\end{equation}

\noindent
where: \\
$P_t$ = Transition Probability Matrix (TPM) of year $t$;\\
$P_{ijt}$ = probability that the condition will deteriorate from state $i$ to state $j$ in year $t$ if $i < j$, or stay in the same state if $i = j$.

\medskip

\noindent
Under the Markovian assumption, the future condition states of a bridge depend only on its current condition state. Previous states do not impact future condition. Therefore, to calculate the future condition state probability, only the present condition state probability vector and the TPM are needed. 

Since a bridge cannot improve to a better condition state on its own, $P_{ijt} = 0$ when $i > j$. Furthermore, the value of 1 in the last row of the TPM corresponding to state $n$ indicates that the bridge condition cannot deteriorate further.

\medskip

\noindent
From the above, the deterioration process of bridge condition without maintenance intervention can be expressed as:

\begin{equation}
	X_{t+1} = X_t P_t
\end{equation}

\noindent
An advanced integrated method using state-/time-based models to build a reliable transition probability for predicting long-term performance of bridge elements has been studied \citep{bu2012performance}.

\paragraph{\citet{jiang1989bridge}}

\citet{jiang1989bridge} discussed the results of regression analysis and Markov chain analysis to estimate the average rating of a group of bridges. They considered Interstate and non-Interstate bridges, steel and concrete bridges. At the beginning the authors considered traffic volume and geographic location, but later did not consider them as separate categories because they did not seem to influence the regression analysis. A relatively small sample was used in the regression analysis. Thus, the results may have been influenced by the limited amount of data available and used. The results of the regression analysis were coefficients for a third-order polynomial describing the NBI condition rating as a function of bridge age. Coefficients were determined for the different bridge types and the deck, superstructure, and substructure. The constant term in the prediction equation is always 9, which represents that the bridge component was in perfect condition when new. The end of service life is assumed to occur when the condition rating reaches a value of 3. The authors developed the following polynomial model for a concrete bridge superstructure:

\begin{equation}
	CS(t) = 9.0 - 0.28877329\,t + 0.0093685\,t^2 - 0.00008877\,t^3
\end{equation}

\noindent
where $CS(t)$ is the condition rating of the bridge at time $t$, and $t$ is the age of the bridge in years.

\paragraph{\citet{zhang2003determination}}

This paper presents the results of a study using Louisiana's National Bridge Inventory (NBI) to determine matrices of the deterioration of bridge components (elements) that can be used in Pontis software. These matrices are for a simplified three-element bridge preservation model the authors developed for the Louisiana Department of Transportation and Development. The results indicate that the deterioration of bridge components does not correlate well with the age of bridges and that the Markov transition probability is a good tool for predicting bridge deterioration. When using state NBI data to generate Markov matrices of bridge components’ deterioration, the resulting probabilities will be affected by the average bridge age in the database and the time intervals depending upon which NBI data are initially analyzed. This fact should be considered in the use of these deterioration matrices for predicting future bridge deterioration. An element deterioration matrix is of the form:

\begin{equation}
	D_t =
	\begin{pmatrix}
		d_{11} & d_{12} & d_{13} & d_{14} & d_{15} \\
		0      & d_{22} & d_{23} & d_{24} & d_{25} \\
		0      & 0      & d_{33} & d_{34} & d_{35} \\
		0      & 0      & 0      & d_{44} & d_{45} \\
		0      & 0      & 0      & 0      & d_{55}
	\end{pmatrix}
\end{equation}

\noindent
where $d_{ij}$ is the probability with which a bridge component (element) deteriorates from a rating $i$ to a rating $j$ over a $t$-year interval.

\medskip

\noindent
The deterioration matrix $D_t$ was developed in this study from the transition probability matrix:

\begin{equation}
	D_t =
	\begin{pmatrix}
		p_{11} & p_{12} & p_{13} & p_{14} & p_{15} \\
		0 & p_{11} + \frac{1}{2}p_{21} & p_{23} + \frac{1}{6}p_{21} & p_{24} + \frac{1}{6}p_{21} & p_{25} + \frac{1}{6}p_{21} \\
		0 & 0 & p_{33} + \frac{3}{5} \sum_{j=1}^{2} p_{3j} & p_{34} + \frac{1}{5} \sum_{j=1}^{2} p_{3j} & p_{35} + \frac{1}{5} \sum_{j=1}^{2} p_{3j} \\
		0 & 0 & 0 & p_{44} + \frac{7}{10} \sum_{j=1}^{3} p_{4j} & p_{45} + \frac{3}{10} \sum_{j=1}^{3} p_{4j} \\
		0 & 0 & 0 & 0 & p_{55} + \sum_{j=1}^{4} p_{5j}
	\end{pmatrix}
\end{equation}

\noindent
This is an empirical formula, which is based upon the consideration that historically, Louisiana had no bridge preservation program in place, and most actions on bridges were conducted through the federal bridge replacement program or by local maintenance crews. In most cases, bridges were replaced when they were in very poor condition (NBI rating 3 or 4).

\paragraph{\citep{hallberg2005development}}

Lifetime Engineering (or Life Cycle Engineering) is a technical approach for meeting the current objective of sustainable development. The approach is aimed to turn today’s reactive and short-term design, management, and maintenance planning towards an optimized and long-term technical approach. The life cycle based management and maintenance planning approach include condition assessment, predictive modeling of performance changes, maintenance, repair and refurbishment planning and decisions. The Life Cycle Management System (LMS) is a predictive and generic life cycle based management system aimed to support all types of decision making and planning of optimal maintenance, repair, and refurbishment activities of any constructed works. The system takes into account some aspects in sustainable and conscious development such as human requirements, life cycle economy, life cycle ecology and cultural requirements. The LMS is a system by which the complete system or parts thereof, works in co-operation or as a complement to existing business support systems. The system is module based where each module represents a subprocess within the maintenance management process. The scope of this thesis is focused on development and adaptation of the predictive characteristic of LMS towards a presumptive user. The objective is to develop and adapt a Service Life Performance Analysis module applicable for condition based Facility Management System in general and for condition based Bridge Management System in particular. Emphasis is placed on development and adaptation of a conditional probability based Service Life Performance Analysis model in which degradation models and Markov chains play a decisive role. The thesis deals also with development and adaptation of environmental exposure data recording and processing, with special emphasis on quantitative environmental classification to provide a simplified method of Service Life Performance Analysis \citep{hallberg2005development}. The input data for the model consisted in inspection, material, and environmental data. The authors assumed that a known degradation function is denoted by Y(n), then,

\begin{equation}
	E(X(n, P)) \approx Y(n)
\end{equation}

\noindent
where $n$ is time, $P$ is the transition matrix, and the elements $p_{ij}$ in the transition matrix $P$ are determined numerically by an iterative process such that the sum of the differences between the known degradation function $Y(n)$ and the Markov chain function $E(X(n, P))$ is minimized.

\begin{equation}
	\min \sum_{n=1}^{N} \left| Y(n) - E(X(n, P)) \right|
\end{equation}

\noindent
where $N$ is the total time period.

\paragraph{\citep{ertekin2008lcca}}

Bridge Life Cycle Cost Analysis (BLCCA) has received a great deal of attention, especially in the last decade. Currently, there is no consensus on the required level of detail in performing a BLCCA concerning the number of elements that should be studied to attain a certain level of accuracy. However, some analysts, as mentioned in the NCHRP-483 report, suggest that considering three elements (substructure, deck, and superstructure) yields an adequately detailed description of most highway bridges.

\medskip

\noindent
The majority of previous studies have focused more on developing algorithms for one individual element of the bridge or treating the bridge itself as a single element, rather than considering all components as a system. Such a system-based approach would make the empirical results more realistic. Thus, the main objective of this paper is to develop a comprehensive BLCCA methodology using real data available in the National Bridge Inventory (NBI). In this study, the inventory is primarily used for modeling the deterioration behavior of bridges.

\medskip

\noindent
The proposed methodology predicts both agency and user costs using a Genetic Algorithm for cost optimization and a Markov-chain approach for deterioration modeling. Monte Carlo Simulation is used to address uncertainties. The BLCCA algorithm developed by the authors can serve as a valuable tool for efficiently allocating limited public resources and for maintaining all parts of the bridges at acceptable performance levels.

\medskip

\noindent
To validate the effectiveness of the proposed algorithm, results from a hypothetical case study are presented. These results verify that the methodology can be successfully applied to steel and concrete superstructure bridges, which constitute the majority of bridges.

\medskip

\noindent
The initial condition state of the bridge is denoted as $CS_i$. As the bridge deteriorates, its condition state changes. The final state of the bridge is denoted as $CS_f$. The probability that the condition state of a bridge will change from $CS_i$ to $CS_j$ over a given time interval is denoted as $T_{ij}$. Assuming there are 10 unique condition states, the transition probability matrix $T$ will be a $10 \times 10$ matrix. The entire transition matrix $T$ can be represented by a vector $P$ that contains its diagonal elements:

\begin{equation}
	P = [P_{11},\ P_{22},\ P_{33},\ P_{44},\ P_{55},\ P_{66},\ P_{77},\ P_{88},\ P_{99},\ P_{10,10}]
\end{equation}

\noindent
The final condition state probability distribution $CS_f$ for a bridge component with an initial distribution $CS_i$ after $N$ transitions is given by:

\begin{equation}
	CS_f = CS_i \cdot T^N
\end{equation}

\subsubsection{Semi-Markov Models}

These models come under a class of stochastic process which moves from one state to another with successive states visited forming a Markov chain. Semi-Markov models make use of random holding time rather than fixed regular intervals. Random holding time helps to incorporate ‘time factor’ into the model. Probabilistic distribution function is employed to predict future condition state of bridge elements. The model assumes that transition probability depends on time spent in initial condition state which gives more realistic results.

\subsubsection{Stochastic Gamma Process Deterioration Models}

In this type of model, increments are independent, non-negative random variables, such as loss of steel section due to corrosion. The variables follow a gamma distribution with a time-dependent shape function and a constant scale parameter. These models are also suitable for modeling gradual damage that monotonically accumulates over time, such as corrosion, wear, erosion, and creep of materials.

\begin{equation}
	f(t) = \frac{\beta^{\lambda(t)}}{\Gamma(\lambda(t))} \, x^{\lambda(t) - 1} \, e^{-\beta x}
\end{equation}

\noindent
where: \\
$t$ = the time interval (or age); \\
$x(t)$ = random variable denoting the cumulative amount of deterioration; \\
$\Gamma(\lambda(t))$ = Gamma function.

\paragraph{Nebraska \citep{hatami2011developing}}

\citet{hatami2011developing} presented in a report the results of a project for the Nebraska Department of Roads in which they developed deterioration models for Nebraska bridges. The models used NBI condition ratings for bridge decks, superstructures, and substructures based on data from 1998 to 2010. The authors used only data for state bridges in the analysis because they believed that inspections done by state inspectors have stricter requirements. These models were defined using deterministic and stochastic approaches. The following factors were considered in the development of the deterioration models:
\begin{itemize}
	\item Structure type
	\item Deck type
	\item Wearing surface
	\item Deck protection
	\item Average daily traffic (ADT)
	\item Average daily truck traffic (ADTT)
	\item Location
\end{itemize}

The authors developed several equations for the deterioration model. For example, they used the following piecewise equations in which deterioration is related to Average Daily Truck Traffic (ADTT):

\begin{equation}
	R(T) =
	\begin{cases}
		10.189 - 0.233\,T + 0.0092\,T^2 - 0.0002\,T^3, & \text{if } \text{ADTT} < 100 \\
		10.754 - 0.342\,T + 0.0127\,T^2 - 0.0002\,T^3, & \text{if } 100 < \text{ADTT} < 500 \\
		10.372 - 0.2311\,T + 0.0039\,T^2 - 0.00004\,T^3, & \text{if } \text{ADTT} > 500
	\end{cases}
\end{equation}

\subsubsection{Artificial Intelligence Models}

Artificial intelligence (AI) models or machine learning models make use of computer techniques that aim to automate intelligent behaviors. AI models consist of expert systems, artificial neural networks (ANN) and case-based reasoning (CBR). ANN models are non-linear, adaptive models which predict conditions based on what it has learnt from past data. The most common learning technique is a non-linear form of three-stage least squares regression, where instruments are estimated to predict future events. ANN models are based on a Bayesian network approach which updates estimates by applying weighted averages based on previous estimates. The weights are based on the number of observations. These models can be better used for smaller databases; however, the ability to adapt makes these models particularly useful in applications where adjustment of predictions due to new data, such as inputs from a monitoring system, can be made in real-time. Recently, deep leanring models have been widely used in infrastructure performance modeling \citep{gao2025modeling,peng2025evaluating,lebaku2024deep,gao2024considering,gao2023deep,yu2023pavement,gao2022missing,gao2021detection,xu2021development,vemuri2020pavement}.

\paragraph{\citet{sobanjo1997neural}}

\citet{sobanjo1997neural} used a multi-layer perceptron model (MPL) to predict condition ratings of bridge superstructures using the age of the bridge parameter as the only input. The network was able to identify 79\% of the ratings correctly by testing a set of 38 ratings.

\paragraph{\citet{cattan1997analysis}}

\citet{cattan1997analysis} developed a neural network approach to predict the condition rating of railway bridges in the metropolitan area of Chicago. The output of the ANN model was the overall condition of the bridge on a rating scale of 1 to 5. The input vector contained physical characteristics of the bridge. The input parameters used were bridge height, bridge length, number of tracks, number of spans, span length, span built, substructure built, deck built, span condition, substructure condition, and deck condition. Many combinations of input parameters were tested and as a result performing network was able to precisely identify 73\% of the ratings with testing data that was not previously seen.

\paragraph{\citet{li2010using}}

\citet{li2010using} used ANN to predict the rating of certain bridge elements. They compared different ANN methods on predicting bridge abutment condition ratings in the state of Michigan. The ANN models were responsible for predicting the discrete National Bridge Inventory (NBI) abutment rating based on physical and operational bridge parameters. The models compared were the MLP, radial basis function (RBF), support vector machine (SVM), supervised self-organizing map (SSOM), fuzzy neural network (FNN), and the ensemble neural network (ENN). Based on predictions from the MLP model, lifetime deterioration curves were created. The results were acceptable, 56\% of the time the rating was predicted correctly, identifying the rating within  1 of the true rating 87\% of the time, and successfully identifying a damaged abutment 65\% of the time. A damaged abutment is assumed to happen when the condition rating reaches a value of 4 or lower.

\paragraph{\citet{morcous2002comparing}}

\citet{morcous2002comparing} compared an ANN model and a case-based reasoning (CBR) model to predict concrete bridge deck condition ratings. The physical and operational parameters of the bridges were taken from the Quebec Ministry of Transportation database. As a result, 33\% of the predicted ratings fell within a tolerance of 0.1 of the original rating, and 100\% fell within 1.0 of the original rating. The rating scale ranges from 1 to 6.

\paragraph{\citet{lee2008improving}}
\citet{lee2008improving} developed ANNs to generate past condition ratings and used them to predict future NBI deck ratings. The networks predicted the condition ratings based on non-bridge parameters related to climate, traffic, and population changes. The networks were allowed to make non-discrete rating predictions, but the NBI rating scale was used as the condition rating. The model predicted 79\% of the historical data within 10\% of the actual ratings. Using these generated ratings, future condition rating predictions were made based only on past condition ratings, and the average error was 3%.

\paragraph{\citet{melhem2003prediction}}
The authors demonstrated that it is feasible to use either the k-nearest-neighbor instance-based learning (IBL) technique or the inductive learning (IL) technique for engineering applications in classification and prediction problems such as estimating the remaining service life of bridge decks. It is shown that IBL is more efficient than IL: The best achieved percentages of correctly classified instances are 50\% as generated by k-nearest-neighbor IBL and 41.8\% when generated by the C4.5/IL learning algorithm. From a machine learning (ML) standpoint both these values are considered low, but this is attributed to the fact that the deterioration model used to compute the remaining service life turned out to be inadequate. It is based on a methodology developed under the Strategic Highway Research Program (SHRP) for life-cost analysis of concrete bridges relative to reinforcement corrosion. Actual bridge deck surveys were obtained from the Kansas Department of Transportation that include the type of attributes needed for the SHRP methodology. The experimentation with the ML algorithms reported in this paper also describes the experience of facing an imperfect model, or with incomplete data or missing attributes. The input to a ML program that learns to diagnose bridge failures may consist of descriptions of bridge structure, traffic pattern, and historical performance data. The output may be a knowledge structure that can concisely explain the reasons for previous failures, or predict when another bridge component can be expected to fail or may need to be replaced. 

\paragraph{\citet{narasinghe2006service}}
This paper presents a methodology for predicting reliability based remaining service lives and estimation of serviceability conditions for masonry arch bridges using Artificial Neural Networks (ANNs). In this ANNs analysis, training was processed by Back-Propagation (BP) Algorithm with corresponding parameters. The critical failure mode of the masonry arch bridge is based on axle loads. The parameters for Back-Propagation are mean values ($M \mu$) and standard deviations ($M\sigma$) of proposed safety margin of the masonry arch bridge. Those parameters were used to predict the serviceability condition of the masonry arch bridges. Finally, the remaining service life of the masonry arch bridge was determined using a target failure probability, while assuming that the current rate of loading magnitude and frequency are constant for future prediction. Proposed methodology is illustrated with a case study bridge selected in the national road network of Sri Lanka.

\subsection{Case Study Using Texas Bridge Data}

\subsubsection{Data}

Raw inventory data (1994 \~{} 2016) retrieved from the TxDOT NBI database was used to develop the condition prediction model. Data from the year 2000 was missing. In order to exclude repair or reconstruction effects on the modeling, data records where condition improved between consecutive years were removed from the database. In total, 289,529 data points were used in the estimation. 

\subsubsection{Variables}

Several sets of prediction variables were applied to develop prediction models. The 9 input variables are (the numbers represent their index in the NBI database): 

\begin{itemize}
	\item 2 District
	\item 3 County
	\item 27 Yr Built
	\item 29 AADT
	\item 31 Design Load
	\item 34 Skew
	\item 43 1 Mn Span Ty
	\item 49 Str Lgth
	\item 52 Deck Width
	\item Lagged 67 Str Eval
\end{itemize}

While 27 Yr Built 29 AADT, 49 Str Lgth, 52 Deck Width, and Lagged 67 Str Eval are continuous variables, the rests are treated as categorical variables. The variable 27 Yr Built is converted to age of the bridge (as in 2016). The variable “67 Str Eval” was designated as the response variable.

\subsubsection{Models}

The following models were tested using the sklearn machine learning package . 

Classification models:
\begin{itemize}
	\item Naïve bayes
	\item Logistic regression
	\item Decision Tree Classifier
\end{itemize}

Regression models:
\begin{itemize}
	\item Linear regression
	\item Ridge regression
	\item Lasso regression
	\item Decision tree regressor
\end{itemize}

The models were evaluated by splitting the dataset into training (75\%) and testing (25\%) parts. For classification models, we used precision rate, recall rate, F1 score, accuracy, and confusion matrix to evaluate different models. For regression models, we used r2 score to evaluate different models. 

Confusion matrix is a table that reports the number of false positives, false negatives, true
positives, and true negatives.

\begin{table}[h!]
	\centering
	\caption{Table 6: Confusion Matrix}
	\begin{tabular}{|c|c|c|}
		\hline
		\textbf{True Value} & \textbf{Predicted Positive} & \textbf{Predicted Negative} \\
		\hline
		\textbf{Actual Positive} & True Positive (TP) & False Negative (FN) \\
		\textbf{Actual Negative} & False Positive (FP) & True Negative (TN) \\
		\hline
	\end{tabular}
\end{table}

\medskip

Precision rate ($P$) is the ratio of the number of records which were correctly classified as positive by the classifier to the total number of records classified as positive:

\begin{equation}
	P = \frac{TP}{TP + FP}
\end{equation}

Recall rate ($R$) is the ratio of the number of records correctly classified as positive by the classifier to the total number of actual positives:

\begin{equation}
	R = \frac{TP}{TP + FN}
\end{equation}

Accuracy ($A$) is the ratio of the number of records correctly classified by the classifier to the total number of records:

\begin{equation}
	A = \frac{TP + TN}{TP + FP + FN + TN}
\end{equation}

F1 score is defined as the harmonic mean of precision and recall:

\begin{equation}
	F1 = \frac{2 \cdot P \cdot R}{P + R}
\end{equation}

$R^2$ score indicates the proportion of the variance in the dependent variable that is predictable from the independent variables:

\begin{equation}
	R^2 = 1 - \frac{\sum_{i=1}^{n} (y_i - \hat{y}_i)^2}{\sum_{i=1}^{n} (y_i - \bar{y})^2}
\end{equation}

\medskip

\noindent
where:
\begin{itemize}
	\item $y_i$ = the true value of the dependent variable for the $i$th data point;
	\item $\hat{y}_i$ = the predicted value of the dependent variable for the $i$th data point;
	\item $\bar{y}$ = the average value of the dependent variable;
	\item $n$ = the number of data points.
\end{itemize}

The performance of these models are presented below. The best classification model is Decision Tree Classification. The best regression model is Ridge Regression. 

\begin{table}[H]
	\centering
	\caption{Table 7: Performance Comparison of Classification Models}
	\begin{tabular}{|l|c|c|c|c|}
		\hline
		\textbf{Model} & \textbf{Accuracy} & \textbf{Precision} & \textbf{Recall} & \textbf{R\textsuperscript{2} Score} \\
		\hline
		Naïve Bayes & 0.076 & 0.450 & 0.080 & -- \\
		Decision Tree Classification & 0.950 & 0.950 & 0.950 & -- \\
		Logistic Regression & 0.685 & 0.570 & 0.690 & -- \\
		Linear Regression & -- & -- & -- & 0.925 \\
		Lasso Regression & -- & -- & -- & 0.898 \\
		Decision Tree Regression & -- & -- & -- & 0.878 \\
		Ridge Regression & -- & -- & -- & 0.925 \\
		\hline
	\end{tabular}
\end{table}

\begin{figure}[H]
	\centering
	\includegraphics[width=0.7\linewidth]{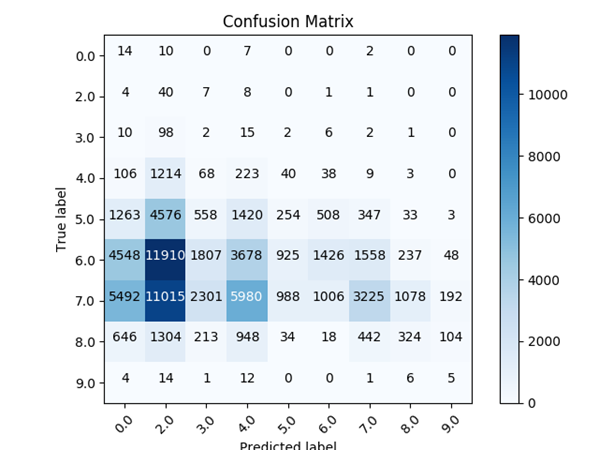}
	\caption{Confusion Matrix of Naïve Bayes Classification Model}
	\label{fig:figure20}
\end{figure}

\begin{figure}[H]
	\centering
	\includegraphics[width=0.7\linewidth]{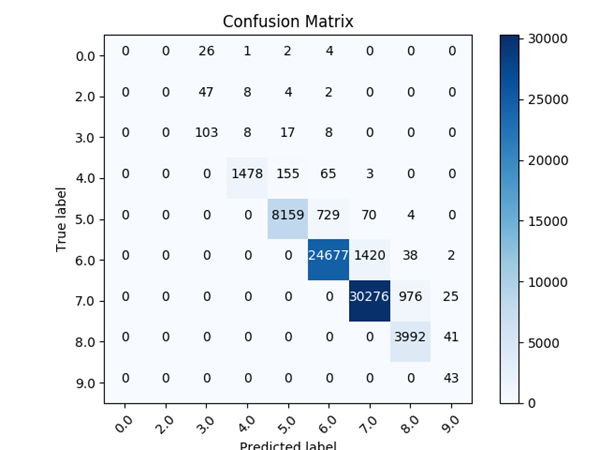}
	\caption{}
	\label{fig:figure21}
\end{figure}

\begin{figure}[H]
	\centering
	\includegraphics[width=0.7\linewidth]{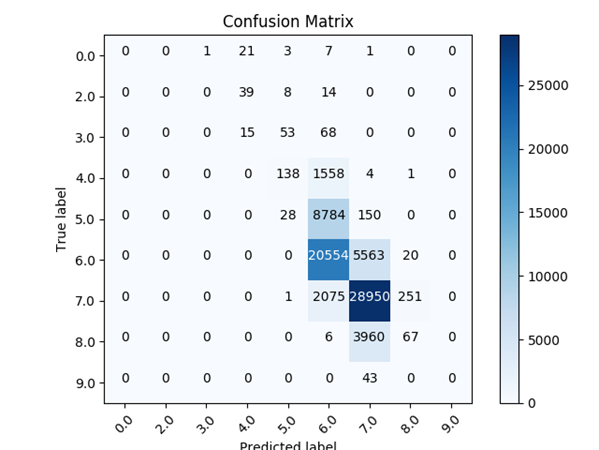}
	\caption{}
	\label{fig:figure22}
\end{figure}

\subsection{Review of Bridge Infrastructure Maintenance Planning Models}
Bridge maintenance planning models try to find the optimal balance between costs and benefits of maintenance treatments and the most appropriate time to execute maintenance. Parameters considered in the planning include the cost of failure, the cost of preventive maintenance and the cost of rehabilitation and reconstruction. The foundation of any maintenance planning model relies on the underlying deterioration process and failure behavior of the structure. Maintenance planning optimization is one of the most critical issues in bridge management since the failure of a bridge during actual operation can be a costly and dangerous event. Numerous efforts have been made to develop optimization models as maintenance strategy decision-making aids for infrastructure management \citep{gao2012network,gao2008robust,gao2013management,gao2009approximate,gao2010network,gao2012approximation,gao2013augmented,gao2010optimal,dhatrak2020considering,gao2010effect}. In the rest of this section, some popular maintenance planning models are discussed.  

The Markov Chain-based linear programming (LP) model is a discrete-time setting model, commonly used for network-level bridge maintenance planning problems. In the LP model, facilities with similar deterioration patterns are grouped together. The solution of this model determines the percentage of a group’s maintenance strategy rather than the strategy for each individual facility \citep{gao2010effect}.

\medskip

\noindent
The mathematical expression of the LP model can be explained as follows. Consider a bridge network as a set:

\[
\mathcal{S} = \{1, 2, \dots, S\}
\]

\noindent
representing different groups of facilities with homogeneous properties. Let:

\[
\mathcal{I} = \{1, 2, \dots, I\}
\]

\noindent
denote the set of condition states, where each element represents a specific condition state of a facility.

\medskip

\noindent
In each time period, a decision must be made to determine:
\begin{itemize}
	\item the proportion of the system that should receive maintenance treatment, and
	\item the type of treatment that should be applied.
\end{itemize}

\noindent
A set of basic maintenance treatments is defined as:

\[
\mathcal{M} = \{1, 2, \dots, M\}
\]

\noindent
where the $M$th treatment is considered the most effective and also the most expensive. The planning time horizon is represented by the discrete set of time periods:

\[
\mathcal{T} = \{1, 2, \dots, T\}
\]

\medskip

\noindent
Linear programming problems can be efficiently solved by the simplex algorithm. The simplex algorithm solves LP problems by constructing a feasible solution at a vertex of the polytope, then traversing the edges of the polytope to adjacent vertices with non-decreasing values of the objective function until an optimum is reached. In general, the simplex algorithm is very efficient and can be guaranteed to find the global optimum if precautions against cycling are taken. Multiple optimal solutions are also possible in linear programming problems.

\medskip

\noindent
The sets, parameters, and variables used in the LP model are presented as follows.

\begin{table}[H]
	\centering
	\caption{Table 8: Notation}
	\begin{tabular}{|p{3cm}|p{13cm}|}
		\hline
		\textbf{Category} & \textbf{Notation and Description} \\
		\hline
		\multicolumn{2}{|l|}{\textbf{Sets}} \\
		\hline
		$\mathcal{S}$ & Set of facility groups. \\
		$\mathcal{I}$ & Set of facility condition states, with $I$ representing the worst condition state. \\
		$\mathcal{M}$ & Set of maintenance treatments, with the $M$th treatment being the most effective and expensive. \\
		$\mathcal{T}$ & Set of planning periods. \\
		\hline
		\multicolumn{2}{|l|}{\textbf{Parameters}} \\
		\hline
		$B_t$ & Available budget at time period $t$. \\
		$c_{m,s,t}$ & Unit cost of applying the $m$th treatment to the $s$th facility group at time $t$. \\
		$N_s$ & Number of facilities in the $s$th facility group. \\
		$p_{ij}^{(m,s)}$ & Deterioration transition probability from condition state $i$ to $j$ when the $m$th treatment is applied to the $s$th facility group; satisfies $\sum_j p_{ij}^{(m,s)} = 1$. \\
		$\alpha_{s,i}^0$ & Initial proportion of the $s$th facility group in condition state $i$ at the beginning of time period 1. \\
		$\delta$ & Minimum requirement on the proportion of facilities in the first (best) condition state. \\
		$\boldsymbol{\alpha}_{s,t}$ & Condition state probability vector of facility group $s$ at time period $t$. \\
		$P$ & Transition probability matrix. \\
		$p_{ij}$ & Transition probability from state $i$ to $j$ in one time period: \\
		& \quad $\bullet$ If $i < j$, the probability that the facility deteriorates from $i$ to $j$. \\
		& \quad $\bullet$ If $i = j$, the probability that the facility stays in the same state. \\
		\hline
		\multicolumn{2}{|l|}{\textbf{Variables}} \\
		\hline
		$x_{s,i,m,t}$ & Proportion of the $s$th facility group in condition state $i$ receiving the $m$th treatment at time $t$. \\
		$\alpha_{s,i,t}$ & Proportion of the $s$th facility group in condition state $i$ at time $t$. \\
		$z_{s,m,t}$ & Proportion of the $s$th facility group receiving the $m$th treatment at time $t$. \\
		\hline
	\end{tabular}
\end{table}

\subsubsection{Deterioration Modeling in Markov Chain Maintenance Planning Models}

The concept of bridge condition is developed to quantitatively relate the condition of a facility to its ability to serve its users. Bridge condition is often represented by discrete ratings or states. Using discrete ratings instead of continuous indicators simplifies the computational complexity of the maintenance decision-making process, as details are not necessary at this level of management.

\medskip

\noindent
The basic idea of the discrete-time, state-based model is introduced as follows. Facility condition at different years is represented by a condition state probability vector:

\begin{equation}
	\boldsymbol{\alpha}_t = [\alpha_{1,t}, \alpha_{2,t}, \dots, \alpha_{I,t}]'
\end{equation}

\noindent
where $\boldsymbol{\alpha}_t$ is the condition state probability vector at time $t$, and $\alpha_{i,t}$ is the probability that the facility is in condition state $i$ at time $t$.

\medskip

\noindent
The deterioration process of a facility can be expressed by the change of the elements in the condition state probability vectors. A transition probability matrix $P$ can be used to calculate this change:

\begin{equation}
	\boldsymbol{\alpha}_{t+1} = \boldsymbol{\alpha}_t \cdot P
\end{equation}

\noindent
Because a facility cannot improve to a better condition state by itself, the elements $p_{ij}$ in the matrix $P$ are replaced by 0 for $i > j$. Furthermore, the value of 1 in the last row of $P$ (corresponding to the worst condition state) indicates that the condition cannot deteriorate further. Based on all the above, the future condition can be predicted as:

\begin{equation}
	\boldsymbol{\alpha}_{t+1} = \boldsymbol{\alpha}_t \cdot P
\end{equation}

\paragraph{Markov Chain Based Maintenance Planning Models }

Consider a bridge network as a set $\mathcal{S}$ of facilities. Condition $\mathcal{I}$ is defined as a set of state space with elements representing the facility condition, in which $1$ represents the best condition state and $I$ the worst. A set of basic maintenance treatments is defined as $\mathcal{M}$, where the $M$th maintenance treatment is the most effective and expensive. The planning time horizon is represented by the discrete set of time periods $\mathcal{T} = \{1, 2, \dots, T\}$. 

\medskip

During each time period, the conditions of the facilities deteriorate due to usage, aging, and environmental effects. The maintenance treatment applied at time period $t$ will affect the condition at time period $t + 1$.

\medskip

Using the discrete-time, state-based deterioration model, the infrastructure maintenance planning problem with deterministic budgets is formulated as follows:

\begin{equation}
	\max \left\{ \sum_{s \in \mathcal{S}} \left( \sum_{t = 1}^{T-1} \alpha_{s,1,t} + \alpha_{s,1,T} \right) \right\}
\end{equation}

\begin{equation}
	\alpha_{s,i,1} = \alpha_{s,i}^0, \quad \forall s \in \mathcal{S},\ i \in \mathcal{I}
\end{equation}

\begin{equation}
	\alpha_{s,i,t+1} = \sum_{j \in \mathcal{I}} \sum_{m \in \mathcal{M}} x_{s,j,m,t} \cdot p_{j i}^{(m,s)}, \quad \forall s \in \mathcal{S},\ i \in \mathcal{I},\ t \in \mathcal{T} \setminus \{T\}
\end{equation}

\begin{equation}
	\sum_{i \in \mathcal{I}} \sum_{m \in \mathcal{M}} x_{s,i,m,t} \cdot c_{m,s,t} \cdot N_s \leq B_t, \quad \forall t \in \mathcal{T}
\end{equation}

\begin{equation}
	\sum_{m \in \mathcal{M}} x_{s,i,m,t} = \alpha_{s,i,t}, \quad \forall s \in \mathcal{S},\ i \in \mathcal{I},\ t \in \mathcal{T}
\end{equation}

\medskip

The objective of the planning problem is to maximize the proportion of all facilities in the best condition state over the planning horizon. The first term inside the parentheses in Equation (47) represents the cumulative proportion from time period 1 to time period $T - 1$. The second term represents the proportion in the best condition at time $T$, because a facility's condition at time $T$ is fully determined by its condition and applied maintenance at time $T - 1$.

\medskip

Equation (48) specifies the initial condition of each facility group at the beginning of the planning horizon. Equation (49) models the deterioration process and transition probabilities between condition states. Equation (50) ensures that the annual expenditure of maintenance activities does not exceed the available budget $B_t$. Equation (51) ensures that the proportion of treatments assigned equals the condition state proportions.

\medskip

Once the decision variables are obtained, the condition of each facility group is updated as:

\begin{equation}
	\alpha_{s,i,t} = \sum_{j \in \mathcal{I}} \sum_{m \in \mathcal{M}} x_{s,j,m,t-1} \cdot p_{j i}^{(m,s)}
\end{equation}

\medskip

The total maintenance decision for each group is calculated as:

\begin{equation}
	z_{s,m,t} = \sum_{i \in \mathcal{I}} x_{s,i,m,t}
\end{equation}

\subsubsection{Integer Programming (IP)}
The Integer Programming (IP) model is another discrete-time approach to solving multi-facility maintenance planning problems under budget constraints. The advantage of the IP model over the Linear Programming (LP) model is that it assigns maintenance treatments directly to individual facilities. However, it is generally applied to small-scale systems due to the computational burden caused by combinatorics.

\medskip

\citet{wang2003decision} developed a multi-objective IP model for network-level transportation infrastructure maintenance planning. The authors used the branch-and-bound algorithm to solve the proposed model. However, due to the combinatorial nature of the IP approach, the computational burden increases exponentially as the number of facilities increases. Therefore, approximation techniques are often used in large-scale infrastructure planning problems.

\medskip

The mathematical formulation of the IP approach is as follows. Let $\mathcal{T} = \{1, 2, \dots, T\}$ represent the planning horizon, and let $\mathcal{A}$ be the set of $N$ facilities in the system. A set of basic maintenance treatments is defined as $\mathcal{M} = \{1, 2, \dots, M\}$, where the $M$th treatment is the most effective and most expensive. Given the initial condition of facility $a$, denoted by $s_a^0$, and the deterioration function $f(\cdot)$, the model is formulated as:

\begin{equation}
	\max \sum_{t \in \mathcal{T}} \sum_{a \in \mathcal{A}} s_a^t
\end{equation}

subject to:

\begin{equation}
	\sum_{a \in \mathcal{A}} \sum_{m \in \mathcal{M}} c_{a m t} \cdot u_{a m t} \leq B_t, \quad \forall t \in \mathcal{T}
\end{equation}

\begin{equation}
	s_a^t = f(s_a^{t-1}) + \sum_{m \in \mathcal{M}} u_{a m t} \cdot e_m, \quad \forall a \in \mathcal{A},\ t \in \mathcal{T}
\end{equation}

\begin{equation}
	\sum_{m \in \mathcal{M}} u_{a m t} \leq 1, \quad \forall a \in \mathcal{A},\ t \in \mathcal{T}
\end{equation}

\begin{equation}
	u_{a m t} \in \{0, 1\}, \quad \forall a \in \mathcal{A},\ t \in \mathcal{T},\ m \in \mathcal{M}
\end{equation}

\begin{equation}
	s_a^t > 0, \quad \forall a \in \mathcal{A},\ t \in \mathcal{T}
\end{equation}

\medskip

\noindent
where:
\begin{itemize}
	\item $c_{a m t}$ = maintenance cost of applying the $m$th treatment to facility $a$ at year $t$;
	\item $B_t$ = available budget at year $t$;
	\item $u_{a m t}$ = binary decision variable, equal to 1 if the $m$th treatment is applied to facility $a$ at time $t$, 0 otherwise;
	\item $s_a^t$ = condition of facility $a$ at time $t$;
	\item $f(\cdot)$ = deterioration function;
	\item $e_m$ = maintenance effectiveness of the $m$th treatment.
\end{itemize}

\medskip

\noindent
The objective function (54) seeks to maximize the average annual condition of all facilities over the planning horizon. Constraint (55) ensures that the total maintenance cost does not exceed the available budget at any time. Constraint (56) models the deterioration process and the effect of maintenance. Constraint (57) restricts each facility to receive at most one treatment per year. Constraints (58) and (59) define the binary decision variables and enforce positive condition levels.

\subsubsection{Reliability Model}

\subsection*{Reliability Model for Infrastructure Maintenance}

The reliability model is one of the continuous-time setting models. It is commonly used to plan maintenance for infrastructure facilities with clearly defined failure modes (e.g., bridges).

\medskip

For a single facility, different reliability models can be applied, including:
\begin{itemize}
	\item \textbf{Age replacement models:} These models develop optimal replacement policies based on age-dependent operating costs.
	\item \textbf{Minimal repair models:} These focus on repairing failed units rather than replacing them. They often combine periodic replacement policies with minimal repair activities upon failure.
	\item \textbf{Inspection/maintenance models:} These are used when the current state of a system is not directly known but can be assessed through inspection.
\end{itemize}

\medskip

For multi-facility systems, reliability-based maintenance models aim to identify optimal policies for a network of facilities, which may be interdependent or independent. These models include:
\begin{itemize}
	\item Block or group maintenance models,
	\item Inventory-based models,
	\item Opportunistic maintenance models.
\end{itemize}

\medskip

An example application of the reliability model is to develop an optimal replacement policy that minimizes the total operating and replacement costs per unit time. The policy replaces a facility at fixed time intervals $t_r$, and the objective is to determine the optimal interval $t_r$ that minimizes the total cost. The total cost per unit time is expressed as:

\begin{equation}
	C(t_r) = \frac{1}{t_r} \left( \int_0^{t_r} c(t)\,dt + C_r \right)
\end{equation}

\noindent
where:
\begin{itemize}
	\item $c(t)$ = operating cost per unit time at time $t$ after replacement,
	\item $C_r$ = cost of replacement,
	\item $t_r$ = replacement time interval.
\end{itemize}

\section{Mechanistic Models for Predicting Bridge Service Lie}

\subsection{Chloride-Induced Corrosion for Concrete Bridge}

\subsubsection{Deterioration Process}

Reinforced concrete structures when exposed to chloride ions cause premature corrosion of steel reinforcement. The intrusion of chloride ions, present in deicing salts and seawater, come in contact with reinforced concrete can cause steel corrosion if oxygen and moisture are also available to sustain the reaction. Chlorides dissolved in water can pass through concrete to reach the rebar through cracks. Chloride-containing admixtures can also be one of the causes of corrosion. The risk of corrosion increases as the chloride content of concrete increases. When the chloride content at the surface of the steel exceeds a certain limit i.e. threshold value, corrosion will occur if water and oxygen are also available. Federal Highway Administration (FHWA) studies have found that a threshold limit of 0.20\% total (acid-soluble) chloride by weight of cement could induce corrosion of reinforcing steel in bridge decks \citep{clear1974evaluation}. The following figure shows a typical chloride corrosion deterioration process, which consists of diffusion, rust accumulation, and cracking propagation process.

\begin{figure}[H]
	\centering
	\includegraphics[width=0.7\linewidth]{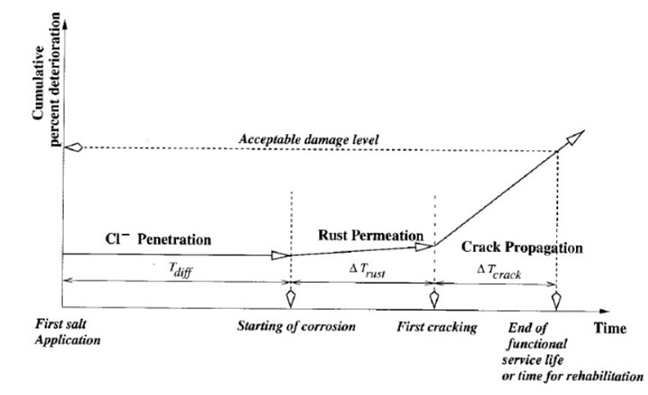}
	\caption{Chloride Corrosion Deterioration Process for a Concrete Element \citet{hu2013development}}
	\label{fig:figure23}
\end{figure}

\paragraph{Analytical Models }

Corrosion is an electrochemical process involving the flow of electrons and ions. At active sites on the reinforcing bar, known as \textit{anodes}, iron atoms lose electrons and move into the surrounding concrete as ferrous ions. This process is referred to as the \textit{half-cell oxidation reaction} or the \textit{anodic reaction}, which is expressed as:

\begin{equation}
	\text{Anodic reaction:} \quad 2\mathrm{Fe} \rightarrow 2\mathrm{Fe}^{2+} + 4e^{-}
	\label{eq:anode}
\end{equation}

The electrons remain in the steel bar and flow to sites called \textit{cathodes}, where they combine with water and oxygen in the concrete. The reaction at the cathode is known as the \textit{reduction reaction}, expressed as:

\begin{equation}
	\text{Cathodic reaction:} \quad 2\mathrm{H}_2\mathrm{O} + \mathrm{O}_2 + 4e^{-} \rightarrow 4\mathrm{OH}^{-}
	\label{eq:cathode}
\end{equation}

To maintain electrical neutrality, the ferrous ions ($\mathrm{Fe}^{2+}$) migrate through the concrete pore water toward the cathodic sites, where they combine with hydroxide ions to form ferrous hydroxide, or rust:

\begin{equation}
	2\mathrm{Fe}^{2+} + 4\mathrm{OH}^{-} \rightarrow 2\mathrm{Fe(OH)}_2
	\label{eq:rust}
\end{equation}

This initial precipitated hydroxide may further react with oxygen to form higher oxides. These oxidation products increase in volume, generating internal stress within the concrete. Over time, this stress can cause cracking and spalling of the concrete cover. The overall electrochemical process of corrosion is illustrated in the following figure.

\begin{figure}[H]
	\centering
	\includegraphics[width=0.7\linewidth]{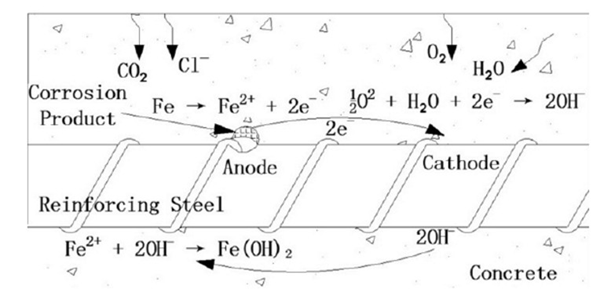}
	\caption{Schematic diagram of reinforcing steel corrosion in concrete as an electrochemical process \citep{zhao2011brillouin}}
	\label{fig:figure24}
\end{figure}

\paragraph{Exposure to Chloride Ions}

Chloride-induced corrosion is caused by the presence of chlorides in the environment. Chloride ions are common in nature and it is possible to find them in the mix ingredients of concrete. They also may be intentionally added as a component of accelerating admixtures. However, chloride ions sources which cause the most problems are deicing salts (usually calcium chloride) and sea water. Moreover, groundwater may be contaminated with chloride ions coming from runoff water for example from bridges or pavements treated with deicing salts \citep{clifton1991predicting}. Steel embedded in concrete develops a passive oxide layer that is highly protective and grows at a very slow rate. As long as the steel remains in good alkaline concrete, the passive layer will prevent corrosion initiation on the surface of the steel. Generally, the chloride ion concentration at some depth just below the surface is often referred to as the surface concentration of chloride ions. It is this concentration of chloride ions that with time diffuses into the concrete element. Previous studies on bridge decks indicated that the chloride concentration in the first 0.25 inch of concrete from the surface is very dependent on exposure conditions \citep{sohanghpurwala2006manual}. The accumulation of chloride ions occurs at about the depth of 0.5 inches because of the exposure of the surface of the concrete to moisture. Rain, snow, and water from other sources that flow over the bridge deck can wash away the chloride ions from the first 0.25 inch of concrete. However, the accumulation occurring a little deeper in the concrete is not affected by such exposure.

\subsubsection{Service Life Estimation}

\subsection*{Service Life Prediction for Corrosion-Induced Damages}

In the literature, service life prediction for corrosion-induced damages is typically modeled in the following three stages:

\begin{itemize}
	\item \textbf{Time to corrosion initiation} ($t_1$): the time required for chloride ions to penetrate the concrete surface and reach the passive film surrounding the reinforcement;
	\item \textbf{Time from initiation to cracking} ($t_2$): the period during which corrosion products accumulate, eventually causing cracking, spalling, or significant structural damage;
	\item \textbf{Time from cracking to limit state} ($t_3$): the time until corrosion damage propagates to a predefined structural or functional limit.
\end{itemize}

The total service life $T$ of a reinforced concrete bridge, with respect to corrosion, is the sum of these three stages:

\begin{equation}
	T = t_1 + t_2 + t_3
	\label{eq:corrosion_service_life}
\end{equation}

\paragraph{Time to Corrosion Initiation ($t_1$)}

\subparagraph{Fick's 2nd Law}

The time of the corrosion initiation is determined by Fick’s second law, assuming that corrosion initiates at the rebar surface when the chloride content reaches a threshold level. Previous studies have shown that chloride diffusion into concrete can be modeled by the error function solution to Fick’s second law of diffusion \citep{hu2013development}.

\begin{equation}
	\frac{\partial C}{\partial t} = D_{ca} \cdot \frac{\partial^2 C}{\partial x^2}
	\label{eq:fick}
\end{equation}

where:  
$D_{ca}$ = apparent diffusion coefficient;  
$C$ = chloride concentration;  
$x$ = depth from surface.

\medskip

The error function solution to Equation~\ref{eq:fick} is:

\begin{equation}
	C(x,t) = C_0 \left(1 - \text{erf}\left(\frac{x}{2\sqrt{D_{ca}t}}\right)\right)
	\label{eq:erfsolution}
\end{equation}

where:  
$C(x, t)$ = chloride concentration at depth $x$ and time $t$;  
$C_0$ = surface chloride concentration;  
$t$ = diffusion time;  
$\text{erf}(\cdot)$ = error function, defined as:  
\[
\text{erf}(x) = \frac{2}{\sqrt{\pi}} \int_0^x e^{-t^2} dt
\]

\medskip

The time to corrosion initiation $t_1$ is estimated by rearranging Equation~\ref{eq:erfsolution} and solving for $t$:

\begin{equation}
	t_1 = \frac{c^2 \cdot \left[\text{erf}^{-1} \left(1 - \frac{C_{\text{th}}}{C_0} \right)\right]^{-2}}{4 D_{ca}}
	\label{eq:t1}
\end{equation}

where:  
$t_1$ = time to corrosion initiation;  
$c$ = depth of concrete cover;  
$C_{\text{th}}$ = threshold level of chloride concentration.

\subparagraph{\citet{bazant1979physical}}

\citet{bazant1979physical} developed a model to estimate the time when the critical chloride concentration at the steel surface will be reached:

\begin{equation}
	t_1 = \frac{c}{12D_c} \left( \frac{c}{1 - \sqrt{\frac{C_{\text{th}}}{C_0}}} \right)^2
	\label{eq:bazant}
\end{equation}

where:  
$t_1$ = time to corrosion initiation;  
$c$ = depth of concrete cover;  
$C_{\text{th}}$ = threshold chloride concentration;  
$C_0$ = surface chloride concentration;  
$D_c$ = chloride diffusion coefficient.

\paragraph{Time from Initiation to Cracking ($t_2$)}

\subparagraph{\citet{bazant1979physical}}

In 1979, Bazant proposed a numerical model for corrosion of reinforcement to assess the time to corrosion that may cause splitting of the concrete cover. The damage due to corrosion considers the expansive volume of hydrated rust forming around the remaining rebar. This accumulated rust is approximately four times the volume of the original steel and exerts outward radial pressure on the surrounding concrete, leading to cracking and splitting.

The time from corrosion initiation to visible cracking, $t_2$, is given by:

\begin{equation}
	t_2 = \frac{q_{\text{cor}} \cdot D \cdot \Delta D}{p \cdot j_r}
	\label{eq:t2}
\end{equation}

where:  
$q_{\text{cor}}$ = Combined density factor for steel and rust = 3600 kg/m\textsuperscript{3}  
$D$ = Diameter of rebar (mm)  
$\Delta D$ = Increase in diameter of rebar due to rust formation (cm)  
$p$ = Perimeter of bar (mm)  
$j_r$ = Instantaneous corrosion rate of rust (g/m\textsuperscript{2}-s)

\vspace{1em}

The instantaneous corrosion rate of rust $j_r$ can be calculated as:

\begin{equation}
	j_r = \frac{W}{F} \cdot i_{\text{corr}}
	\label{eq:jr}
\end{equation}

where:  
$W$ = Equivalent weight of steel = 27.925  
$F$ = Faraday’s constant = 96,847 C  
$i_{\text{corr}}$ = Corrosion current density ($\mu$A/cm\textsuperscript{2})

\vspace{1em}

The critical time $t_{\text{cr}}$, defined as the time at which corrosion produces cracks through the concrete cover, is expressed as:

\begin{equation}
	t_{\text{cr}} = t_1 + t_2
	\label{eq:tcr}
\end{equation}

where $t_1$ is the time to corrosion initiation and $t_2$ is the steady-state corrosion period until cracking.

\subparagraph{\citet{morinaga1988prediction}}

\citet{morinaga1988prediction} proposed a model to estimate the amount of steel corrosion in concrete at the point when the expansion of rust on the rebar causes cracking of the concrete cover. The corrosion rate plays a key role in determining the time to cover cracking. This time depends on factors such as corrosion rate, concrete cover thickness, and rebar diameter.

The amount of corrosion required to cause cracking, $Q_{\text{cr}}$, is given by:

\begin{equation}
	Q_{\text{cr}} = 0.602D\left(1 + \frac{2C_v}{D}\right)^{0.85}
	\label{eq:qcr}
\end{equation}

where:  
$Q_{\text{cr}}$ = Amount of corrosion when concrete cracks ($\times 10^{-4}$ g/cm\textsuperscript{2})  
$C_v$ = Concrete cover thickness (mm)  
$D$ = Diameter of rebar (mm)

\vspace{1em}

The time from the start of corrosion to cover cracking, $t_2$, can then be calculated as:

\begin{equation}
	t_2 = \frac{Q_{\text{cr}}}{j_r}
	\label{eq:t2_morinaga}
\end{equation}

where:  
$j_r$ = Instantaneous corrosion rate of rust (g/m\textsuperscript{2}-s)

\subparagraph{\citet{wang1993residual}}

The research by \citet{wang1993residual} fcuses on the thickness of the corrosion product and its relation to the time when surface concrete cracks. An expression has been developed to relate the ratio of the thickness of the corrosion product to the penetration depth of the rebar $H$, and to compare it with the cracking in concrete:

\begin{equation}
	\frac{\Delta}{H} = 0.33 \left( \frac{D}{C_v} \right)^{0.565} f_{cu}^{1.436}
	\label{eq:delta_H}
\end{equation}

where:  
$D$ = Diameter of reinforcing bar (mm)  
$C_v$ = Concrete cover thickness (mm)  
$f_{cu}$ = Cube strength of concrete (kN/cm\textsuperscript{2})  
$\Delta$ = Thickness of corrosion product (cm)  
$H$ = Cracks in concrete cover

\vspace{1em}

Using the thickness of the corrosion product, cracks in cover concrete can be determined. The time required for longitudinal cracking of the concrete cover, $t_2$, can then be expressed as:

\begin{equation}
	t_2 = \frac{H}{\rho_r}
	\label{eq:t2}
\end{equation}

where:  
$\rho_r$ = Penetration rate of rebar due to corrosion

\vspace{1em}

The penetration rate of the rebar due to corrosion, $\rho_r$, can be calculated as:

\begin{equation}
	\rho_r = \frac{W}{F \rho_{st}} i_{\text{corr}}
	\label{eq:rho_r}
\end{equation}

where:  
$W$ = Equivalent weight of steel = 270,925  
$F$ = Faraday’s constant = 96,847 C  
$i_{\text{corr}}$ = Corrosion current density ($\mu$A/cm\textsuperscript{2})  
$\rho_{st}$ = Density of steel (kg/m\textsuperscript{3})

\subparagraph{\citet{liu1998modeling}}

\citet{liu1998modeling} proposed a method to estimate the critical amount of rust required to produce a crack and the time needed to generate that volume of rust. In this method, the corrosion cracking process of a thick-wall concrete cylinder is assumed to follow three stages, as illustrated in the following figure:

\begin{enumerate}
	\item \textbf{Free Expansion}:  
	As corrosion takes place on the surface of the rebar, the porous zone around the steel-concrete interface becomes filled with corrosion products. This stage corresponds to the condition when the total amount of corrosion products $W_T$ is less than the amount required to fill the porous region, denoted as $W_P$.
	
	\item \textbf{Stress Initiation}:  
	When $W_T > W_P$, the accumulation of corrosion products begins to generate expansive pressure on the surrounding concrete. The pressure increases as corrosion continues.
	
	\item \textbf{Cracking}:  
	Cracking of the concrete occurs when the total amount of corrosion products $W_T$ exceeds a critical threshold $W_{\text{crit}}$, which is the amount required to initiate cracking in the concrete cover.
\end{enumerate}

\begin{figure}[H]
	\centering
	\includegraphics[width=0.7\linewidth]{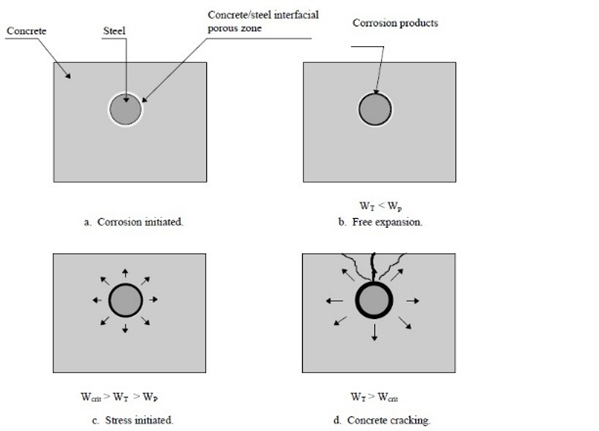}
	\caption{Schematic diagram of corrosion cracking processes \citep{liu1996modeling}}
	\label{fig:figure25}
\end{figure}

The critical amount of rust required to produce a crack can be calculated by solving the following equation:

\begin{equation}
	W_{\text{crit}} = \rho_{\text{rust}} \left[ \pi D (d_s + d_0) + \frac{W_{\text{st}}}{\rho_{\text{st}}} \right]
	\label{eq:rust_volume}
\end{equation}

\noindent where:
\begin{itemize}
	\item $W_{\text{crit}}$ = critical volume of corrosion product required to induce a crack
	\item $\rho_{\text{rust}}$ = density of rust (226 lb/ft\textsuperscript{3} in Liu and Weyers, 1998)
	\item $d_0$ = thickness of the pore band around the steel-concrete interface (4.9 mils in Liu and Weyers, 1998)
	\item $\rho_{\text{st}}$ = density of steel
	\item $D$ = diameter of the steel
	\item $W_{\text{st}}$ = amount of steel corroded, equals to $\alpha W_{\text{crit}}$
	\item $\alpha$ = ratio of molecular weight of steel to molecular weight of corrosion products (0.523 for Fe(OH)\textsubscript{3}, 0.622 for Fe(OH)\textsubscript{2})
	\item $d_s$ = thickness of corrosion products that generates pressure equivalent to concrete tensile strain
\end{itemize}

\citet{liu1998modeling} showed that $d_s$ can be expressed as:

\begin{equation}
	d_s = \frac{c f_t}{E_{\text{ef}}} \left( \frac{a^2 + b^2}{b^2 - a^2} + \nu_c \right)
	\label{eq:ds_equation}
\end{equation}

\noindent where:
\begin{itemize}
	\item $c$ = depth of concrete cover
	\item $f_t$ = tensile strength of concrete
	\item $E_{\text{ef}}$ = effective elastic modulus of concrete
	\item $\nu_c$ = Poisson’s ratio for concrete
	\item $a$ = inner radius of the thick-walled cylinder, $a = \frac{D + 2d_0}{2}$
	\item $b$ = outer radius of the thick-walled cylinder, $b = c + \frac{D + 2d_0}{2}$
\end{itemize}

\begin{figure}[H]
	\centering
	\includegraphics[width=0.7\linewidth]{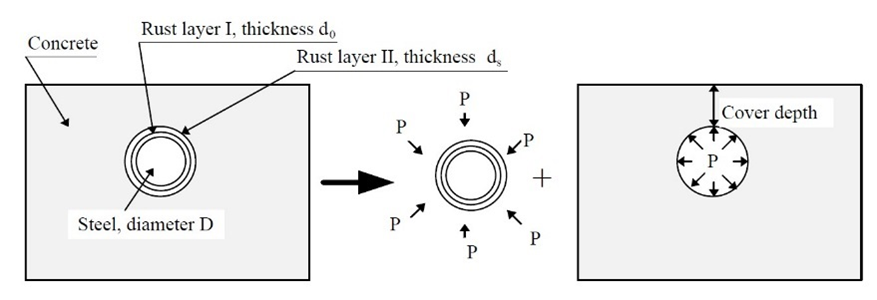}
	\caption{Expansive pressure on surrounding concrete due to formation of rust products \citep{liu1996modeling}}
	\label{fig:figure26}
\end{figure}

\begin{figure}[H]
	\centering
	\includegraphics[width=0.7\linewidth]{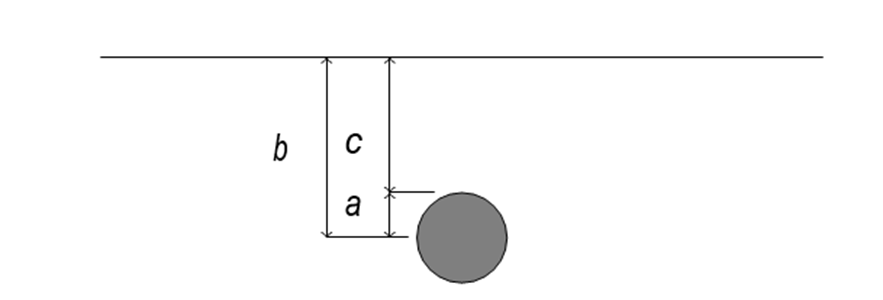}
	\caption{Input parameters}
	\label{fig:figure27}
\end{figure}

Based on the previous equation, the time to cracking $t_2$ can be estimated by:

\begin{equation}
	t_2 = \frac{W_{\text{crit}}^2}{2k_p}
	\label{eq:t2_cracking}
\end{equation}

\noindent where $k_p$ is the rate of rust production, calculated as:

\begin{equation}
	k_p = 0.098 \cdot \left(\frac{1}{\alpha}\right) \pi D i_{\text{corr}}
	\label{eq:kp_equation}
\end{equation}

\noindent in which:
\begin{itemize}
	\item $W_{\text{crit}}$ = critical volume of corrosion product required to induce cracking
	\item $\alpha$ = molecular weight ratio between steel and rust products
	\item $D$ = diameter of steel (mm)
	\item $i_{\text{corr}}$ = annual mean corrosion rate (mA/ft\textsuperscript{2})
\end{itemize}

Various techniques such as linear polarization resistance (LPR) and AC impedance are available to estimate the corrosion rate of steel in concrete. To assess the current corrosion status of steel, methods like the half-cell potential test can be used.

According to \citet{stewart1998time}, the mean corrosion current density $i_{\text{corr}}$ is typically taken as 1~$\mu$A/cm\textsuperscript{2}. However, they also noted that the maximum current density in uncracked concrete under aggressive environments may be several orders of magnitude higher, and may increase further if cracking in the concrete cover occurs.

\paragraph{Time for Corrosion Damage Propagating to a Limit State ($t_3$)}

\subparagraph{Reduction in Reinforcement Section \citep{andrade1990initial}}

The measurement of corrosion currents of steel reinforcement in concrete have been used in estimating the remaining service life of reinforced concrete in which corrosion is the main deterioration mechanism. The polarization resistance technique was used to measure corrosion currents.

\citet{andrade1990initial} used measurements of corrosion current to estimate the remaining service life. The model considers reduction in reinforcement section as the major consequence of corrosion, instead of cracking or spalling of the concrete as seen in other models.

\begin{figure}[H]
	\centering
	\includegraphics[width=0.7\linewidth]{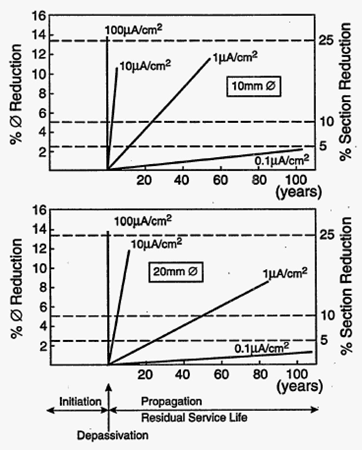}
	\caption{Effect of corrosion on the diameter and cross-section of reinforcing steel bars, with diameters of 10mm and 20mm \citep{andrade1990initial}}
	\label{fig:figure28}
\end{figure}

The corrosion current was converted to reductions in the diameter of reinforcing steel using the following relationship:

\begin{equation}
	\theta(t) = \theta_i - 0.023 \cdot i_{\text{corr}} \cdot t
	\label{eq:rebar_diameter}
\end{equation}

\noindent where:
\begin{itemize}
	\item $\theta(t)$ = diameter of rebar at time $t$ (mm),
	\item $\theta_i$ = initial diameter of the rebar (mm),
	\item $i_{\text{corr}}$ = corrosion rate ($\mu$A/cm\textsuperscript{2}),
	\item $t$ = time since the beginning of the propagation period (years),
	\item $0.023$ = conversion factor from $\mu$A/cm\textsuperscript{2} to mm/year.
\end{itemize}

The service life predictions were estimated by modeling the effect of this reduction in the rebar cross-section on the load-carrying capacity of the concrete structure.

\subparagraph{Percentage of Damaged Area \citep{williamson2007bridge}}

\citet{williamson2007bridge} used a deterioration level of 12\% as the End of Functional Service Life (EFSL) for a bridge deck, which is most commonly rehabilitated at the 12\% damage level. The authors developed the following estimation model based on surveyed data of seven bridge decks in Virginia:

\begin{equation}
	t_3 = 8.61 \left( \sqrt{\% \text{Deterioration} + 1.38} - 1.45 \right) - 3.34
	\label{eq:efsl_model}
\end{equation}

\noindent where:
\begin{itemize}
	\item $\% \text{Deterioration}$ is the specified damage level (e.g., 12\%).
\end{itemize}

\noindent It takes approximately 16 years for the damaged area to propagate from 2\% to 12\%.

\subparagraph{Percentage of Damaged Area \citep{hu2013development}}
\citet{hu2013development} used Monte Carlo simulation to estimate the cracking area of bridge decks in Michigan. The authors divided the whole deck into many cells with each cell containing only one rebar. Initial conditions such as chloride concentration, concrete strength, and diffusion coefficient are assumed to follow normal distributions among different cells. Simulations were conducted to estimate the time 25\% of the deck area has been damaged by corrosion-induced cracks.

\subparagraph{Crack Width \citep{hu2013development}}

A crack width of 0.3 mm (0.013 in) is frequently recommended as the maximum limit. \citep{rahim2006concrete} reported that some European countries set 0.2 mm as the limit crack width for service life. A crack width limit state of 0.01 in (0.3 mm) was selected because it has been suggested that crack widths of less than 0.01 in (0.3 mm) have little to no effect on the ingress of chlorides into the concrete \citep{atimtay1974early}  (Atimay and Ferguson, 1974). Williamson et al. (2007) estimated that the total crack propagation time (until 0.3 mm) in Virginia is around 6 years. Hu et al. (2013) used the following equation to calculate the crack width at the surface of the concrete:

\begin{equation}
	w_s = 0.1916 \Delta A_s + 0.164
	\label{eq:crackwidth}
\end{equation}

\noindent where:
\begin{itemize}
	\item $w_s$ = surface crack width;
	\item $\Delta A_s = D - D(t)$ = average loss of rebar cross-section;
	\item $D(t)$ = diameter of rebar at time $t$.
\end{itemize}

\noindent The function $D(t)$ is defined piecewise as:

\[
D(t) =
\begin{cases}
	D, & t \leq t_1 \\
	D - 2\lambda(t - t_1), & t_1 < t \leq t_1 + \dfrac{D}{2\lambda} \\
	0, & t \geq t_1 + \dfrac{D}{2\lambda}
\end{cases}
\]

\noindent where:
\begin{itemize}
	\item $D$ = initial diameter of rebar;
	\item $\lambda$ = corrosion rate, $\lambda \approx 0.0116 R i_{\text{corr}}$;
	\item $R$ = pitting factor accounting for localized corrosion (typically $4 \sim 6$);
	\item $i_{\text{corr}}$ = corrosion current density;
	\item $t_1$ = corrosion initiation time;
	\item $t$ = elapsed time since corrosion started.
\end{itemize}

\subparagraph{Crack Width \citep{vu2005predicting}}

\citet{vu2005predicting} proposed an empirical equation to calculate the time to excessive cracking for reinforced concrete structures:

\begin{equation}
	t_3 = t_1 + k_R \times 0.0114 \, i_{\text{corr}} \left[ A \left( \frac{c}{w_c} \right)^B \right]
	\label{eq:vu_t3}
\end{equation}

\noindent where:
\begin{itemize}
	\item $t_3$ = time to excessive cracking (years),
	\item $t_1$ = time to crack initiation (years),
	\item $k_R$ = rate of loading correction factor,
	\item $i_{\text{corr}}$ = corrosion current density ($\mu$A/cm$^2$),
	\item $A$, $B$ = empirical constants based on the selected crack width limit state,
	\item $c$ = concrete cover depth (mm),
	\item $w_c$ = water-to-cement ratio.
\end{itemize}

The correction factor $k_R$ is calculated as:

\begin{equation}
	k_R = 0.95 \left[ \exp\left(-\frac{0.3 \, i_{\text{corr(exp)}}}{i_{\text{corr(real)}}} \right) - \frac{i_{\text{corr(exp)}}}{2500 \, i_{\text{corr(real)}}} \right] + 0.3
	\label{eq:vu_kr}
\end{equation}

\noindent where:
\begin{itemize}
	\item $i_{\text{corr(exp)}}$ = corrosion rate used in experiments ($\mu$A/cm$^2$),
	\item $i_{\text{corr(real)}}$ = corrosion rate observed in the structure ($\mu$A/cm$^2$).
\end{itemize}

\begin{table}[H]
	\centering
	\caption{Constants $A$ and $B$ Based on Crack Width Limit}
	\begin{tabular}{ccc}
		\toprule
		\textbf{Limit Crack Width} & \textbf{A} & \textbf{B} \\
		\midrule
		0.3 mm & 65  & 0.45 \\
		0.5 mm & 225 & 0.29 \\
		1.0 mm & 700 & 0.23 \\
		\bottomrule
	\end{tabular}
	\label{tab:vu_constants}
\end{table}

\paragraph{Remaining Life Prediction}

\subparagraph{\citet{widyawati2014remaining}}

The remaining life prediction for a reinforced concrete bridge, in the case of section loss due to steel corrosion, can be expressed as the number of expected service years if the section loss is left uncorrected. In other words, the remaining life $R$ can be expressed using the life expectancy ($T$) and the period of service ($t$) as:

\begin{equation}
	R = T - t
	\label{eq:remaining_life}
\end{equation}

\subparagraph{\citet{clear1989measuring}}

Another model proposed by \citet{clear1989measuring} suggested the use of relationships between corrosion rates and remaining service life. These relationships were derived from laboratory tests, outdoor exposure, and field studies, and are shown in the following table.

\begin{table}[H]
	\centering
	\caption{Corrosion Rates and Remaining Service Life \citep{clear1989measuring}	}
	\begin{tabular}{|c|c|}
		\hline
		\textbf{Corrosion rate ($\mu$A/cm\textsuperscript{2})} & \textbf{Remaining service life} \\
		\hline
		$ i_{\text{corr}} < 0.5 $ & No corrosion damage \\
		\hline
		$ 0.5 < i_{\text{corr}} < 2.7 $ & Corrosion damage in the range of 10 to 15 years \\
		\hline
		$ 2.7 < i_{\text{corr}} < 27 $ & Corrosion damage in the range of 2 to 10 years \\
		\hline
		$ i_{\text{corr}} > 27 $ & Corrosion damage less than 2 years \\
		\hline
	\end{tabular}
	\label{tab:corrosion_life}
\end{table}

\paragraph{Important Variables}

\subparagraph{Critical Chloride Content ($C_{th}$) }

The presence of corrosion of reinforcements only occurs once a threshold value of chloride ion content adjacent to the bars is reached. Previous research has recommended that the critical chloride content to initiate corrosion ranges from 0.2\% to 1.5\% by weight of cement. The following table shows various critical chloride contents recommended by previous studies.

\begin{table}[H]
	\centering
	\caption{Chloride Threshold Levels}
	\begin{tabular}{|l|c|l|}
		\hline
		\textbf{Steel Bars} & \textbf{Chloride Threshold Level} & \textbf{Reference} \\
		\hline
		Black steel bars & 1.2 kg/m\textsuperscript{3} and 2.0 kg/m\textsuperscript{3} & Page 46 of (Hu et al., 2013) \\
		\hline
		General rebar & 0.6--1.2 kg/m\textsuperscript{3} & (Stewart and Rosowsky, 1998) \\
		\hline
	\end{tabular}
	\label{tab:chloride_threshold}
\end{table}

\subparagraph{Chloride Diffusion Coefficient (Dca) }

The chloride diffusion coefficient can be back calculated based on the diffusion model. Using the chloride ion concentrations at various depths determined from the laboratory data, \( D_{ca} \) can be calculated using a least sum of squared error curve fitting analysis. There are several ASTM standards for chloride ion concentration determination. ASTM C1152 is the standard test method for acid-soluble chlorides in concrete and mortar. ASTM C1218 is the standard test method for water-soluble chloride in mortar and concrete. Tex-617-J is the method used by TxDOT to determine the concentration of water-soluble chloride and sulfate ions in concrete.

According to \citet{hoffman1994predicting}, the mean diffusion coefficient in the US ranged from \( 0.6 \times 10^{-8} \, \text{cm}^2/\text{s} \) to \( 7.5 \times 10^{-8} \, \text{cm}^2/\text{s} \), with an overall mean around \( 2.0 \times 10^{-8} \, \text{cm}^2/\text{s} \) and a coefficient of variation around 0.75.

William et al. (2007) recommended the following sampling rate and a minimum of five chloride concentrations at various depths (0.5 in, 0.75 in, 1.0 in, 1.25 in, 1.5 in), at the depth of the reinforcement, and below the reinforcement. The background chloride concentration was calculated to be the average of the chloride concentrations taken from below the reinforcement.

\begin{equation}
	\text{Number of } D_{ca} \text{ values} = 20 + \frac{L - 150}{7}
	\label{eq:Dca_values}
\end{equation}

where \( L \) is the length of the bridge deck in feet.

\subparagraph{Surface Chloride Concentrations(C0)}

It was suggested by previous studies (e.g., Williamson et al., 2007) that the chloride concentration profiles for bridge deck cores reach a maximum at a depth of approximately 0.5 inch after 5 to 10 years of exposure to deicing salts, due to the propensity for chlorides to be washed out of the surface of the deck, resulting in lower concentrations at the surface than at a depth just below the surface. Thus, it is suggested to use the chloride concentration at a depth of 0.5 inches (or 0.25 to 0.75 in) as the \( C_0 \).

The mean surface chloride content in the US varied from \( 1.2 \, \text{kg/m}^3 \) to \( 8.2 \, \text{kg/m}^3 \). The mean and coefficient of variation of surface chloride content are \( 3.5 \, \text{kg/m}^3 \) and 0.5, respectively (Hoffman and Weyers, 1994). The chloride concentration in concrete (\( C_0 \) and \( C(x, t) \)) can be determined through analysis of concrete samples, which can be collected on-site at different depths up to and beyond the depth of the reinforcing steel using a hammer drill. The chloride ion content of concrete is usually measured in a laboratory using wet chemical analysis \citep{sohanghpurwala2006manual}.

\subparagraph{Depth of Cover }

The location of a reinforcement bar and its depth of cover ($c$) can be obtained by using a pachometer or a covermeter. These devices measure variations in magnetic flux caused by the presence of reinforcement bars to locate their presence and depth \citep{sohanghpurwala2006manual}. Another method consists in drilling small holes into the concrete to measure the depth from surface. This method can be more accurate, but it also introduces defects into the structure. The number of cover depth measurements are to be determined as \citep{williamson2007bridge}:

\begin{equation}
	\text{Number of cover depths} = 40 + \frac{L - 20}{3}
	\label{eq:88}
\end{equation}

where,  
$L$ = length of the bridge deck in feet.

\paragraph{Commercial Software}

STADIUM (SIMCO) is powerful software developed to predict the service life of a concrete structure. It can take into account the effect of concrete and reinforcement type, exposure condition, repair history and can evaluate the performance of a concrete structure by estimating the transport of chloride ions based on experimentally obtained (or user provided) parameters. It uses advanced models to estimate the transfer of chloride ions, which can account for the interaction of multiple ions (Nernst-Planck equation), water movement, and temperature (Hu et al. , 2013).

Life-365 is a program used to predict the service life of concrete structures. The service life is assumed to be the sum of the initiation period of corrosion process and the propagation period. The initiation period is estimated by solving Fick’s second law using the finite difference method while assuming the diffusion coefficient of concrete to be a function of both time and temperature. The propagation period is assumed to be constant (6 or 20 yrs.) (Hu et al. , 2013).

\subsection{Carbonation-Induced Corrosion for Concrete Bridge}

\subsubsection{Deterioration Process}

Carbonation occurs when carbon dioxide from the air penetrates into concrete and reacts with hydroxides, such as calcium hydroxide, to form carbonates. In the reaction with calcium hydroxide, calcium carbonate is formed. Carbonation is generally a slow process and involves a physiochemical reaction between atmospheric carbon dioxide and the calcium hydroxide generated in cement hydration. The production of calcium carbonate reduces the pH level of the concrete:
\begin{equation}
	\text{Ca(OH)}_2 + \text{CO}_2 \rightarrow \text{CaCO}_3 + \text{H}_2\text{O}
	\label{eq:89}
\end{equation}

Corrosion of steel due to carbonation usually occurs particularly in an urban area which has a high level of carbon dioxide, emitted from vehicles and industrial factories. Carbonation can be defined as the chemical reaction between carbon dioxide present in the air and cement hydration products such as mainly calcium hydroxide and the CSH gel phase, which results in the formation of calcium carbonate. Thus, the risk of carbonation is more severe in urban or/and industrial area. 

Carbonation of concrete itself does not do harm in view of the performance of the structure, adversely a marginal enhancement of the compressive strength was observed. However, when carbonation reaches the depth of the steel, the high alkalinity of the concrete pore solution is neutralized, and hydration products are dissolved then to lower the buffering capacity of hydrations against a pH fall. At this moment, the passivation layer on the steel surface, which otherwise would protect the steel embedment from a corrosive environment, is destroyed, and steel is directly exposed to oxygen and water, eventually to corrode \citep{ann2010service}.

In high-quality concrete, it has been estimated that carbonation will proceed at a rate up to 1.0 mm (0.04in) per year. Commonly, the time needed to carbonate 20mm of high-quality concrete is estimated to be of the order of tens of years, and the penetration rate of carbon dioxide falls quickly below a mm/year soon after construction \citep{page1982aspects,locke1983mechanism}.

The amount of carbonation is affected by a high water-cement ratio, low cement content, short curing period, low strength, and highly permeable or porous paste. These factors are responsible for increasing carbonation level in concrete. Carbonation is highly dependent on the relative humidity of the concrete. The highest rates of carbonation occur when the relative humidity is maintained between 50\% and 75\%. Below 25\% relative humidity, the degree of carbonation that takes place is considered insignificant. Above 75\% relative humidity, moisture in the pores restricts CO2 penetration (Institute, 1992). Carbonation induced corrosion often occurs on areas of structures that are exposed to rainfall, shaded from sunlight, and have a low concrete cover over the reinforcing steel.

Carbonation of concrete can reduce its alkalinity sufficiently to depassivate the steel and initiate corrosion. Carbonation involves the reaction of gaseous carbon dioxide with calcium hydroxide of concrete to form calcium carbonate. If fully carbonated, the pH of the concrete is reduced to around 9 at which pH steel is susceptible to corrosion. Also, if the depth of carbonation extends to the reinforcing steel, then the chloride ion threshold concentration can be significantly reduced. The rate and extent of carbonation depend on the environmental relative humidity, reaching a maximum at 50\% relative humidity. Diffusion of gaseous carbon dioxide takes place several orders of magnitude more rapidly through air than through water. If the pores of concrete are saturated with water, the amount of carbonation occurring will be negligible \citep{clifton1991predicting}.

Most damage to concrete bridges occurs due to rebar corrosion. Fresh concrete around the steel embedded creates a corrosion resistant barrier because of the high alkalinity (pH > 13) within the concrete. However, over time the presence of chlorides, carbonation, acid attack or combination of all these reduces pH of concrete. At this point, the natural corrosion barrier is lost and the reinforcement steel starts to corrode (INC., 2005). The following figure shows the influence of pH on steel corrosion occurrence.

\begin{figure}[H]
	\centering
	\includegraphics[width=0.7\linewidth]{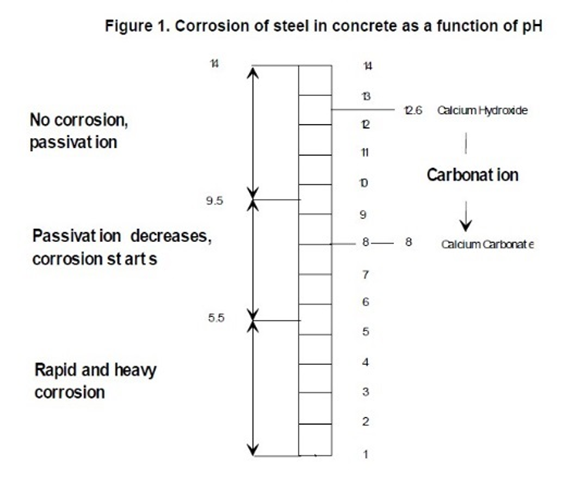}
	\caption{Corrosion of steel in concrete as a function of pH (INC., 2005)}
	\label{fig:figure29}
\end{figure}

The potential for rebar corrosion in bridges can be predicted in advance by measuring concrete pH and chloride content at the first rebar level. Once the concrete pH reaches level 11 or the water-soluble chloride content in the cement reaches 300ppm, corrosion will occur.

The most common method for measuring the pH of concrete is the extraction of pore solution. However, this method is a time consuming and destructive process. Researchers have worked continuously to develop non-destructive methods using embedded sensors with the purpose of measuring concrete pH based on real-time monitoring. The different sensors that have been developed can be categorized into different types such as ion-sensitive-field-effect transistor (iSFET), fiber optic, hydrogel film and solid-state pH sensors \citep{behnood2016methods}.

\subsubsection{Carbonation Service Life Prediction}

\paragraph{\citet{hookman1990rehabilitation}}

A carbonation model proposed by \citet{hookman1990rehabilitation}, was used to predict the service life of an ore dock constructed in 1909. The model has the following formulation:
\begin{equation}
	t_c = \frac{L}{R_c}
	\label{eq:90}
\end{equation}

where,  
$t_c$ = time to full cover carbonation  
$L$ = remaining uncarbonated cover  
$R_c$ = rate of carbonation  

The prediction of remaining service life was modeled using the relationship:
\begin{equation}
	t_2 = k_c \cdot k_e \cdot L^2 + k_a \cdot L
	\label{eq:91}
\end{equation}

where,  
$t_2$ = service life in years  
$L$ = thickness of concrete cover  
$k_c$ = quality coefficient of the concrete  
$k_e$ = coefficient of environment  
$k_a$ = coefficient of active corrosion

\paragraph{\citet{clifton1991predicting}}

Another approach for predicting the remaining service life when carbonation is the main deterioration process is to use the square root of time principle, which is given by the following equation:
\begin{equation}
	x = \sqrt{2 D_c (c_1 - c_2)t}
	\label{eq:92}
\end{equation}

where,  
$x$ = the carbonation depth (mm);  
$D_c$ = the diffusion coefficient for CO$_2$ of concrete of a given composition and moisture condition;  
$c_1 - c_2$ = the concentration difference of CO$_2$ between air and the carbonation front;  
$t$ = time.

\paragraph{IAEA (2002)}

The time required for carbonation can be estimated knowing the concrete grade and using the following equation:
\begin{equation}
	t = \left( \frac{d}{k} \right)^2
	\label{eq:93}
\end{equation}

where,  
$t$ = the time for carbonation;  
$d$ = the concrete cover;  
$k$ = the permeability.

Typical permeability values are shown in the following table:

\begin{table}[H]
	\centering
	\caption{Typical permeability values (IAEA, 2002)}
	\begin{tabular}{|c|c|}
		\hline
		\textbf{Concrete Grade} & \textbf{Permeability} \\
		\hline
		15 & 17 \\
		20 & 10 \\
		25 & 6  \\
		30 & 5  \\
		35 & 4  \\
		40 & 3.5 \\
		\hline
	\end{tabular}
\end{table}

Another formula used to estimate the depth of carbonation utilizes the age of the structure, the water-cement ratio, and a constant that varies depending on the surface coating on the concrete:

\begin{equation}
	C = \frac{\sqrt{y} \cdot R(4.6x - 1.76)}{\sqrt{7.2}}
	\label{eq:94}
\end{equation}

where,  
$y$ = age of structure in years;  
$x$ = water-to-cement ratio;  
$C$ = carbonation depth;  
$R = \alpha \beta$ is a constant.  

$R$ varies depending on the surface coating on the concrete ($\beta$) and whether the concrete has been in external or internal service ($\alpha$).  
$\alpha$ is 1.7 for indoor concrete and 1.0 for outdoor concrete.  
$\beta$ values are shown in the following table:

\begin{table}[H]
	\centering
	\caption{Values of $\beta$ (IAEA, 2002)}
	\begin{tabular}{|l|c|c|}
		\hline
		\textbf{Finished Condition} & \textbf{Indoor} & \textbf{Outdoor} \\
		\hline
		No Layer         & 1.7  & 1.0 \\
		Plaster          & 0.79 &     \\
		Mortar + Plaster & 0.41 &     \\
		Mortar           & 0.29 & 0.28 \\
		Mortar + Paint   & 0.15 &     \\
		Tiles            & 0.21 & 0.07 \\
		Paint            & 0.57 & 0.8  \\
		\hline
	\end{tabular}
\end{table}

\paragraph{\citet{papadakis2005estimation}}

For constant values of parameters and one-dimensional geometry, the progress of carbonation depth, $x_c$ (m), with time $t$ (s), is given by the following equation \citep{papadakis2005estimation}:
\begin{equation}
	x_c = \sqrt{\frac{2D_{e,\text{CO}_2} \left(\frac{\text{CO}_2}{100}\right)t}{0.33CH + 0.214CSH}}
	\label{eq:95}
\end{equation}

where:  
$\text{CO}_2$ = CO$_2$ content in the ambient air at concrete surface (varies between 0.03\%–0.15\%);  
$D_{e,\text{CO}_2}$ = effective diffusivity of CO$_2$ in carbonated concrete (m/s$^2$);  
$CSH$ = calcium silicate hydrate content in concrete volume (kg/m$^3$);  
$CH$ = calcium hydroxide content in concrete volume (kg/m$^3$).

In an ambient relative humidity, $RH$ (\%), the diffusivity is given by the following empirical equation (Papadakis, 1999):
\begin{equation}
	D_{e,\text{CO}_2} = 6.1 \times 10^{-6} \left[ \frac{\varepsilon_c - \varepsilon_{\text{air}}}{1 - \frac{A}{d_A} - \varepsilon_{\text{air}}} \right]^3 \left(1 - \frac{RH}{100}\right)^{2.2}
	\label{eq:96}
\end{equation}

where:  
$\varepsilon_{\text{air}}$ = volume fraction of entrapped air per concrete volume (m$^3$/m$^3$);  
$\varepsilon_c$ = porosity of carbonated concrete;  
$d_A$ = aggregate density (kg/m$^3$);  
$A$ = aggregate content in concrete volume (kg/m$^3$).

The equations presented are valid for Portland and blended cements \citep{papadakis2005estimation}. The critical time, $t_{\text{cr,carb}}$ (s), required for carbonation to reach the reinforcement placed at a distance $c$ (concrete cover, m) from the outer surface, can be estimated with the following equation:
\begin{equation}
	t_{\text{cr,carb}} = \frac{(0.33CH + 0.214CSH) \cdot c^2}{2D_{e,\text{CO}_2} \left( \frac{\text{CO}_2}{100} \right)}
	\label{eq:97}
\end{equation}

As a general conclusion from various works, the propagation period depends strongly on relative humidity. According to \citet{morinaga1988prediction}, for a typical environmental temperature of 20 oC and a relative humidity between 55\% and 95\%, the rate of corrosion, $q_c$ ($10^{-4}$ g/cm$^2$/year), of the rebar in concrete can be estimated with the following formula:
\begin{equation}
	q_c = 65\left( \frac{RH}{100} \right) - 35
	\label{eq:98}
\end{equation}

The critical amount of corrosion that causes cracking of the concrete cover, for typical concrete strength and a reinforcing bar of 10 mm diameter, can be estimated by \citep{morinaga1988prediction}:
\begin{equation}
	Q_{\text{cr}} = 6(1 + 0.2c)^{0.85}
	\label{eq:99}
\end{equation}

where:  
$Q_{\text{cr}}$ in $10^{-4}$ g/cm$^2$;  
$c$ = concrete cover (mm).

As a result, the propagation period in years can be estimated by the ratio $Q_{\text{cr}}/q_c$:
\begin{equation}
	t_{\text{pr,carb}} = \frac{6(1 + 0.2c)^{0.85}}{65(RH/100) - 35}
	\label{eq:100}
\end{equation}

Finally, the service lifetime, $Z_{\text{carb}}$ (in years), as regards the carbonation-induced corrosion of the concrete reinforcement, is the total sum of the two periods ($t_{\text{cr,carb}}$ must be converted to years by dividing by 31,557,600 s/year):
\begin{equation}
	Z_{\text{carb}} = \frac{t_{\text{cr,carb}}}{31,\!557,\!600} + t_{\text{pr,carb}}
	\label{eq:101}
\end{equation}

\subsection{Sulfate Attack Deterioration for Concrete Bridge}

\subsubsection{Deterioration Process}

Sulfates are present in most cements, some aggregates, soils, groundwater, sea water, industrial wastes and acid rain. Sulfates can attack concrete by reacting with hydrated components in the cement. These reactions can induce enough pressure to break the concrete and result in loss of strength and also accelerates the corrosion of the reinforcement. Environmental conditions have a significant influence on sulfate attack: the attack is greater in concrete subject to wet-dry cycles. Sulfates can accumulate at the concrete surface of a structure when water evaporates. This situation leads sulfates to increase in concentration becoming a potential for causing deterioration in concrete. If the concrete structure is continuously immersed in water containing sulfates, a softening process takes place as the sulfates ingress into the concrete. The effect area is behind the sulfate-moving interface and its depth is proportional to the depth of the interface. As a result, of the softening process the mechanical properties of the concrete are reduced. Assuming that the concrete is continuously immersed in the sulfate water, the process is likely to be controlled by diffusion \citep{clifton1991predicting}. Moreover, porous concrete is susceptible to weathering caused by salt crystallization. For example, sodium sulfate and sodium carbonate are salts known to cause weathering of concrete. As a result of surface evaporation, salt solutions can migrate to the surface by capillary action and the solution phase becomes supersaturated and salt crystallization occurs. Occasionally this crystallization produce pressures big enough to cause cracks and scaling in concrete \citep{mehta2000sulfate}. Sulfate attack is commonly seen in arid areas of the United States like the Northern Great Plains and parts of the Western. Sulfates are also present in seawater but are not as severe as an exposure of sulfates in groundwater. Sulfate attack can be external or internal. The external attack is the most common type of attack and is caused by penetration of sulfates in solution for example in groundwater. The internal attack is due to the presence of a soluble source in the concrete at the time of mixing.

\subsection{Sulfate Attack Service Life Prediction}
\paragraph{\citet{atkinson1989mechanistic}}

The service life model for sulfate attack of concrete is based on a model developed by \citet{atkinson1989mechanistic}. The model is based on the following assumptions:

\begin{itemize}
	\item Sulfate ions from the environment penetrate the concrete by diffusion;
	\item Sulfate ions react expansively with aluminates in the concrete; and
	\item Cracking and delamination of concrete surfaces result from the expansive reactions.
\end{itemize}

The model considers diffusion as the main mode of sulfate ion transport into the concrete. The basic equation developed by Atkinson and Hearne is:

\begin{equation}
	R = \frac{X_{\text{spall}}}{t_{\text{spall}}} = \frac{E B^2 c_0 C_E D_i}{\alpha \gamma (1 - \nu)}
	\label{eq:102}
\end{equation}

where:  
\begin{tabular}{ll}
	$R$ & = degradation rate of concrete by sulfate ions (mm/s) \\
	$X_{\text{spall}}$ & = thickness of the concrete layer that spalls off (mm) \\
	$t_{\text{spall}}$ & = time to spalling (s) \\
	$E$ & = elastic modulus of concrete (20 GPa) \\
	$B$ & = stress generated by 1 mol sulfate reaction in 1 m$^3$ (1.8$\times$10$^{-6}$ m$^3$/mol) \\
	$c_0$ & = concentration of sulfate in solution (mol/m$^3$) \\
	$C_E$ & = concentration of reacted sulfate as ettringite (mol/m$^3$) \\
	$D_i$ & = diffusion coefficient of sulfate ions in concrete (m$^2$/s) \\
	$\alpha$ & = roughness factor of the degrading area (assumed 1.0) \\
	$\gamma$ & = energy required to destroy concrete surface (10 J/m$^2$) \\
	$\nu$ & = Poisson’s ratio (0.2) \\
\end{tabular}

The basic assumption of the model is that damaging expansion and cracking is due to the formation of expansive ettringite within the concrete. This leads to failure when the tension caused by the growing ettringite crystals exceeds the concrete strength and a layer $X_{\text{spall}}$ thick spalls from the concrete surface.

When cracking and delamination of the concrete surface occur, a new surface is exposed to a concentration of sulfate ions similar to the groundwater sulfate concentration, rather than the smaller concentration from diffusion. The model predicts that the sulfate attack rate will be mainly controlled by the concentration of sulfate ions and aluminates, diffusion and reaction rates, and the fracture energy of concrete \citep{clifton1991predicting}.

\subsection{Freeze-Thaw Deterioration for Concrete Bridge}

\subsection{Deterioration Process}

Abrupt changes in temperature are one of the most destructive factors affecting concrete. These changes can cause cycles of freezing and thawing. Freezing temperatures can cause water to frozen and expand around 9\%. Water contained in concrete also freezes and produces pressure in the pores and capillaries of the concrete. If such pressure is higher than the tensile strength of concrete it will cause a rupture in the structure. Concrete does not have to be completely saturated with water for damage to occur, for most concretes the critical level of saturation is approximately 85\%. The accumulative effect of consecutive freeze-thaw cycles can eventually cause significant damage to concrete like cracking, scaling, and crumbling. The following figure shows the frequency of freeze-thaw exposure of different regions of the United States. It is notable that in the majority of Texas’ territory it is rare to find freeze-thaw cycles and in the north part of the state it is occasional to have exposure to freeze-thaw cycles.

\begin{figure}[H]
	\centering
	\includegraphics[width=0.7\linewidth]{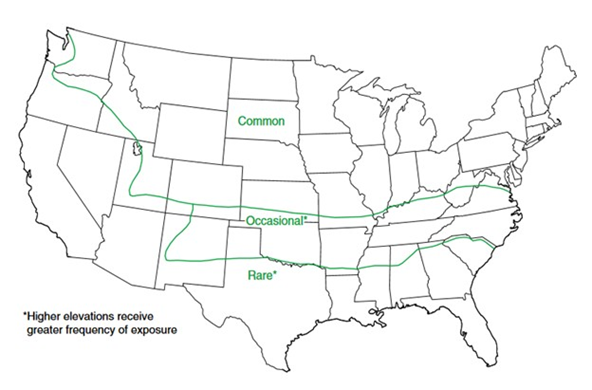}
	\caption{Frequency of freeze-thaw exposure typically encountered in different areas of the United States (PCA, 2002)}
	\label{fig:figure30}
\end{figure}

\subsubsection{Service Life Prediction}
\paragraph{Chen and Qiao (2015) }

A service life prediction model is based on the previous experimental studies from \citep{russell1943freeze}. According to the authors, an equivalent coefficient ($C_{\text{Equivalent}}$) for the number of freeze/thaw (F/T) cycles in the indoor laboratory ($N_{\text{indoor}}$) is approximately 6.5 in their study. They assumed that the annual F/T cycles ($N_{\text{Annual}}$) in some severe environments are 200. Thus, the service life $T$ (in years) can be calculated as:

\begin{equation}
	T = \frac{C_{\text{Equivalent}} \cdot N_{\text{Indoor}}}{N_{\text{Annual}}}
	\label{eq:103}
\end{equation}

It is important to consider that the equivalent coefficient for the number of laboratory F/T cycles with respect to the field F/T cycles varies depending on the region and its environmental conditions. Moreover, several important factors should be noted for this service life prediction model for concrete subject to cyclic F/T degradation as far as the equivalent coefficient is concerned:

\begin{itemize}
	\item Critical level of saturation plays a significant role in affecting the F/T concrete resistance.
	\item Air entrainment is an important factor that will affect the F/T deterioration.
	\item Permeability of concrete may considerably impact the F/T durability of concrete.
\end{itemize}

The accuracy of the predicted service life depends on the accuracy of the equivalent coefficient for the number of F/T cycles calculated in the laboratory, where the critical saturation level, air entrainment, and permeability of the studied concrete are required \citep{chen2015probabilistic}.

\paragraph{\citet{shuman1989barrier}}

Another empirical model described by \citet{shuman1989barrier}, frost durability is modeled by relating the decrease in the dynamic modulus of elasticity of concrete to the percentage of entrained air, water-cement ratio, and number of F/T cycles using ASTM C666. According to the experimental results, changes in dynamic modulus were assumed linear when the number of F/T cycles was over 50. The amount of annual degradation, $R_{\text{ft}}$, based on the fraction of strength loss, is given by:

\begin{equation}
	R_{\text{ft}} = \left(\frac{N}{T_c}\right) \left[\frac{0.05}{\theta^{1/2}} - 0.21T_r\right]
	\label{eq:104}
\end{equation}

where,  
$N$ = number of F/T cycles;  
$T_c$ = amount of experimental time required to reach a 50\% decrease in dynamic modulus of elasticity;  
$\theta$ = water content of the concrete;  
$T_r$ = unsaturated pore content of the concrete.

Finally, it was recognized that ASTM C666 usually significantly overestimates the field damage caused by frost attack. In conclusion, this model will likely considerably underestimate the service life of concretes exposed to freeze-thaw cycles.

\subsubsection{Commercial Software}

CONLIFE can be used to predict the service life of concrete structures due to sulfate attack and freeze-thaw effects. It assumes that sorption is the primary transport mechanism in concrete. A test method for sorptivity is proposed (Hu et al. , 2013).

\subsection{Alkali Silica Reaction Deterioration for Concrete Bridge}
\subsubsection{Deterioration Process}

\section*{Alkali-Aggregate Reaction}

Normally, aggregates in concrete are chemically inert. However, some aggregates react with the alkali hydroxides in concrete, causing expansion and cracking over time. There are two types of alkali-aggregate reactions: \textbf{alkali-silica reaction (ASR)} and \textbf{alkali-carbonate reaction (ACR)}. The most common type is ASR, which is of more concern than ACR because aggregates containing reactive silica materials are more common. Cases of ACR are very rare and restricted to a few isolated regions (FHWA).

In the Alkali-Silica Reaction, aggregates composed of certain forms of silica can react with alkali hydroxide in concrete to form a gel that expands as it absorbs water from the surrounding cement paste or the environment. As the gel grows, it can produce enough pressure to cause concrete damage, as represented by the following equations:

\begin{align}
	\text{Alkalis} + \text{Reactive Silica} &\rightarrow \text{Gel Reaction Product} \\
	\text{Gel Reaction Product} + \text{Moisture} &\rightarrow \text{Expansion}
\end{align}

For the alkali-silica reaction (ASR) to occur, three essential components must be present:
\begin{itemize}
	\item Reactive forms of silica in the aggregate
	\item High-alkali pore solution
	\item Sufficient moisture
\end{itemize}

The use of reactive aggregates in concrete is one of the conditions for ASR to happen. These aggregates tend to break down under exposure to highly alkaline pore solution in concrete and consequently react with alkali hydroxides (sodium and potassium) to form the ASR gel.

Another condition is the high-alkali-content pore solution. Alkali hydroxides in solution will react readily with reactive forms of silica in the aggregate. As the aggregate reactivity increases, gel reaction products can be formed with lower concentrations of alkali. That is why the use of low-alkali cements alone may not be sufficient to control ASR with highly reactive aggregates.

The potential for ASR increases as the alkalinity of the pore solution increases. If the alkali concentration is high enough, alkali hydroxides can break stronger silicon bonds found in less reactive aggregates to form the gel. This is the reason why some non-reactive aggregates occasionally show ASR.

The third condition for ASR to occur is sufficient moisture. It has been found that concrete with highly reactive aggregates and high-alkali cements has shown little or no expansion in very dry environments (FHWA). Similarly, differences in moisture levels in various parts of the same concrete structure can result in different outcomes. Parts of the structure exposed to constant moisture have shown significant ASR damage, whereas dry areas showed little or no damage.

Consequently, the exposure conditions and the presence of moisture in the concrete structure play a key role in ASR. The most common indicators of ASR are:
\begin{itemize}
	\item Map cracking
	\item Closed joints
	\item Concrete surface spalling
\end{itemize}

Usually, the cracks appear in areas with constant moisture. Therefore, to avoid ASR, it is recommended to keep concrete structures as dry as possible. The reaction can be virtually stopped if the internal relative humidity of the concrete is kept below 80\% (FHWA), though this is difficult to achieve and maintain.

Potentially reactive aggregates are present throughout the country. However, ASR damage to concrete is not common due to measures taken to control it. Moreover, not all ASR gel reaction products experience excessive and destructive swelling.

\subsubsection{Service Life Prediction}

\section*{Service Life of Concrete with Alkali-Aggregate Reactivity}

There is not much literature on the service life of concrete experiencing alkali-aggregate reactivity (ASR). The Delaware Department of Transportation (DelDOT) developed a method for determining the rating of ASR with the purpose of generating an approximate ASR reaction rate to predict concrete pavement life. The ASR rating method was developed using an ASR test that tracked the progression of ASR over time. The evaluation timeframe was set to 20 years.

The ASR rating consists of six levels from 0 to 6: none, low, average, moderate, heavy, and severe, respectively. The ASR rating data from DelDOT were plotted against time, and the ASR reaction rate was derived from the curve. Since DelDOT does not regularly inspect pavement conditions, ASR testing is often limited to a single core sample. Consequently, the ASR reaction rate was assumed to follow a linear trend, defined by two points: the time the core is tested, $t_t$, and the initial placement time of the pavement, $t_0$.

The ASR reaction rate was defined as follows:

\begin{equation}
	\text{ASR Reaction Rate} = \frac{\text{ASR Rating \#}}{t_t - t_0}
	\label{eq:asr_rate}
\end{equation}

where:
\begin{itemize}
	\item $t_t$ = date of coring (years),
	\item $t_0$ = date pavement was placed (years),
	\item ASR Rating \# = ASR rating of the core at $t_t$.
\end{itemize}

Once the ASR reaction rate is estimated, the approximate prediction of the remaining service life can be computed using the equation:

\begin{equation}
	\text{Years Remaining} = \frac{5.0 - \text{ASR Rating \#}}{\text{ASR Reaction Rate}}
	\label{eq:years_remaining}
\end{equation}

This model allows for a basic estimation of future deterioration due to ASR based on limited historical data.

\subsection{Fatigue Life Prediction for Steel Bridge}
\subsubsection{Fatigue Life Prediction}
Fatigue is the process of cumulative damage of a material subjected to cyclic loading resulting in microscopic cumulative damage until a crack appears. There are two prevalent methods for characterizing the fatigue resistance of a structure component. The resistance can be determined experimentally following a stress-based approach (S-N curves) or numerically from a fracture-mechanics model.

\paragraph{Stress-Based Approach}

The classical experimental method to characterize fatigue resistance is to test a given detail in tension at a constant-amplitude stress range and count the number of cycles to failure. Nominally identical details are tested at different constant-amplitude stress ranges until failure. The stress range, $S_r$, is defined as the difference between the highest and lowest values in a stress history.

After testing the detail at various stress ranges, the data can be plotted on a graph of the number of cycles to failure ($N_f$) for various cyclic stress ranges ($S_r$). The data are typically plotted on a log-log plot. It was determined empirically that the stress range and type of detail are the primary variables affecting fatigue resistance. 

It is important to note that the stress range is typically the nominal stress range. The flow of stresses at discontinuities and/or welds creates locations of stress concentrations that lead to higher magnitudes than those calculated from engineering mechanics (e.g., bending stress, axial stress). Because the stress concentrations vary with the detail, the nominal stress (stress calculated away from the discontinuities and/or welds) is typically used to characterize the S-N curves.

\begin{figure}[H]
	\centering
	\includegraphics[width=0.7\linewidth]{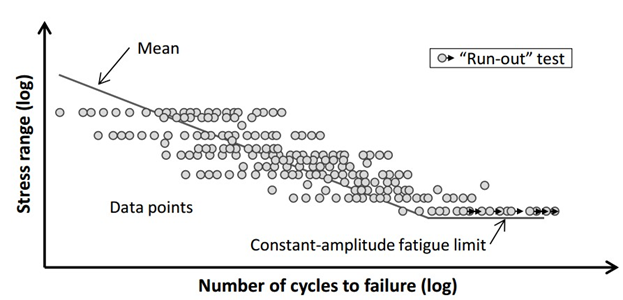}
	\caption{Example of data from a representative fatigue test (Fasl, 2013)}
	\label{fig:figure31}
\end{figure}

\section*{Fatigue Life Prediction Models}

By considering a single type of detail, the fatigue resistance can be described by:

\begin{equation}
	N_f = C \times S_r^B
	\label{eq:SN_general}
\end{equation}

where  
\begin{itemize}
	\item $N_f$ = number of cycles until failure at $S_r$;
	\item $C$ = empirical constant for specific detail from fatigue data (ksi$^B$);
	\item $S_r$ = nominal constant-amplitude stress range (ksi);
	\item $B$ = slope of the S-N curve.
\end{itemize}

The constants $B$ and $C$ are determined empirically from fatigue test data. Regression techniques are used to define the line that passes through the mean of the data. For most metals, $B$ typically ranges from $-2$ to $-4$ (with $-3$ commonly used in codes). The code value is determined by shifting the mean failure line approximately two standard deviations downward, corresponding to a 95\% confidence interval with approximately 5\% probability of failure.

There is a stress range, known as the constant-amplitude fatigue limit (CAFL), below which the detail is assumed to have infinite fatigue life. The CAFL is typically defined at 2 million cycles, beyond which specimens are considered “run-out” tests. Although this limit was originally based on older equipment limitations, modern testing allows higher cycle counts.

The fatigue life equation used in the AASHTO LRFD Specifications is:

\begin{equation}
	N_f = \frac{A}{S_r^3}
	\label{eq:SN_AASHTO}
\end{equation}

where  
\begin{itemize}
	\item $N_f$ = number of cycles until failure at $S_r$;
	\item $A$ = fatigue constant for the detail category (ksi$^3$);
	\item $S_r$ = constant-amplitude stress range (ksi).
\end{itemize}

AASHTO classifies fatigue details into categories.  
For example:
\begin{itemize}
	\item Category A: Base metal (flat plate without welded attachments);
	\item Category B: Continuous longitudinal fillet welds.
\end{itemize}

These categories have high CAFLs and large $A$ values, so they rarely govern fatigue design. Instead, fatigue considerations are more critical for details with discontinuities or attachments, such as fillet or groove welds, especially those aligned parallel or perpendicular to applied stresses. These details typically fall into Categories C through E$'$.

If the effective stress range is less than the CAFL, fatigue failure will not occur. If it is greater, failure may still happen. For stress ranges below the CAFL, a straight-line extension of the S-N curve can be conservatively applied.

\subparagraph{Miner’s Rule }

The S-N curve uses a constant-amplitude stress range. However, real structures are subjected to varying-amplitude stress ranges. A cumulative damage theory is needed to relate the varying-amplitude cycles to the constant-amplitude fatigue data. Miner’s rule is the most commonly used cumulative damage theory because it is simple and agrees well with historic fatigue data. The rule follows a linear-damage hypothesis and is expressed by the following equations:

\begin{equation}
	D_j = \frac{n_j}{N_{f,j}}
\end{equation}
\begin{equation}
	D = \sum_{j=1}^{k} D_j \tag{109}
\end{equation}

where,  
$D_j$ = contribution of cycles $n_j$ to Miner’s damage accumulation index;  
$D$ = Miner’s damage accumulation index;  
$n_j$ = number of cycles measured at $S_{r,j}$;  
$k$ = number of different stress ranges;  
$N_{f,j}$ = number of cycles until failure at $S_{r,j}$.

The damage from a spectrum of stress ranges equals the damage from a single, effective stress range as shown in the equation below:

\begin{equation}
	S_{re} = \left( \sum_{j=1}^{k} \frac{n_j}{N_m} S_{r,j}^3 \right)^{1/3} \tag{110}
\end{equation}

where,  
$S_{re}$ = effective stress range (ksi);  
$N_m$ = total number of cycles measured.

\subparagraph{Cycle-Counting }

When field measurements are utilized, a cycle-counting method is needed to transform the stress history into a histogram of stress amplitudes. Four types of counting methods for fatigue analysis are included in ASTM E1049 (ASTM, 1999): (1) level-crossing counting, (2) peak counting, (3) simple-range counting, and (4) rain flow counting or related methods.

\subparagraph{AASHTO Manual for Bridge Evaluation (2011) }
In 2011, AASHTO released the second edition of the \textit{Manual for Bridge Evaluation} (AASHTO, 2011) for evaluating fatigue in steel bridges. The following equation can be used to estimate the remaining fatigue life:

\begin{equation}
	Y = \frac{R_R \times A}{365 \times n \times \left(\text{AADT}_{SL} \times (R_s \times S_r)^3 \right)} - a \tag{111}
\end{equation}

where,  
$Y$ = fatigue life in years;  
$R_R$ = resistance factor;  
$A$ = detail constant ($\text{ksi}^3$);  
$n$ = stress cycles per truck passage;  
$\text{AADT}_{SL}$ = average number of trucks per day in a single lane, averaged over the entire fatigue life;  
$S_r$ = stress range (ksi);  
$R_s$ = partial load factor.

% Bibliography
\bibliographystyle{unsrtnat}
\bibliography{references}

\end{document}